\documentclass[12pt]{article}

\usepackage{amsmath}
\numberwithin{equation}{section}
\usepackage{amssymb}
\usepackage[margin=3cm]{geometry}
\usepackage{graphicx}
\usepackage{feynmp}
\usepackage{cite}
\usepackage{hyperref}

\def\be{\begin{equation}}
\def\ee{\end{equation}}
\def\ba{\begin{eqnarray}}
\def\ea{\end{eqnarray}}	
\def\l{\left}
\def\r{\right}
\def\fr{\frac}
\def\la{\label}
\def\d{\partial}

\newcommand{\sfrac}[2]{{\textstyle\frac{#1}{#2}}}

\begin{document}
   
\begin{fmffile}{graphs}

\setcounter{page}0
\def\ppnumber{\vbox{\baselineskip14pt
%\hbox{hep-th/0000000}
}}
\def\ppdate{\footnotesize{}} \date{}

\author{Bart Horn,\thanks{bh2478@columbia.edu}\, Alberto Nicolis,\thanks{a.nicolis@columbia.edu}\, Riccardo Penco\thanks{penco@phys.columbia.edu}\\
[7mm]
{\normalsize \it Physics Department and Institute for Strings, Cosmology, and Astroparticle Physics,}\\
{\normalsize  \it Columbia University, New York, NY, 10027, USA}\\
[3mm]}

\bigskip
\title{\bf  Effective string theory for vortex lines \\ in fluids and superfluids
\vskip 0.5cm}
\maketitle

\thispagestyle{empty}

\begin{abstract}
\noindent We discuss the effective string theory of vortex lines in ordinary fluids and low-temperature superfluids, by describing the bulk fluid flow in terms of a two-form field to which vortex lines can couple. We derive the most general low-energy effective Lagrangian that is compatible with (spontaneously broken) Poincar\'e invariance and worldsheet reparameterization invariance.  This generalizes the effective action developed in \cite{Lund:1976ze, Endlich:2013dma}.  
By applying standard field-theoretical techniques, we show that certain low-energy coupling constants---most notably the string {\em tension}---exhibit RG running already at the classical level. We discuss applications of our techniques to the study of Kelvin waves, vortex rings, and the coupling to bulk sound modes.

%We discuss the coupling of vorticose sources to sound in the limit of small vortex rings and indicate how this procedure can be systematically generalized to include higher multipoles.
\end{abstract}
\bigskip
\newpage

\tableofcontents

\vskip 1cm

\section{Introduction}

%
%Effective field theory and string theory have been usefully applied to a wide range of problems in classical physics, both for constructing a systematic perturbation theory and for making manifest the role of symmetry.  One useful application is to the dynamics of vortex lines and rings in superfluid helium, and their relationship to microscopic excitations such as phonons and rotons.  Such studies constitute a direct application of weakly coupled strings to experimentally accessible low-energy systems, complementing the application of string theory to condensed matter via holography and the AdS/CFT correspondence.

For zero-temperature superfluids the only allowed vortex configurations are string-like objects, the so-called vortex lines. These have a quantized circulation, and a microscopic, atomic-size thickness \cite{landau9}. For ordinary fluids one can have much more general vortex configurations, but it is still possible (and fairly easy) to set up a long-lived string-like vortex, with a thickness that is much smaller than its other typical length scales. In both cases, the position and shape of a vortex line is a placeholder for a fairly complicated bulk fluid flow: vorticity is localized on the line, but the velocity field away from it is non-trivial (albeit irrotational). For instance, for a straight line it circulates with a $1/r$ profile. 

As a result, the equation of motion for the line itself is a complicated integro-differential (and therefore non-local) equation, which is typically attacked numerically (see e.g.\cite{barenghi2009review, barenghi06}).
It is natural to expect, however, that such a complicated equation of motion can be replaced by a {\em local} effective action for a string coupled to the bulk modes of the fluid, with the usual conceptual and practical advantages that such a transition---from equations of motion to Lagrangian and from non-local action at a distance to local interaction with fields---entails.

In the present paper we systematically develop such an effective string theory, generalizing and completing the program begun in \cite{Lund:1976ze, Endlich:2013dma}.  This Lagrangian was also studied in the nonrelativistic limit in the related work \cite{lund1991defect,2014PhRvA..90e3617L,Gubser:2014yma}, where it was derived from the Gross-Pitaevskii model, and in particular it was used in~\cite{Gubser:2014yma} to analytically study the instability modes of propagating vortex rings.  As we will see, the irrotational bulk fluid flow can be described in terms of a two-form field, whose excitations decompose under an appropriate choice of gauge into sound waves and the non-dynamical `hydrophoton' field.  The two-form can be coupled to (1+1)-dimensional vortex line defects through a Kalb-Ramond term, and the energetics and other microscopic properties of the string can be encoded in a derivative expansion generalizing the familiar Nambu-Goto action.  Our first result is a general effective action compatible with the symmetries of the system:  Poincar\'e invariance, which is spontaneously broken by the medium, reparameterization invariance on the string worldsheet, and the gauge invariance associated with the two-form field.  
%Using this we can systematically enumerate the couplings of vorticose sources to sound as a function of fluid and source velocity.

For strings living in empty space rather than in a medium,
the `bottom-up' picture for effective string theory was investigated in \cite{Polchinski:1991ax} and more recently considered in the context of QCD flux tubes in \cite{Dubovsky:2012sh, Hellerman:2014cba}.  It is worth stressing that while a fundamental (i.e., UV-complete) theory of strings is only consistent with Lorentz invariance at the quantum level in 26D (or 10D with worldsheet supersymmetry), a low-energy effective theory of strings is possible in any number of dimensions.  This effective theory will be valid at distances much greater than the core size, and higher derivative corrections will encode information about the microscopic degrees of freedom making up the string.

Once the system is described in terms of a local effective action, one can apply standard field theoretical ideas and techniques to analyze it. We will discuss how renormalization works in our case, and how to use the resulting RG running of  low-energy couplings to streamline certain classic computations as well as to  efficiently perform new ones. 
In fact, we will see that for many purposes the bulk modes can be integrated out; since they are gapless, however, the resulting worldsheet effective theory is not local. The non-locality is very mild though, and can be phrased as a simple RG evolution for certain couplings localized on the worldsheet. In particular, the classic logarithms appearing in a number of physical quantities concerning vortex lines \cite{barenghi2009review, donnelly1991quantized}---from their energy per unit length to the spectrum of Kelvin waves---can be understood in this way.  Such logarithmic running at the classical level arises quite generally when the dynamics of a codimension-two brane---which in our case is just the worldsheet spanned by the vortex line---couples to fields in the bulk\cite{Goldberger:2001tn, deRham:2007dg}.

%In \S 2 we dualize the $P(X)$ Lagrangian for superfluids to a two-form field description, and we recover the actions of \cite{Lund:1976ze, Endlich:2013dma} in the appropriate limit.  In \S 3 we derive the higher derivative corrections to the vortex string action from the perspective of symmetry, using the coset construction in a particular gauge and then restoring Lorentz and worldsheet reparameterization invariance.  In \S 4 we discuss this action in the limit of small vortex rings, and compute the leading $v^2/c^2_s$ corrections to the potential interaction between two vortex rings.  We conclude in \S 5 and indicate further directions.
%
We should emphasize that although for this paper we will be mostly interested in the classical dynamics of vortex lines, it is straightforward to apply our formalism to problems at the quantum level as well, and we will present a sample quantum computation in sect.~\ref{sec:phonon absorption}.
Moreover, in the following we will refer almost exclusively to the superfluid case, but everything we say (apart from quantum effects) applies to ordinary fluids as well. In particular, for irrotational fluid flows there is a duality between superfluids and ordinary fluids directly at the level of the Lagrangian \cite{Dubovsky:2005xd}, and so in sect.~\ref{two form} we start directly with the superfluid Lagrangian, with the understanding that that covers both cases. 
It should be mentioned that, for ordinary fluids, vortex lines will eventually decay away due to viscosity (like any other type of fluid flow). Within the regime of validity of the hydrodynamical description, however, viscosity effects are of higher order in the derivative expansion, and thus negligible in the first approximation. In this limit Kelvin's theorem holds, and as a result the circulation of a vortex line is conserved in time and along the line, and the line thickness stays small.

Our paper is long, and not all readers will be interested in all of it. We feel that sects.~\ref{reparameterization}, \ref{sec: power counting}, and \ref{sec:small} can be omitted without impacting the general flow, although sect.~\ref{sec:non-relativistic limit} will be particularly relevant for readers interested in the non-relativistic case.

\vspace{.5cm}
\noindent
{\em Conventions:}
We use the  $(-+++)$ metric signature and $\hbar = c =1$ units throughout the paper.

%%%%%%%%%%%%%%%%%%%%
%%%%%%%%%%%%%%%%%%%%
\section{Two-form description of superfluids} \la{two form}
From a QFT standpoint, superfluids are systems in which a spontaneously broken $U(1)$ charge $Q$ is at finite density \cite{Nicolis:2011pv}. At zero temperature and at sufficiently large distances and time-scales, their dynamics are dominated by the single Goldstone excitation (the \emph{phonon}) that follows from the breaking of $Q$. In the relativistic case, the low-energy effective action for the Goldstone $\pi$ can be written in the following compact form~\cite{Son:2002zn}
\be \la{F(X)}
S = \int d^4 x\, P(X), \qquad \quad X= - \d_\mu \phi \d^\mu \phi, \qquad \quad \phi = \bar \mu  t + \pi,
\ee
where the sign in the definition of $X$ is chosen as to make $X$ positive for our choice of signature, $\bar \mu$ is the equilibrium chemical potential  for $Q$,\footnote{Throughout the paper we will denote all equilibrium quantities with a bar, to distinguish them from the local values the same quantities can take in the presence of fluctuations.} and $P$ is an a priori arbitrary function whose precise form is determined by the superfluid equation of state. Notice that the background value $\langle \phi \rangle = \bar \mu t$  breaks Lorentz invariance. This is a consequence of the fact that for the superfluid---like for any other condensed matter system---there is a preferred reference frame:\ the one in which the system is at rest. Starting from the action (\ref{F(X)}), it is easy to check that the stress-energy tensor is that of a fluid with energy density, pressure and 4-velocity given respectively by
\be  \la{hydroX}
\rho = 2 X P'(X) - P(X), \qquad \qquad p = P(X), \qquad \qquad u_\mu = -\fr{\d_\mu \phi}{\sqrt{X}}.
\ee
Since the 4-velocity $u_\mu$ is the gradient of a scalar (up to a normalization factor), it obeys a relativistic version of the irrotationality condition and thus describes potential flow---as befits a superfluid.

The effective description of a relativistic superfluid provided in (\ref{F(X)}) is very economical, in that it makes use of a {single} scalar field $\phi$ to describe the dynamics of a Goldstone $\pi$. However, this is not the only possibile description. It is in fact known~\cite{Cremmer:1973mg}  that in $3+1$ dimensions, the theory of a scalar field whose action is invariant under a {global} shift symmetry $\phi \to \phi + c$ admits a dual formulation based on a 2-form field $A_{\mu\nu}$ whose action is invariant under {local} gauge transformations of the form
\be \la{gauge transf}
A_{\mu\nu} \to A_{\mu\nu} + \d_\mu \xi _\nu - \d_\nu \xi_\mu
\ee
(see also~\cite{Buchbinder:1992rb} for a pedagogical derivation).\footnote{In $d+1$ dimensions, the scalar field is dual to a $(d-1)$-form $A_{\mu_1 ... \mu_{d-1}}$.} 
For a non-linear theory such as the one we are considering, the effective action for the dual theory is  
\be \la{G(Y)}
S = \int d^4 x \, G(Y), \qquad \qquad Y = - F_\mu F^\mu, \qquad \qquad F^\mu = \sfrac{1}{2}\epsilon^{\mu\nu\lambda\rho}\d_\nu A_{\lambda \rho} \; .
\ee
The gauge-invariant quantity $F^\mu$ is the analogue for a 2-form field of the dual electromagnetic field strength $\tilde F^{\mu\nu} = \epsilon^{\mu\nu\lambda\rho} \d_\lambda A_\rho$, and the function $G$ is in one-to-one correspondence with our function $P$, as we now explain.

Starting from (\ref{G(Y)}),  it is easy to show that the stress-energy tensor is still that of a fluid, but now with
\be \la{hydroY}
\rho = -G(Y), \qquad \qquad p = G(Y) - 2 Y G'(Y), \qquad \qquad u_\mu = -\fr{F_\mu}{\sqrt{Y}}.
\ee
The precise relation between the scalar $\phi$ and the 2-form $A_{\mu\nu}$ then can be obtained by comparing the expressions for the 4-velocity in (\ref{hydroX}) and (\ref{hydroY}),
\be \la{relation}
-\fr{\d_\mu \phi}{\sqrt{X}} = -\fr{F_\mu}{\sqrt{Y}},
\ee
as well as the expressions for the density and pressure. We find that the functions $P(X)$ and $G(Y)$ are related simply by a Legendre transform (when expressed as functions of $\sqrt{X}$ and $\sqrt{Y}$): 
\begin{align} 
\sqrt{Y} = \frac{dP}{d\sqrt{X}} \; , \qquad & G(Y) = P (X) - \sqrt{X} \frac{dP}{d\sqrt{X}}        	\la{legendre1} \\
\sqrt{X} = - \frac{dG}{d\sqrt{Y}}\; , \qquad & P (X) = G (Y) -  \sqrt{Y}  \frac{dG}{d\sqrt{Y}}  	\; .	\la{legendre2}
\end{align}
From a thermodynamical viewpoint, $\sqrt X$ and $\sqrt Y$ are the most natural variables to use:\ they are  the local chemical potential and  number density,
\be \la{mun}
\mu = \sqrt X \; ,\qquad  n = \sqrt Y \; .
\ee
Then, $P$ expresses the pressure as a function of $\mu$, and $G$ expresses (minus) the energy density as a function of $n$. Our Legendre-transform relations above correspond to the standard zero-temperature thermodynamic identities
\be
dp = n \, d \mu \; , \qquad  d \rho = \mu \,  dn \; , \qquad \rho + p = \mu n \; . 
\ee 

From a field-theoretical viewpoint, the important point to stress is that although eqs.~\eqref{relation}--\eqref{legendre2} provide a local relation between the \emph{derivatives} of $\phi$ and $A_{\mu\nu}$,  the corresponding relation between the two fields is  highly non-local. This is a standard feature of dualities in field theory, and means that if one picture admits local terms in the Lagrangian in which some fields appear without derivatives, then these very same terms will look highly non-local in the dual picture. This is the reason why we are introducing the two-form formulation of the superfluid effective theory in the first place: it turns out that the most relevant local coupling between phonons and vortex lines involves an undifferentiated $A_{\mu\nu}$~\cite{Lund:1976ze,Zee:1994qw}, and thus cannot be easily rewritten in terms of the scalar~$\phi$.\footnote{Recently, the dual language was also used to write down a Wess-Zumino term for superfluids in 2+1 dimensions~\cite{Golkar:2014paa}. This term does not have a local counterpart in the $\phi$ language~\cite{Delacretaz:2014jka}.} 
From a microscopic perspective, this fact has a simple explanation: a vortex line is just a low-energy proxy for a topological defect in the $\phi$ effective theory, and in the presence of such a defect, $\phi$ becomes the winding angle and is not single-valued.
Something similar happens in electromagnetism, where the gauge potential is not single-valued in the presence of a magnetic monopole, but it is possible to write a local coupling between the monopole and the dual gauge potential $\tilde A_\mu$, which is related to the dual field strength  by $\tilde F_{\mu\nu} \equiv \d_\mu \tilde A_\nu - \d_\nu \tilde A_\mu$.

Before turning to the study of vortex lines though, we will explicate the duality between $\phi$ and $A_{\mu\nu}$ by showing explicitly that the effective action (\ref{G(Y)}) describes a single gapless degree of freedom---the superfluid phonon.
In order to study the spectrum of excitations in the 2-form language, we first need to determine what background $\langle  A_{\mu\nu} \rangle$ corresponds to the Lorentz-violating background $\langle \phi \rangle = \bar \mu t$. To this end, we can evaluate equation (\ref{relation}) on the background to obtain (for the $\mu=0$ and $\mu=i$ components respectively)
\ba \la{bkgdconditions}
&\frac{1}{2}\epsilon^{ijk}  \d_i \langle  A_{jk} \rangle = - \bar n, \\
& \frac{1}{2}\epsilon^{ijk} ( 2 \d_j \langle A_{0k} \rangle - \d_0 \langle  A_{jk} \rangle) = 0,
\ea
where the parameter $\bar n \equiv \langle \sqrt{Y} \rangle$ is, according to (\ref{mun}), simply the background number density. The most general solution to the equations (\ref{bkgdconditions}) then is
\be \la{Abackground}
  \langle A_{0i} \rangle = -\sfrac{1}{2}\dot f_{ij}(t)x^j \, , \qquad \qquad \quad \langle A_{ij} \rangle = -\sfrac13 \bar n \,  \epsilon_{ijk} x^k + f_{jk}(t) \ ,
\ee
where $f_{ij}(t)$ is an arbitrary function of time only. On the one hand, it should not come as a surprise that the background (\ref{Abackground}) is not completely specified given that the action (\ref{G(Y)}) for $A_{\mu\nu}$ has a gauge invariance. On the other hand, we can use this gauge invariance to further simplify our result and set $f_{ij}(t)$ to zero via a large gauge transformation with parameters $\xi_i = \tfrac{1}{2} f_{ij}(t) x^j$. In conclusion, we find that a suitable background value for $A_{\mu\nu}$ is 
\be \la{bkgd}
 \langle A_{0i} \rangle = 0\, , \qquad \qquad \quad \langle A_{ij} \rangle= -\sfrac13 \bar n \,  \epsilon_{ijk} x^k \ .
\ee

Let us now study the linear behavior of fluctuations around such a background. We can parameterize them using two 3-vectors $\vec A$ and $\vec B$,
\be
 A_{0i} = \bar n \, A_i (t, \vec{x}), \qquad \qquad \quad  A_{ij} =  \bar n \, \epsilon_{ijk} \l[ - \sfrac13 x^k + B^k (t, \vec{x})\r] \; ,
\ee
where the normalization chosen is convenient for what follows.
Then, we will add to our action (\ref{G(Y)}) a gauge fixing term of the form
\be \la{gf}
S_{\rm gf} \propto -\fr{1}{2\xi} \int d^4 x \, (\partial_i A^{i \mu})^2 \; .
\ee
Notice that there is no real disadvantage in choosing a gauge fixing term that is not covariant, given that Lorentz invariance is spontaneously broken by the background (\ref{bkgd}) anyway. On the contrary, we will see that this gauge fixing term proves to be particularly advantageous in the  $\xi \to 0$ limit, where $\vec A$ and $\vec B$ lend themselves to a simple physical interpretation.
If we now expand in perturbations the action (\ref{G(Y)}) supplemented with the gauge fixing term (\ref{gf}), and we use the expression $Y = \bar{n}^2\big( (1-\vec \nabla \cdot\vec B)^2 - (\dot{\vec B} - \vec \nabla \times \vec A)^2 \big)$, we find the quadratic part of the action to be
\begin{align} 
S_{(2)} = \bar w \int \! d^4 {x}  \, & \Big\{ \sfrac12  (\vec{\nabla} \times \vec{A})^2 
+  \sfrac12  \big[ \dot{\vec{B}}^2 - c^{2}_{s} (\vec{\nabla} \cdot \vec{B})^2 \big]
\la{S2} \\
& -  \dot{\vec{B}}\cdot(\vec{\nabla} \times \vec{A}) - \sfrac{1}{2\xi} (\vec{\nabla} \times \vec{B})^2 
+ \sfrac{1}{2\xi} (\vec{\nabla} \cdot \vec{A})^2 \Big\} \; ,
\nonumber
\end{align}
where $\bar w = (\bar \rho + \bar p) = -2\bar{n}^2 G'(\bar{n}^2)$ is the background enthalpy density, and we have introduced the speed of sound squared $c_s^2 \equiv (2 Y G'' + G')/ G' = dp / d \rho$, also evaluated on the background. 

After switching to Fourier space, it is fairly straightforward to  invert the kinetic term to find the propagators. For arbitrary values of $\xi$, the final result looks quite complicated and involves non-vanishing mixed propagators of the form $\langle A^i B^j \rangle$. The explicit expressions as well as more details on the derivation can be found in Appendix~\ref{appa}. In the rest of the paper we will restrict ourselves to the  $\xi \to 0$ limit, in which case great simplifications occur and the propagator matrix becomes diagonal. The propagators for $\vec A$ and $\vec B$ are then respectively
\be \la{propagators}
%\parbox[t][3.6mm][b]{20mm}{
%\begin{fmfgraph*}(65,25) 
%\fmfleft{i1}
%\fmfright{o1}
%\fmfv{label=$A_i$,label.angle=-90}{i1}
%\fmfv{label=$A_j$,label.angle=-90}{o1}
%\fmfv{decor.shape=circle, decor.filled=full, decor.size=1.5thick}{i1,o1}
%\fmf{dashes}{i1,o1}
%\end{fmfgraph*}
%} \quad
\includegraphics[scale = 0.2]{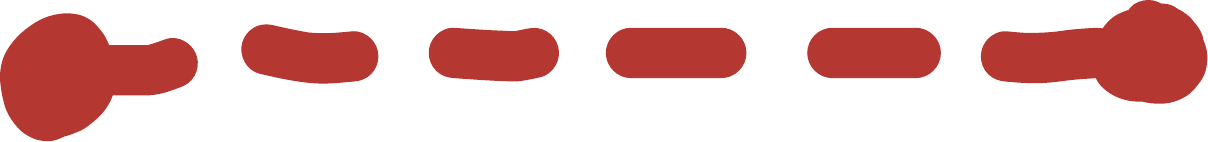} \, =\frac{1}{\bar w}\frac{i}{k^2 }\big(\delta^{ij} - {\hat k^i \hat k^j} \big), \qquad \quad 
%\parbox[t][3.6mm][b]{20mm}{
%\begin{fmfgraph*}(65,25) 
%\fmfleft{i1}
%\fmfright{o1}
%\fmfv{label=$B_i$,label.angle=-90}{i1}
%\fmfv{label=$B_j$,label.angle=-90}{o1}
%\fmfv{decor.shape=circle, decor.filled=full, decor.size=1.5thick}{i1,o1}
%\fmf{wiggly}{i1,o1}
%\end{fmfgraph*}
%} \quad phonon-propagator.pdf
\includegraphics[scale = 0.2]{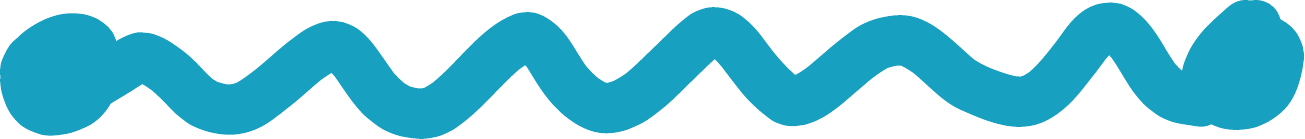} \,= \frac{1}{\bar w} \frac{i }{\omega^2 - c^2_s k^2} {\hat k^i \hat k^j}  \; ,
\ee
\vspace{.05cm}
where the usual $i \epsilon$ prescription is understood.

Taking the  $\xi \to 0$ limit amounts to imposing $\partial_i A^{i \mu} = 0$, which is the 2-form analogue of the Coulomb gauge in electromagnetism. In terms of our 3-vectors $\vec A$ and $\vec B$, this is equivalent to demanding that $\vec{\nabla} \times \vec{B} =  \vec{\nabla} \cdot \vec{A} = 0$. Thus, in this gauge, $\vec B$ is purely longitudinal and $\vec A$ is purely transverse. Moreover, we see from (\ref{propagators})  that in this gauge $\vec B$ describes the only propagating degree of freedom---the phonon---whereas $\vec A$ is a Coulomb-type constrained field that does not propagate any additional degree of freedom:\ it is (minus) the `hydrophoton' of ref.~\cite{Endlich:2013dma}. 

When switching to a different gauge, $\vec A$ and $\vec B$ change as
\be
\vec A \to \vec A +  \dot {\vec \xi} - \vec \nabla \xi_0 \equiv{\vec A} +  \dot {\vec \xi} \; , \qquad \vec B \to \vec B + \vec \nabla \times \vec \xi \; ,
\ee
where we have reabsorbed the $\xi_0$ contribution to the gauge variation of $\vec A$ into the longitudinal part of $\vec \xi$; such a redefinition does not affect the gauge variation of $\vec B$.  Physical quantities like those of eq.~\eqref{hydroY} are gauge invariant, and thus must be functions of the gauge-invariant combinations
\be
\vec \nabla\cdot \vec B \; , \qquad\quad \dot{\vec B} - \vec \nabla \times \vec A \; .
\ee
For instance, the four-velocity field is given by
\be
u^\mu(x) \propto \big(1 - \vec \nabla\cdot \vec B, \, \dot{\vec B}-\vec \nabla \times \vec A \, \big)  \; ,
\ee
suitably normalized.
By comparing with the standard form $u^{\mu} = \gamma_u \, (1, \vec{u} \,)$, we see that the three-velocity is
\be\label{fluid3velocity}
\vec{u} = \frac{\dot{\vec B}-\vec \nabla \times \vec A}{1 - \vec \nabla\cdot \vec B} \; .
\ee

%%%%%%%%%%%%%%%%%%%%%%%%%%%%
%%%%%%%%%%%%%%%%%%%%%%%%%%%%
\section{Effective action for vortex lines}\label{action for lines}

As we have seen, to lowest order in derivatives the bulk dynamics of a superfluid are described by the action \eqref{G(Y)} for a two-form field $A_{\mu\nu}$. If we now consider a thin string-like object living in the superfluid---such as a vortex line---to zeroth order in the core thickness we can parameterize its dynamics by the positions of its line-elements $\vec X(t, \sigma)$, where $\sigma$ is a coordinate along the string. As is the case for standard relativistic string theory, here too it is convenient to introduce reparameterization invariance for the time coordinate and use $X^\mu(\tau, \sigma)$ instead of the more physical---i.e., less redundant---$\vec X(t, \sigma)$, where $\tau$ and $\sigma$ are now two arbitrary world-sheet coordinates (more on this in the following section).

With $X^\mu$ we can now construct Lagrangian terms localized on the world sheet of the string.  As we did for the $\phi$ and $A_{\mu\nu}$ of last section, we can write arbitrary powers of $\d X$, but we need to work perturbatively in further derivatives. The reason is that the typical background configurations we will be interested in are of the form
\be
\phi \sim \bar \mu t \; , \qquad A_{\mu\nu} \sim \bar n x \; , \qquad X^\mu \sim (\tau, \sigma) \; ,
\ee
which have {\it large} first derivatives, but vanishing second derivatives, and so our power-counting scheme is the correct one for perturbation theory about these backgrounds. The symmetries that we have to impose are Poincar\'e-invariance, gauge-invariance for $A_{\mu\nu}$, and world-sheet reparameterization invariance. In fact, there is no a priori reason why we should impose this last symmetry (apart from the fact that we are used to it). We will explicitly address the physics behind it in the next section, but for the moment we will take it for granted and move on. 

The whole point of trading $\phi$ for $A_{\mu\nu}$ in the last section was to write a local coupling between bulk modes and a string-like defect directly at the level of the potential (rather than the field-strength). Here it is:
\be
S \supset \lambda \int d \tau d \sigma A_{\mu\nu} \, \d_\tau X^\mu  \d_\sigma X^\nu    \; ,
\ee
where $\lambda$ is a coupling constant. Such a term is the only invariant that involves $A_{\mu\nu}$ without derivatives. It is the analogue of $  \int A_\mu dx^\mu$ for a point-charge in electromagnetism, and $\lambda$ thus plays the role of  a charge per unit length.

Furthermore, we can also write Nambu-Goto (NG)-type terms:
\be \label{NG with G}
S_{\rm NG} \propto \int   d \tau d \sigma \sqrt{-\det \big( G_{\mu\nu}(X) \, \d_\alpha X^\mu \d_\beta X^\nu \big) } \; ,
\ee
where $\alpha$ and $\beta$ run over $(\tau, \sigma)$, and $G_{\mu\nu}(x)$ is {\it any} bulk tensor that can play the role of a spacetime metric.
For standard relativistic strings in empty  space the only available structure is $G_{\mu\nu}(x) = \eta_{\mu\nu}$, which leads to the NG action. Here instead the underlying medium spontaneously breaks Lorentz-invariance, and we can use its four-velocity
\be \la{UY}
u^\mu(x) = -\frac{F^\mu}{\sqrt{Y}}
\ee
as well as the scalar $Y = -F_\mu F^\mu$ to construct new tensors. For instance,
\be
G_{\mu\nu}(x) = \eta_{\mu\nu} + C(Y) \, u_\mu u_\nu 
\ee
is a perfectly fine ``metric" to use in \eqref{NG with G}. 
By varying $C(Y)$ we can apparently generate infinitely many inequivalent terms of the form \eqref{NG with G}, and it is not immediately clear what kind of general structure can emerge from taking their sum.
To address this question, we will use a standard result of bi-gravity theories~\cite{Hassan:2011zd}: given two metric tensors $g_{\alpha\beta}$ and $h_{\alpha\beta}$ in $D$ spacetime dimensions, the most general diff-invariant, zero-derivative Lagrangian one can write down is
\be \label{bigravity}
\int d^D x \sqrt{- \det g} \, f\big( (g^{-1} \cdot h)^\alpha {}_\beta \big) \; ,
\ee
where $f (M^\alpha {}_\beta)$ is a function that is invariant under similarity transformations $M \to S \cdot M \cdot S^{-1}$ but is otherwise generic. For our world-sheet, $D=2$ and we can take the two independent induced metrics to be
\be
g_{\alpha\beta} = \eta_{\mu\nu} \cdot \d_\alpha X^\mu \d_\beta X^\nu \; , \qquad h_{\alpha\beta} = u_{\mu} u_{\nu} \cdot \d_\alpha X^\mu \d_\beta X^\nu \; .
\ee
The fact that $h_{\alpha\beta}$ is degenerate (it is a rank-one matrix) does not impair the argument.

To figure out how many independent invariants of $(g^{-1} \cdot h)^\alpha {}_\beta$ we have, we can completely specify the coordinates of spacetime and of the string worldsheet. The number of independent nonzero components of $(g^{-1} \cdot h)^\alpha {}_\beta$ in any such basis is the number of independent invariants. For any given worldsheet point $({\tau_0, \sigma_0})$, we can perform a Lorentz boost and align the Minkowski time with the underlying fluid flow at that point, $u^\mu = \delta^\mu _0$. Then, we can perform a Lorentz rotation and make the $X^1$ direction tangent to the world-sheet at that point, and via a world-sheet diff we can make $\tau$ and $\sigma$ locally the same as $X^0$ and $X^1$, so that we have $\d _\alpha X^\mu = \delta_\alpha^\mu$. In conclusion, at that point, in these coordinates we have
\be
g_{\alpha\beta} = \eta_{\alpha\beta} \; , \qquad h_{\alpha\beta} = \delta_\alpha^0 \delta_\beta^0 \; ,
\ee
and $g^{-1} \cdot h$ has only one nonzero entry, the $00$ one. Therefore, we only have one independent invariant, which in a general coordinate system we can take to be the trace,
\be
g^{\alpha\beta} h_{\alpha\beta} \; ,
\ee
where $g^{\alpha\beta}$ is the inverse of $g_{\alpha\beta}$.
We should note that in our case the function $f$ in \eqref{bigravity} can also be supplemented with a scalar argument, our $Y = -F_\mu F^\mu$, which is invariant under all the symmetries. 

To summarize, the interactions of a relativistic superfluid and of a string-like defect living in it  can be modeled in terms of a two-form field $A_{\mu\nu}(x)$ and of the embedding coordinates of the string $X^{\mu}(\tau,\sigma)$. To lowest order in the derivative expansion and up to gauge-fixing terms, the low-energy effective action is the sum of three distinct structures:
\be\la{effS}
S = S_{\rm bulk} + S_{\rm KR} + S_{\rm NG'} \ ,
\ee
with
\begin{subequations} \la{SSS}
\begin{align}
S_{\rm bulk} & = \int d^4 x \, G(Y) \la{Sbulk} \\
S_{\rm KR} & = \lambda \int d \tau d \sigma  \, A_{\mu\nu} \, \d_\tau X^\mu  \d_\sigma X^\nu \la{KR} \\
S_{\rm NG'} & = - \int d \tau d \sigma \,  \sqrt{- \det g} \, \mathcal{T}\big( g^{\alpha\beta} h_{\alpha\beta}, Y \big) \; . \label{NG'}
\end{align}
\end{subequations}
The first piece encodes the bulk dynamics of the superfluid. The a priori generic function $G(Y)$ is completely determined by the equation of state, via the relation
\be
\rho = - G(n^2) \; ,
\ee
where $\rho$ and $n$ are the densities of energy and charge for the superfluid.
The second piece is a Kalb-Ramond-type interaction between the bulk degrees of freedom and the string. The third piece generalizes the Nambu-Goto action to our case, where the underlying medium breaks Lorentz invariance (spontaneously).
In particular, the function $\mathcal{T}$ is a functional generalization of the string tension. When expanded in perturbations about a background configuration, it yields a finite number of couplings at each order in perturbation theory, which in principle can be fixed by experiment.

Before using the action above for a number of concrete computations, we would like to pause for a moment and go back to the question of world-sheet reparameterization invariance. Readers uninterested in this technical detour can safely skip to sect.~\ref{expansion}.

%%%%%%%%%%%%%%%%%%%%%%
%%%%%%%%%%%%%%%%%%%%%%
\section{World-sheet reparameterization invariance} \la{reparameterization}

As is the case for gauge symmetries in general, reparameterization invariance is more properly thought of as a statement of redundancy rather than one of symmetry. In most physically relevant situations,  gauge redundancy is a property of a gauge field. This may or may not be a dynamical field. For example, the former class includes the physical electromagnetic  and gravitational fields, and the latter class includes the non-dynamical metric we introduce to describe non-gravitational field theories in curved space in a coordinate-independent fashion.
However, for a string with tension $T$ described by the Nambu-Goto action,
\be \label{NG}
-T\int   d \tau d \sigma \sqrt{-\det g} \; , \qquad\qquad  g_{\alpha \beta} \equiv \eta_{\mu\nu} \, \d_\alpha X^\mu \d_\beta X^\nu  \; ,
\ee
the role of such a gauge field is played by the world-sheet induced metric. Since the induced metric is a given functional of the embedding fields $X^\mu ({\tau, \sigma})$ and nothing else,
%\be
%g_{\alpha\beta} = \eta_{\mu\nu} \cdot \d_\alpha X^\mu \d_\beta X^\nu \; ,
%\ee
the redundancy associated with reparameterization invariance must now be a property of those fields. And it is a {\it physical} property, in the sense that it relies on certain (implicit) physical assumptions about the object  that the NG action is supposed to  describe. To identify what these are, it is convenient to analyze reparameterizations of $\tau$ and $\sigma$ in turn.

Reparameterization invariance for $\tau$ is just a  convenient technical trick to implement {\it manifest} Lorentz invariance. This is already evident for a relativistic point particle with mass $m$ and trajectory $\vec X(t)$, whose action can be written in two physically equivalent forms:
\be
 -m \int dt \sqrt{1- \big(\sfrac{d \vec X}{dt}\big)^2} =  -m\int d\tau \sqrt{-\eta_{\mu\nu} \, \sfrac{d X^\mu}{d\tau} \sfrac{d X^\nu}{d \tau}} \; .
\ee
The first form only involves the physical field $\vec X(t)$, but at the same time obscures Lorentz invariance, which acts by mixing such a field with its argument---time. The second form accomplishes manifest Lorentz invariance, but it does so at the expense of introducing an arbitrary, Lorentz-scalar parameter $\tau$ along with a redundant field $X^0 (\tau)$. To work with the non-redundant degrees of freedom only, one can always choose the ``physical'' gauge, $\tau = X^0$, and end up with the first form of the action.

Reparameterization invariance for $\sigma$ is more interesting. To appreciate why, it is convenient to work in physical gauge for $\tau$ ($\tau = X^0 = t$), in which case the Nambu-Goto action reduces to a functional of the spatial position field $\vec X(t , \sigma)$:
\be
-T\int   d t d \sigma \sqrt{\big( \d_\sigma \vec X \big)^2-  \big(\d_\sigma \vec X \cdot \d_t \vec X \big)^2} \; .
\ee
This is still invariant under general, {\it time-dependent} reparameterizations of $\sigma$,
\be
\sigma \to \sigma'(\sigma, t) = \sigma + \xi(\sigma, t)\; ,
\ee
under which our dynamical field $\vec X(t, \sigma)$ shifts by
\be
\vec X \to \vec X + \xi  \, \d_\sigma \vec X \; ,
\ee
where we have kept terms up to first order in the transformation parameter $\xi$. 
At any given point along the string, $\d_\sigma \vec X$ is a vector locally tangent to the string, and $\xi$ is an arbitrary function of time and $\sigma$. Invariance under the transformation above thus means that {\em motion and deformations along the string are unphysical}. This is a very physical, concrete statement. For instance, an infinite straight string formally oscillating in a longitudinal mode,
\be
\vec X (t, \sigma) = (\sigma + A  \cos(k \sigma - \omega t), 0, 0) \; ,
\ee
carries {\em no} energy associated with such oscillations: its action and energy are identical to those of the unperturbed configuration, $\vec X = (\sigma,0,0)$. Only the transverse oscillations carry energy. 

The physical origin of this behavior can be traced to the fact that the string in question does not break spacetime symmetries---in particular, Lorentz invariance---{\em along itself}. For instance, its stress-energy tensor is the lower-dimensional analogue of that of a cosmological constant. As a consequence, motion or deformations like compression or dilation along the string are unphysical:\ there is no way to move, compress, or dilate a cosmological constant. The fields $\vec X(t, \sigma)$, which transform non-trivially under spacetime symmetries, parameterize the physical Goldstone modes for the symmetries broken by the string---transverse translations, rotations, and boosts---but {\it have} to be redundant when it comes to parameterizing the Goldstone of a symmetry that is not broken in the first place.

Perhaps the best way to appreciate all this is through a counterexample---a concrete example of a string system that does not feature the properties above. We do not have to look very far: consider any ordinary string-like object in the real world, such as a violin string.
The material that makes up such a string is  a {\it solid}, and as a consequence, the symmetry breaking pattern of spacetime symmetries along the world-sheet must be that of a solid. In 1+1 dimensions, boosts and translations are broken along with an internal shift symmetry down to an unbroken combination that ensures the homogeneity of physical properties in the ground state (see e.g.~\cite{Nicolis:2013lma, Nicolis:2015sra} for recent reviews). The resulting string action to lowest orders in derivatives is:
\be \label{solid string}
\int   d \tau d \sigma \sqrt{-\det g } \, F\big(( \d \phi)^2 \big)\; ,
\ee
where $g_{\alpha\beta}$ is the same induced metric appearing in the NG action (\ref{NG}), and $\phi(\tau,\sigma)$ keeps track of the comoving coordinates of the solid: it is a world-sheet field specifying which solid line-element occupies position $\sigma$ at time $\tau$.
The implicit contraction in $( \d \phi)^2$ is done through $g^{\alpha\beta}$,  and $F$ is a function determined by the equation of state for the solid: in coordinates locally comoving with the solid ($\d_\alpha \phi \propto \delta_{\alpha}^1$), the world-sheet stress-energy tensor has~\cite{Endlich:2012pz}
\be \label{ws rho p}
\rho = -F \; , \qquad p = F  - 2 F' \cdot ( \d \phi)^2 \; .
\ee

Eq.~(\ref{solid string}) is clearly reparameterization invariant, with $\phi$ transforming as a world-sheet scalar, but it involves one extra degree of freedom $\phi$ compared to the Nambu-Goto action. Alternatively, one can work in so-called unitary gauge,
\be
\phi (\sigma, \tau) = \sigma \; ,
\ee 
which is a just a specific choice for the coordinate $\sigma$, and end up with an action that depends on the induced metric only,
\be
\int   d \tau d \sigma \sqrt{-\det g } \, F\big(g^{11}\big)\; .
\ee
However, now this action is not reparameterization invariant for $\sigma$. Either way, one has one more physical degree of freedom compared to a string described by the Nambu-Goto action. This additional degree of freedom corresponds to the longitudinal mode of motion or deformation of the string, which for a solid string is as physical as the transverse ones: it is the Goldstone mode associated with the spacetime symmetries broken {\it along} the string.

We can thus conclude that a string parameterized by the NG action can be thought of as a string made up of {\it cosmological constant}, which is quite different from being made up of an ordinary solid material. As a check, notice that if in (\ref{ws rho p}) we take the cosmological-constant limit, $p \to - \rho$, we get $F ' \to 0$, $\phi$ disappears from the action, and we recover the NG action. With hindsight, now we can also appreciate that reparameterization invariance for $\tau$ is not that automatic after all:\ it relies on the  implicit assumption that time-translations are unbroken on the world-sheet. A string made up of {\it supersolid} material \cite{Son:2005ak, Nicolis:2013lma, Nicolis:2015sra} would violate this assumption. Its low-energy effective action would be
\be
\int   d \tau d \sigma \sqrt{-\det g } \, F\big(( \d \phi)^2, (\d \psi)^2, \d \phi \cdot \d \psi \big)\; ,
\ee
where $\phi$ and $\psi$ are two world-sheet scalar fields, playing the roles of the solid comoving coordinate discussed above and of a superfluid scalar phase like that  discussed in sect.~\ref{two form}. The action is reparameterization invariant, but we have two additional degrees of freedom compared to the NG action: they correspond to the two types of gapless phonons---solid and superfluid---one can have in a supersolid. Alternatively, one can choose unitary gauge,
\be
\phi (\sigma, \tau) = \sigma \; , \qquad \psi(\sigma, \tau) = \tau \; ,
\ee
and work with an action that depends on the induced metric only, but that is not reparameterization invariant anymore,
\be
\int   d \tau d \sigma \sqrt{-\det g } \, F\big(g^{11},g^{00}, g^{01})\; .
\ee
In fact, it is the most general function of the induced metric.

In conclusion:\ as usual, reparameterization invariance can always be achieved by adding redundant degrees of freedom. However, there are situations in which it can be achieved by using only the degrees of freedom that are already at one's disposal, such as, for instance, our embedding fields $X^\mu(\tau,\sigma)$. In those situations,  reparameterization invariance is equivalent to the statement that some of {\it those} degrees of freedom are redundant---a property that may or may not be featured by the system under consideration.

So, what about our vortex lines? Vortex lines in fluids and superfluids are defined as loci of non-zero vorticity. The only quantitative measure of how large vorticity is on them is the circulation $\Gamma$, which is constant along each line.
% and---in the absence of viscosity---in time as well.
Such a geometric characterization does not associate any physical meaning to motion or deformation of the lines along themselves. In particular, only their shapes and their overall $\Gamma$'s are sufficient to reconstruct everywhere the incompressible part of the surrounding velocity field~\cite{donnelly1991quantized}. It is thus natural to postulate that their world-sheet action is reparameterization invariant already when using just the embedding fields $X^\mu(\tau,\sigma)$, precisely the same as for the NG action. 

We would like to stress that---however consistent and natural-sounding---this is still an {\it assumption}. It is not clear how to ascertain which symmetries are broken by the string along its worldsheet, since the surrounding medium is already breaking some of these longitudinal symmetries---e.g.~Lorentz boosts.
Certainly, given the discussion above, if we were to describe a solid string moving in a superfluid, we should give up $\sigma$-reparameterization invariance.  That is because we have an idea of what a solid is on its own, in the absence of the surrounding superfluid.  In the case of our vortex lines instead, there is no such thing as a vortex line without the surrounding superfluid. However, vortex lines in superfluids are intrinsically quantum objects, and their characterization purely in terms of semi-classical concepts like the vorticity of the fluid flow might well turn out to be incomplete. For instance, it is not obvious to us why quantum effects could not endow the string with some ``materiality", that is, solid-like physical properties which would imply the existence of  gapless longitudinal degrees of freedom. Ultimately, this is a question that has to be settled by experiment. For the time being, we content ourselves by noticing that for vortex lines in non-relativistic {\em classical} fluids, the reparameterization invariant action of \cite{Endlich:2013dma} (and reproduced here) yields the correct equations of motion as implied by the Euler equation.
We thus postulate reparameterization invariance for vortex lines in superfluids as well.

%%%%%%%%%%%%%%%%%%%%%%
%%%%%%%%%%%%%%%%%%%%%%

\section{Expansion of the action} \la{expansion}

To use the action we derived in sect.~\ref{action for lines} for concrete computations in perturbation theory, we should expand it in powers of the fields.
To begin with, it is important to realize that, within the regime of validity of the effective theory, the string can only move slowly. This is not to say that we should take the fully non-relativistic limit: the speed of sound $c_s$ can still be relativistic; but the local speed of the string has to be much slower than $c_s$. To see this, recall that the circulation $\Gamma$ is the line integral of the velocity field taken around the string. So, at distances $r$ from the string the fluid has a typical velocity $v \sim \Gamma/r$. Imposing that this be  sub-sonic all the way down to distances of order of the string core radius $r \sim r_c$, we get $\Gamma \lesssim c_s \, r_c$. Now, suppose that the string is perturbed---that is, curved---with some typical wavelength $\ell$. Up to logarithmic factors, the typical velocity of the string will be~\cite{donnelly1991quantized}
\be
\d_t \vec X \sim {\Gamma}/{\ell} \ll c_s \; ,
\ee
where we have used that, for our effective field theory to be valid, $\ell $ has to be much bigger than the string thickness $r_c$. 

We can thus expand our action \eqref{effS} in powers of $\d_t \vec X$. Notice that, thanks to the Kalb-Ramond coupling $S_{\rm KR}$, the expansion starts at {\it first} order in the velocities. This is quite different from standard mechanical systems in empty space, for which the kinetic action starts at quadratic order in the velocities, and it is the reason behind the peculiar mechanical behavior of a vortex line (see \cite{Endlich:2013dma} for a recent review). We also expand in powers of the fluctuations of $A_{\mu\nu}$,  parameterized by the $\vec A$ and $\vec B$ fields of sect.~\ref{two form}. The expansion is easy for the bulk and Kalb-Ramond terms in the action.
Working in physical gauge for $\tau$ ($\tau=X^0 = t$), we get
\begin{align}\label{actionslowstring}
S_{\rm bulk} \to \bar w & \int d^ 4x \, \Big[ \sfrac12 (\vec{\nabla} \times \vec{A})^2 + \sfrac12 \big( \dot{\vec{B}} \, ^2 - c^{2}_{s} (\vec{\nabla} \cdot \vec{B})^2 \big)  \nonumber \\ 
& \qquad \qquad + \sfrac12 (1 - c_s^2) \vec \nabla \cdot \vec B (\dot {\vec B} - \vec \nabla \times \vec A)^2 \nonumber \\
& \qquad \qquad + \left(\sfrac12 (1-c^2_s)-\sfrac23 g_3\right)(\vec \nabla \cdot \vec B)^3 + \dots  \Big]  \\
S_{\rm KR} \to \bar n \lambda & \int d t d \sigma \Big[ \epsilon_{ijk}\left(- \sfrac13 \, X^k + B^k \right) \d_t X^i \d_\sigma X^j +  A_i  \, \d_\sigma X^i \Big] \; , \label{KR expanded}
\end{align}
where we have kept up to first order in $\d_t \vec X$, up to first order in $\vec A$ or $\vec B$ on the world-sheet, and up to cubic order in $\vec A$ or $\vec B$ in the bulk.  
%The  $B \, \partial_t X \partial_{\sigma} X$ term is formally of second order in this expansion, but we keep it since it will be important for some of the processes we will consider below.  
(In section \ref{sec: power counting} we will study how to expand the action in a more systematic way.) The $\xi \to 0$ gauge fixing terms are understood, and we have introduced a shorthand notation for derivatives of the function $G(Y)$ evaluated on the background: 
\be
g_n \equiv Y^{n-1}\frac{G^{(n)}(Y)}{G'(Y)} \bigg|_{Y = \bar n^2}\; .
\ee
Without prior knowledge of the function $G$ (or of the equation state), the $g_n$'s should be taken as independent coupling constants, to be fixed by experiment.

Expanding the generalized Nambu-Goto term $S_{\rm NG'}$ requires more work. We first need to derive how the arguments of the function $\mathcal T(g^{\alpha \beta}h_{\alpha \beta}, Y)$ depend on the fields. For the second argument, we already know that 
\be
Y = \bar{n}^2 \Big[ \big(1 - \vec\nabla \cdot \vec B \, \big)^2 - \big(\dot{\vec B} - \vec \nabla \times \vec A \, \big)^2 \Big] \; .
\ee
On the other hand, the explicit form of the first argument is quite complicated, and it is more convenient to express it in terms of the fluid velocity $\vec u$ defined in eq.\ ~\eqref{fluid3velocity}. Introducing for notational simplicity the string velocity $\vec v = \partial_t \vec X$, we can write the first argument as follows:
\begin{equation}\label{ginversehinformuv}
\begin{split}
g^{\alpha \beta} h_{\alpha \beta} &= - \frac{u_{\mu}u_{\nu}\epsilon^{\alpha \gamma}\epsilon^{\beta \delta}\partial_{\alpha} X^{\mu} \partial_{\beta}X^{\nu}  \partial_{\gamma}X^{\lambda} \partial_{\delta}X_{\lambda}}{\det (\partial_{\sigma}X^{\rho}\partial_{\tau}X_{\rho})}\\
&= \frac{\gamma^2_u((1-\vec u \cdot \vec v)^2 + (\vec u \cdot \d_\sigma\vec X)^2 (-1 + v^2) - 2 (-1 + \vec u \cdot \vec v)(\vec u \cdot \d_\sigma\vec X)(\vec v \cdot \d_\sigma\vec X))}{1 - v^2 + (\vec v \cdot \d_\sigma\vec X)^2}\\
&= \frac{(1- \vec u_{\bot} \cdot \vec v_{\bot})^2 - u^2_{\|}(1-v^2_{\bot})}{(1-u^2)(1-v^2_{\bot})}
\end{split}
\end{equation}
In the last line we have decomposed $\vec u, \vec v$ into components which are locally parallel and perpendicular to the string.

If we now expand the generalized Nambu-Goto terms out to linear order in the fields $\d_t \vec X$, $\vec A$, or $\vec B$, we find 
\be
S_{\rm NG'} \to \int dt d\sigma |\partial_\sigma \vec X| \big[-T + 2T_{(01)}\vec\nabla \cdot \vec B +2 T_{(10)} \, (\dot{\vec B} - \vec\nabla \times \vec A)_{\bot} \cdot \vec{v}_{\bot} + \dots \big] \, , \label{NG' expanded}
\ee
where the coupling constants $T$, $T_{(01)}$ and $T_{(10)}$ are defined as follows. $T$ just denotes the value of our function ${\cal T}$ when evaluated on the background; it is the string tension. $T_{(mn)}$ denotes ${\cal T}$'s derivatives, also evaluated on the background, and normalized as
\be \label{Tmn}
T_{(mn)} \equiv a^m b^n \frac{\d^m}{\d a^m}\frac{\d^n}{\d b^n} {\cal T}(a,b) \; .
\ee
They all have units of tension, that is, energy per unit length.
Just as for the $g_{n}$'s, with no information on the function ${\cal T}$ in advance, $T$ and the $T_{(mn)}$'s should all be taken as independent coupling constants, to be fixed by experiment.   We will have more to say on the systematics of the effective field theory and the relative importance of various terms in subsequent sections.

As a check of our results, we can see whether we reproduce the vortex-line action derived in \cite{Endlich:2013dma} for ordinary fluids in the near incompressible limit. Recall that $\vec B$ describes sound,  that is, compressional waves, and so the near incompressible limit corresponds to working to lowest order in $\vec B$.
Keeping interactions that are at most linear in $\vec B$ and neglecting for the moment the $S_{\rm NG'}$ piece, our action reduces to
\begin{align}
S_{\rm bulk} + S_{\rm KR} & \to {\bar w}\int dt d^3 x \Big[ \sfrac12 (\vec{\nabla} \times \vec{A})^2 + \sfrac12 \big( \dot{\vec{B}} \, ^2 - c^{2}_{s} (\vec{\nabla} \cdot \vec{B})^2 \big) \Big]  \nonumber \\
 & + \bar n \lambda \int d t d \sigma \Big[ - \sfrac13  \epsilon_{ijk} \, X^k \d_t X^i \d_\sigma X^j +   A_i  \, \d_\sigma X^i \Big]  \label{NR limit} \\
 & + \bar n \lambda  \int d t d \sigma \, \epsilon_{ijk} B^k \, \d_t X^i \d_\sigma X^j  +  {\bar w} \int d^3 x d t \, \sfrac12 (1-c^2_s) \vec \nabla \cdot \vec B (\vec \nabla \times \vec A)^2 \; . \nonumber
\end{align}

We recognize in the first and second line the lowest-order result of \cite{Endlich:2013dma}: $\vec A$ is (minus) the `hydrophoton' field and $\vec B$ is the sound field. Then, our coupling $\lambda$ is related to the vortex circulation $\Gamma = \oint \vec v \cdot d \vec \ell \, $ by
\be \la{lambda}
\lambda = \frac{\bar w}{\bar n} \Gamma \; .
\ee
On other hand, the third line of eq.~\eqref{NR limit} looks more problematic: the first term describes a local coupling of sound to the string which is simply not there in the result of \cite{Endlich:2013dma}, whereas the second term describes a bulk $B A A$ interaction that has the same  field and derivative content as an analogous interaction in \cite{Endlich:2013dma}, but with a different structure of contractions. However, it is straightforward (albeit tedious) to check that, to the order we are working, both discrepancies can be fixed via non-linear field redefinitions:
\be
\vec A  \to \vec A - (\vec \nabla \times \vec{A}) \times \vec B  \; , \qquad \vec X \to \vec X + \vec B (\vec X, t) \; .
\ee
%%
%to cancel the cubic interaction (\ref{int}) and produce a new interaction term of the form
%%
%\be
%S_{int}' = - 2 G'' \int d^3 x d t \, (\nabla \times A)^i (\nabla \times A)^j \d_i B_j.
%\ee
%%
%Once we take into account that $G'' = \frac{1}{2}(c_s^2 - c^2) \to - \frac{1}{2}$ for $c_s \to 0, c=1$, we see that our results agree with \cite{Endlich:2013dma}, with precisely the right coefficient.
For what follows it is more convenient to stick with our original parameterization of the fields and not perform these field redefinitions.

%The first field redefinition sends
%\be
%\sfrac{1}{2}(\vec \nabla \times \vec A)^2 \rightarrow \sfrac{1}{2}(\vec \nabla \times \vec A)^2 - \sfrac{1}{2}(\vec \nabla \cdot \vec B)(\vec \nabla \times \vec A)^2 + (\vec \nabla \times \vec A)^i (\vec \nabla \times \vec A)^{j} \nabla_j B_i\,,
%\ee
%recovering the structure of the cubic term in \cite{Endlich:2013dma}.  Similarly, the second redefinition sends
%\be
%\epsilon_{ijk}\left(-\sfrac{1}{3}X^k + B^k \right)\partial_t X^i \partial_{\sigma}X^j \rightarrow -\sfrac{1}{3}\epsilon_{ijk}\partial_t X^i \partial_{\sigma}X^j\,.
%\ee
%The field redefinitions will also act on the hydrophoton coupling to produce
%\be
%A_i \partial_{\sigma}X^i \rightarrow A_i \partial_{\sigma}X^i + (\vec \nabla \cdot \vec B)(\vec A \cdot \partial_{\sigma}\vec X) \,,
%\ee
%which is absent from the vortex-line action in \cite{Endlich:2013dma}, as this action was derived in the incompressible limit where sound waves decouple.  Our derivation restores these corrections.  For what follows it is more convenient to stick with our original parameterization of the fields, however, and so we will not perform these field redefinitions.

We close this section by noticing that the action of \cite{Endlich:2013dma}---which reproduces the correct dynamics of vortex lines in ordinary fluids as derived from the hydrodynamical equations---lacks our generalized Nambu-Goto piece $S_{\rm NG'}$. This is surprising: such a piece is allowed by all the symmetries, and enters our action at the same order in derivatives as all the other terms that we are keeping; by the standard rules of effective field theory, it should be there. We will see below that its absence in the results of \cite{Endlich:2013dma} is illusory: the zero-thickness limit for a vortex line is notoriously singular, with many physical observables diverging logarithmically already at the classical level; our $S_{\rm NG'}$ term will be needed to correctly renormalize the theory  in that limit. We will now explain in detail how this works, and how in fact we can use such logarithmic divergences to our advantage.
For simplicity, we will restrict ourselves to classical processes, because many  aspects of renormalization and running will be relevant already there; however, the same ideas and techniques can be extended to quantum effects as well.

\section{A local world-sheet theory with running couplings}\label{runningcouplings}

Our vortex-line action, eq.~\eqref{effS}, is not localized on the string worldsheet: it includes terms that are integrated over the volume of the surrounding superfluid. As a result, even for the simplest physical processes such as the free vibration of a infinite straight string, there is a constant interplay between the worldsheet modes ($X^\mu$) and the bulk ones ($A_{\mu\nu}$). This makes computations complicated. The bulk modes can be formally integrated out---at least when they do not appear in the initial or final state---but, being gapless, they yield a non-local effective action for the worldsheet ones~\cite{Endlich:2013dma}. 
This phenomenon is at the origin of the apparent non-locality of the  standard vortex-line equations of motion, which involve Biot-Savart-type integrals~\cite{donnelly1991quantized}: the line is just the `tip of the iceberg'---it is a placeholder for a quite complicated and highly delocalized velocity profile in the surrounding superfluid.

However, it is well known that for many processes taking place at length scales much longer than the vortex-line thickness, things  drastically simplify and one can approximate the equations of motion with local ones~\cite{donnelly1991quantized}. This goes under the name of the LIA---local induction approximation. We will now see that the LIA and similar but more general approximations can be understood in RG terms: for a variety of processes, the leading non-local effect mediated by the bulk modes can be captured by a logarithmic scale-dependence for a worldsheet {\it local} coupling. That is, we will be able to dispose of most of the bulk modes and replace them with running couplings for a purely local worldsheet theory. This classical running of couplings on defects has been studied in the context of braneworld models \cite{Goldberger:2001tn, deRham:2007dg}.  The bulk modes that cannot be gotten rid of in this way are of course the asymptotic states, which can be absorbed, scattered, or emitted by the string. These will be kept explicitly, and will interact with the string via worldsheet localized (running) couplings.

To carry out this program systematically, one should follow for example \cite{Goldberger:2004jt, Goldberger:2007hy} and adapt the technology developed there  to our case. We will discuss some of the systematics of the expansion in \S 7, but we will leave the full development of the effective theory for future work.  For the moment, we will just analyze a few examples of how these ideas can drastically simplify certain computations. 

%%%%%%%%%%%%%%%%%%%%%%%%%%
%%%%%%%%%%%%%%%%%%%%%%%%%%
\subsection{Tension renormalization and the running tension}\label{running tension}
It is well known (see e.g. \cite{donnelly1991quantized}) that the classical energy per unit length of a vortex line is formally divergent, scaling like the log of an infrared cutoff (e.g., the size of the container) over a UV cutoff (e.g., the line thickness). From a field-theoretical viewpoint,  the UV divergence is harmless, since it can be canceled by a local counterterm.  The IR divergence, however, is interesting: it signals that the corresponding coupling will run with scale.

To see how this works, consider a straight, infinite vortex line, at rest in an unperturbed superfluid. We have
\be
\vec X(\sigma, t) = (0, 0, \sigma) \; ,
\ee
where we chose the gauge $\sigma = X^3 = z$. Such a line will source some static $\vec A$ and $\vec B$ fields in the bulk, through the interaction terms
\begin{align}
S & \supset \int d t d \sigma  \big[ \bar n \lambda \,  A_i  \, \d_\sigma X^i + 2 \big| \d_\sigma \vec X \big|  T_{(01)} \vec \nabla \cdot \vec  B \big] \nonumber \\
& \equiv \int d^4 x \big[ \vec J_A \cdot \vec A + \vec J_B \cdot \vec B \big] \label{JA JB} \; ,
\end{align}
where we have rewritten the world-sheet localized interactions formally as bulk sources, which is convenient for the computations that follow. For our simple configuration, these are
\be \label{JA JB straight}
\vec J_A = \bar n \lambda \cdot \hat z \, \delta^2(\vec x_\perp) \; , \qquad \vec J_B = - 2 T_{(01)} \cdot \vec \nabla \delta^2(\vec x_\perp) \, \qquad \qquad \vec x_\perp \equiv (x,y) \; .
\ee
To compute the total energy of this configuration, we can now proceed in two equivalent ways. We can solve the lowest-order equations of motion for $\vec A$ and $\vec B$, and plug the solutions into the energy functional, which for our static configuration to lowest order reads
\be
E  =\int \,  dz \, T + \int \, d^3 x \big[ - \bar w \sfrac12 (\vec{\nabla} \times \vec{A})^2 -  \vec J_A \cdot \vec A  + \bar w \sfrac12 c^2_s(\vec{\nabla} \cdot\vec{B})^2 - \vec J_B \cdot \vec B \big]  \; .
\ee
Such a computation would parallel the standard hydrodynamical one, in which the bulk energy of the string configuration is simply the kinetic energy of the rotating fluid (see e.g.~\cite{donnelly1991quantized}). Alternatively, we can use more field-theoretical methods based on the  effective-action formalism (see e.g.~\cite{Goldberger:2007hy} for a review). This is more general and can be extended straightforwardly to quantum phenomena as well; we will therefore carry out the computation in this way. 

\begin{figure}[htb!]
\centering%
\includegraphics[scale = 0.2]{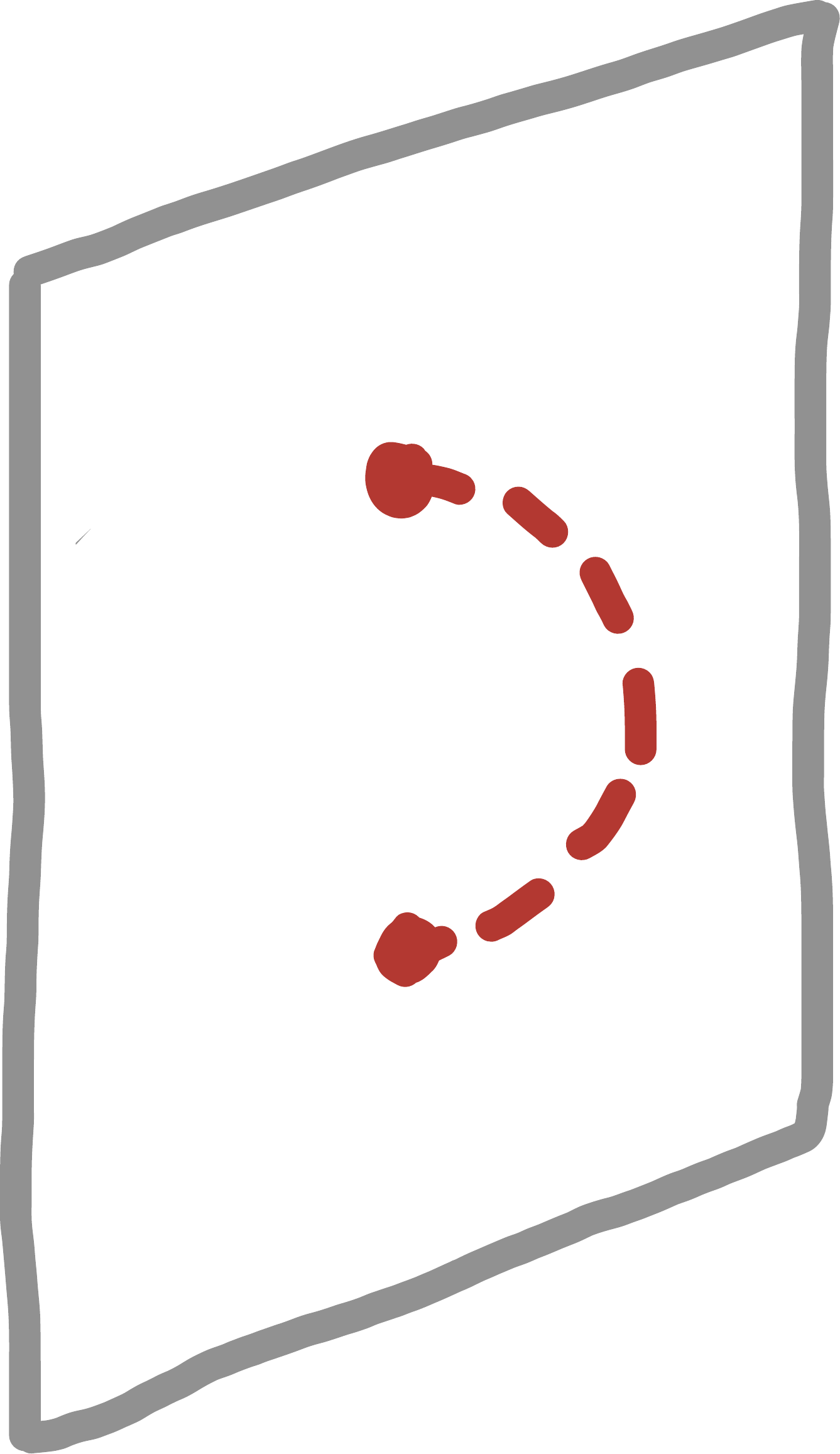} \hspace{3.5cm}
\includegraphics[scale = 0.2]{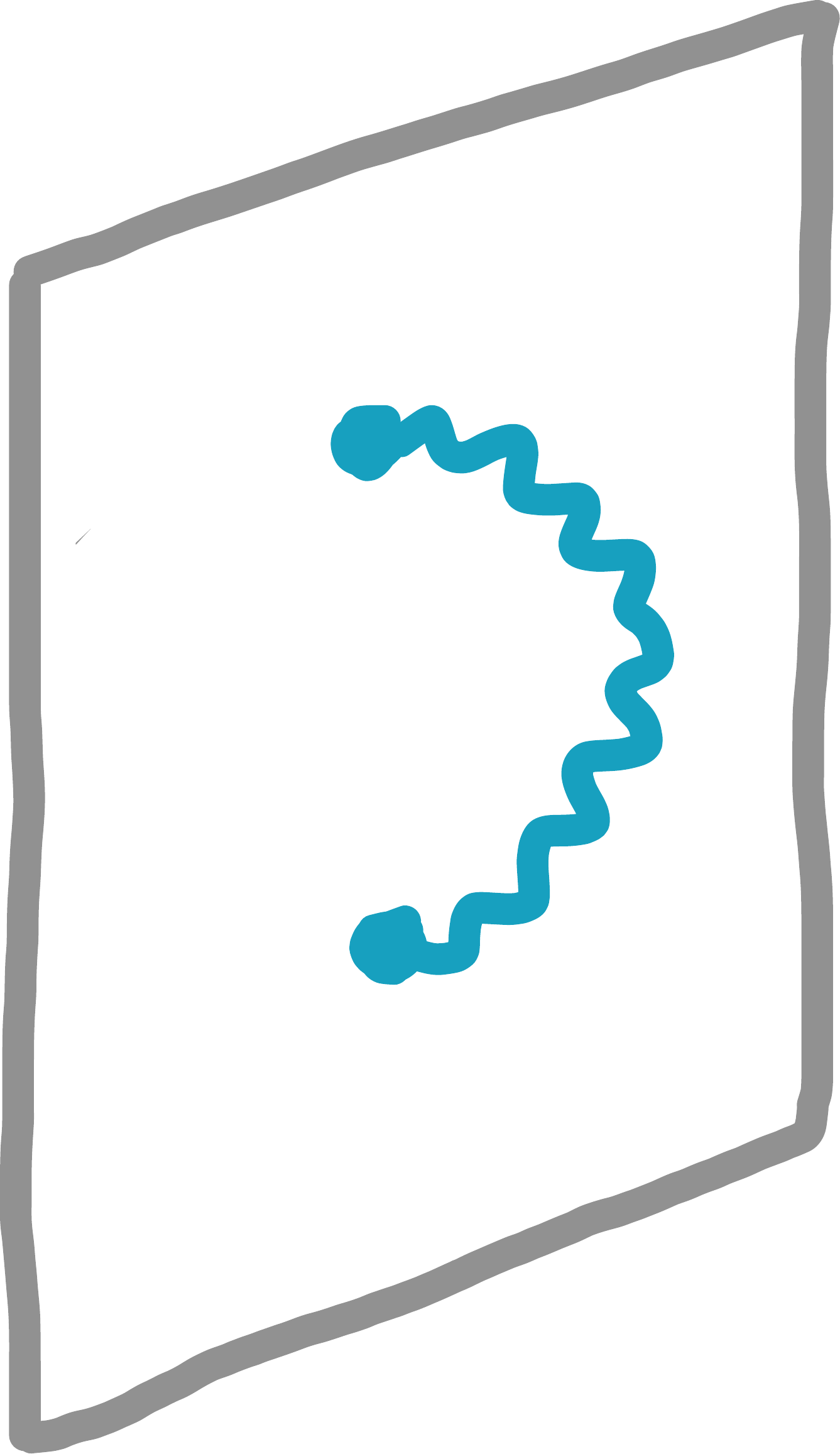}
\caption{\emph{Diagrams contributing to the self-energy of the vortex line. In the first diagram, the hydrophoton (red, dashed line) is emitted and reabsorbed by the worldsheet. In the second diagram, the same process occurs for the phonon (blue, wavy line).}}
\label{fig:energy}
\end{figure}

The idea is to formally perform the path-integral over $\vec A$ and $\vec B$, for given $\vec X$ fields,
\be
e^{i S_{\rm eff}[X]} = \int {\cal D}A \, {\cal D} B \, e^{iS[X, A, B]} \; .
\ee
To lowest order, this shifts the effective action of $\vec X$ by
\be \label{Seff}
S_{\rm eff}[\vec X] \supset \int \! d^4 x d^4 y \, \big[  \sfrac12 J^i_A(x) \,  iG^{ij}_A(x-y) \,  J^j_A(y)  + \sfrac12  J^i_B(x) \, iG^{ij}_B(x-y) \, J^j_B(y) \big] \ ,
\ee
where the $G$'s are the propagators \eqref{propagators}. 
This is depicted by the self-energy diagram of Fig.~\ref{fig:energy}: a string in its ground state exchanges $\vec A$ and $\vec B$ fields with itself; such a process shifts the action of the string, and thus its energy.
For a static configuration such as ours, the energy is defined by $S = - \int dt E$. The shift in the energy thus is
\be
E \supset -\int \! \frac{d^3 \vec p}{(2\pi)^3} \big[  \sfrac12 J^i_A( -\vec p \, ) \,   i G^{ij}_A(\vec p \,) \,  J^j_A(\vec p \, )  +\sfrac12  J^i_B(-\vec p\, ) \,   i G^{ij}_B(\vec p \,) \,  J^j_B(\vec p\, ) \big]  \; ,
\ee 
where the propagators have to be computed at zero frequency, and the $J$'s now stand for purely spatial Fourier transforms.

Including the tension contribution, we get an energy per unit length
%
%
%For such a simple geometric configuration, the only coupling in the action above that can contribute to the bulk solutions for $\vec A$ and $\vec B$ is the $A \cdot \d_\sigma X$ term. We thus get
%{\bf (AN: not sure about the factors)}
%\be
%\vec A(\vec x, t) = \hat z \, \frac{\bar n \lambda}{\bar w} \frac{1}{2\pi} \log r   \; , \qquad \vec B(\vec x, t) = 0 \; ,
%\ee
%where $r$ is the distance from the vortex line. Then, the energy only gets contributions from the $(\nabla \times A)^2$ and $A \cdot \d_\sigma X$ terms---which have the same structure upon integration---and from the tension $T$. Overall, we get an energy per unit length
\begin{align} 
\frac{dE}{dz} & = T + \frac12 \frac{\bar n^2 \lambda^2}{\bar w} \int \frac{d^2 p_\perp}{(2\pi)^2 \, p^2_\perp} - 2\frac{T^2 _{(01)}}{\bar w c_s^2} \int \frac{d^2 p_\perp}{(2\pi)^2} \nonumber \\
& = T + \frac{\bar n^2 \lambda^2}{\bar w} \frac{1}{4 \pi} \log (L \Lambda) -\frac{T^2_{(01)}}{\bar w c_s^2} \frac{1}{2\pi} \Lambda^2 \label{energy} \; ,
\end{align}
where we integrated from an IR momentum cutoff $1/L$---the typical size of the container---to a UV cutoff $\Lambda$.
The second term is the  standard hydrodynamical result, which is due to the kinetic energy of the surrounding fluid. The third term is due to sound-mode exchange and, to the best of our knowledge, has not appeared in the literature.

Notice that we are formally keeping the UV cutoff $\Lambda$ separate from the physical thickness of the line $a$ (always with $\Lambda < 1/a$, to trust our computations), to emphasize its arbitrariness:
the UV cutoff should be interpreted as a formal device to carry out the renormalization program, but ultimately one wants to express long distance observables in terms of physical parameters that can be measured at long distances. Related to this, unless one knows exactly how this `thickness' is realized (is there a step-function in the vorticity? an exponential drop-off?), the precise value of $a$ does not have a well defined meaning anyway. So, it is not wise to have it appear in predictions for long-distance observables.
On the other hand, the IR cutoff $L$ is physical:\ it can be interpreted as the typical scale of the process.

As usual, to remove the UV-cutoff dependence from a physical, measurable quantity---the energy per unit length---we have to move it to a Lagrangian `bare' coupling---the tension. If we parameterize $T$ as
\be
T = T(\mu) + \frac{\bar n^2 \lambda^2}{\bar w} \frac{1}{4 \pi} \log (\mu /\Lambda) +\frac{T^2_{(01)}}{\bar w c_s^2} \frac{1}{2\pi} \Lambda^2 \; ,
\ee
where $\mu$ is an arbitrary renormalization momentum scale and $T(\mu)$ is finite, then eq.~\eqref{energy} is manifestly finite:
\be 
\frac{dE}{dz} = T(\mu) +  \frac{\bar n^2 \lambda^2}{\bar w} \frac{1}{4 \pi} \log (\mu \, L) \; . \label{energy  mu}
\ee

Formally now the energy depends on an arbitrary reference scale $\mu$, but in practice there are unambiguous physical predictions: for instance, for  containers of different sizes the energies per unit length differ by
\be
\frac{dE}{dz} \Big|_{L_1} - \frac{dE}{dz} \Big|_{L_2}  =   \frac{\bar n^2 \lambda^2}{\bar w} \frac{1}{4 \pi} \log (L_1/L_2) \; .
\ee
Following  standard RG ideology, this is conveniently rephrased in terms of a {\it running tension}: for the physical quantity $dE/dz$ to be independent of the arbitrarily chosen scale $\mu$, $T(\mu)$ has to change whenever $\mu$ is changed, in such a way as to leave $dE/dz$ unchanged:
\be \label{running T}
\frac{d}{d \log \mu} T(\mu) = -  \frac{\bar n^2 \lambda^2}{\bar w} \frac{1}{4 \pi} \; , \qquad T(\mu) =  -  \frac{\bar n^2 \lambda^2}{\bar w} \frac{1}{4 \pi}  \log (\mu/\mu_0) \; ,
\ee
where $\mu_0$ is a fixed scale, to be determined by experiment.
Then, when evaluating the string energy for a container of typical size $L$, one can conveniently choose $\mu \sim 1/L$ and simply get 
\be
\frac{dE}{dz} \Big|_{L} = T(1/L) \; .
\ee
Comparing the energies for two different container sizes now corresponds to making the tension `run' from $\mu \sim 1/L_1$ to $\mu \sim 1/L_2$:
\be
\frac{dE}{dz} \Big|_{L_1} - \frac{dE}{dz} \Big|_{L_2} = T(1/L_1) - T(1/L_2) = \frac{\bar n^2 \lambda^2}{\bar w} \frac{1}{4 \pi} \log (L_1/L_2) \; ,
\ee
which is the same result as above. We will see below how this viewpoint can significantly simplify certain computations.

Notice that, upon renormalization, there is no remnant of the quadratic divergence in \eqref{energy} (the term due to sound-exchange): for instance, the energy does not depend on the coupling $T_{(01)}$ anymore.  This is the fate of all power-law divergences: they can renormalize local couplings---in this case, the tension---but they have no measurable consequences at long distances. For this reason, sometimes it is useful to use a regularization procedure in which all power-law divergences are automatically set to zero; dimensional regularization (``dim-reg") has this property. Perhaps more importantly, it also has the property of respecting gauge invariance---something we have not been careful about so far. For these reasons, from now on we will use dim-reg as the UV regulator. The UV-divergent integrals in eq.~\eqref{energy} then become
\begin{align}
\int \frac{d^2 p_\perp}{(2\pi)^2 \, p^2_\perp}\quad  &\to \quad \mu^{2-d} \int_{1/L} \frac{d^d p_\perp}{(2\pi)^d \, p^2_\perp} \simeq -\frac{1}{4\pi} \Big[ \frac{2}{d-2}+ \gamma_E - \log 4 \pi \Big] + \frac1{2\pi} \log(\mu L ) \label{dim-reg} \\
\int \frac{d^2 p_\perp}{(2\pi)^2} \quad  &\to \quad \mu^{2-d} \int \frac{d^d p_\perp}{(2\pi)^d} = 0 \label{dim-reg2}\; ,
\end{align}
where $\mu$ is an arbitrary renormalization scale. Notice that we had to introduce an explicit IR cutoff $1/L$ in the first integral:\ as we discussed, the dependence on this quantity is physical, and we should make sure not to lose it by playing with dim-reg (without the $1/L$ cutoff, the first integral is formally zero as well).
In the $\overline{\rm MS}$ scheme, one then cancels the whole term in brackets via the counterterm $T$,
\be
T = T(\mu) + \frac{\bar n^2 \lambda^2}{\bar w} \frac{1}{8 \pi} \Big[ \frac{2}{d-2}+ \gamma_E - \log 4 \pi \Big] \; ,
\ee
after which one is left with expression \eqref{energy mu} for the energy, exactly as before.

%%%%%%%%%%%%%%%%%%%%%%%%
%%%%%%%%%%%%%%%%%%%%%%%%
\subsection{Running tension for perturbations}
Let us now take our straight string and perturb it a little:
\be
\vec X(t, \sigma) = (0,0, \sigma) + \vec \pi(t, \sigma) \; , \qquad \qquad \pi_z = 0 \; ,
\ee
where we chose the gauge $\sigma = X^3 = z$, and $\vec \pi$ has some typical wavelength $\ell= 2\pi/k$. The displacement of the string will induce perturbations 
in the $\vec A$ and $\vec B$ fields in the bulk, again through the couplings \eqref{JA JB}, which will in turn backreact on $\vec \pi$. To find the spectrum of excitations, one thus has to diagonalize the $(\vec \pi, \vec A, \vec B)$ system. This has been done  in the incompressible limit (i.e., with $\vec B$ set to zero) recently in \cite{Endlich:2013dma}, and long ago via traditional hydrodynamical methods by Lord Kelvin in~\cite{thomson1880xxiv}. Although the modern computation is conceptually much simpler than the original one, it is admittedly still algebraically tedious.
Armed with our running tension, we can now streamline the computation and reduce it to a purely world-sheet one, with no reference to bulk modes anymore. 

\begin{figure}[t!]
\centering%
\includegraphics[scale = 0.2]{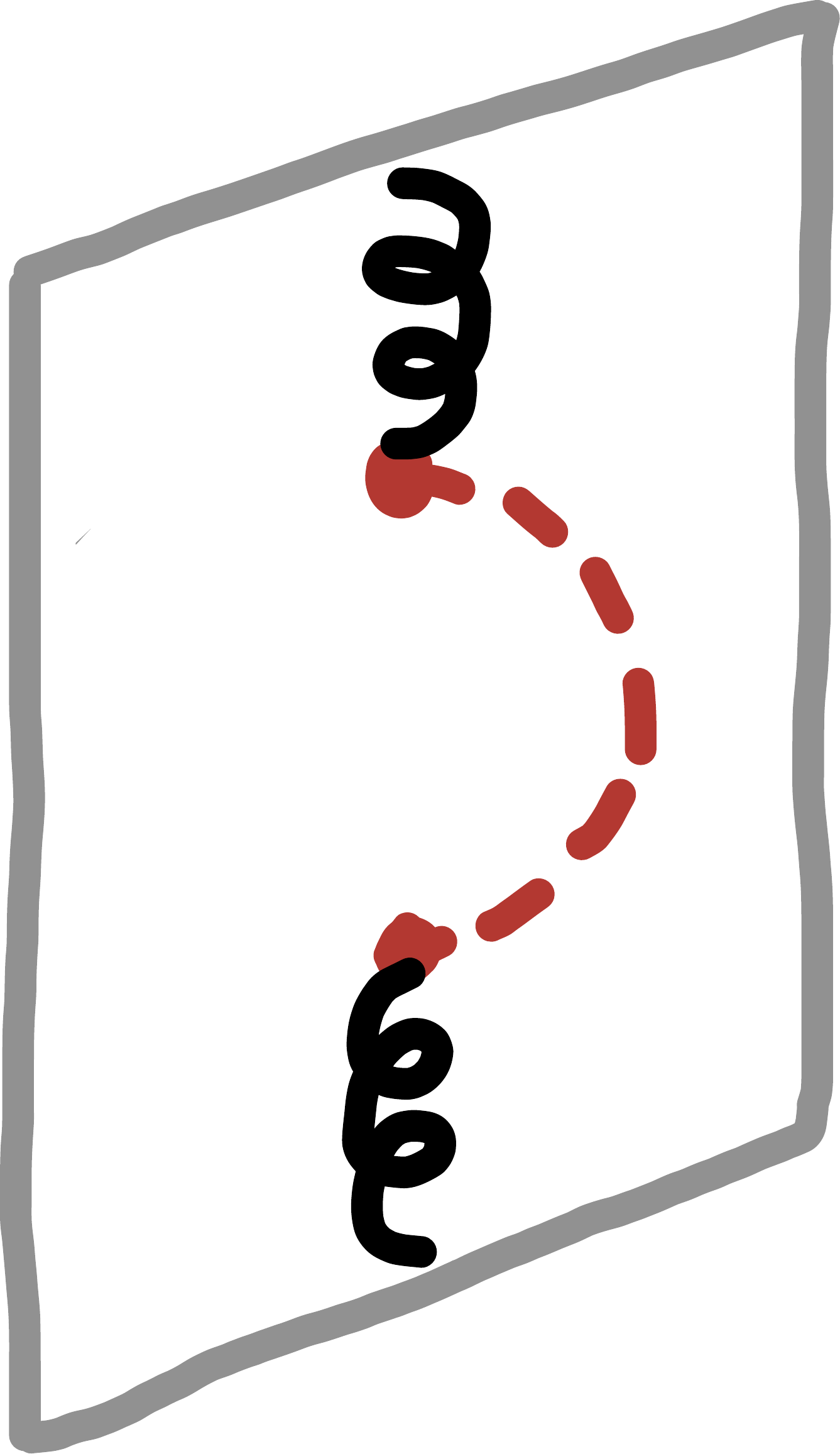} \hspace{1.5cm}
\includegraphics[scale = 0.2]{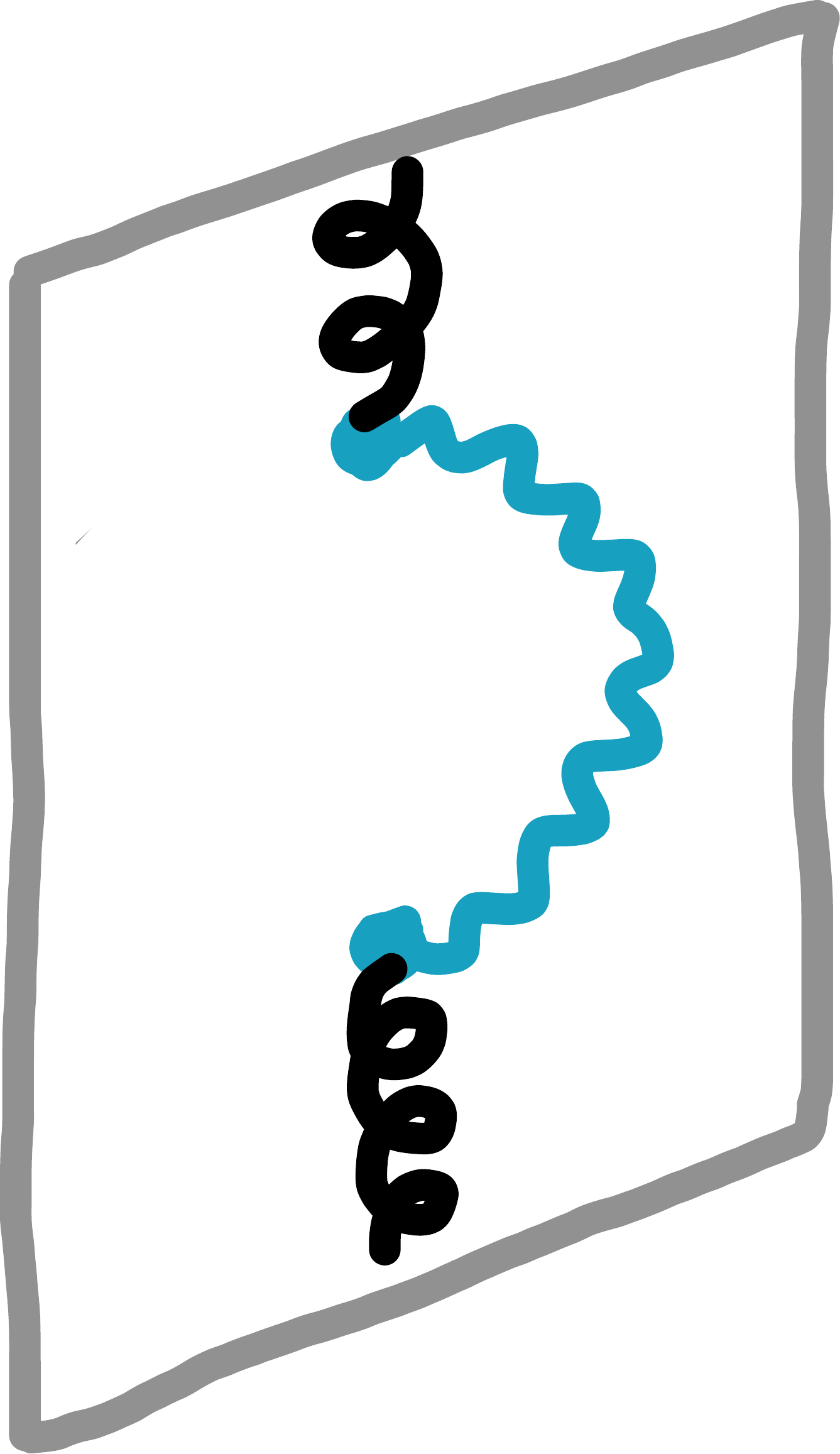}
\caption{\emph{Contribution to the propagator of Kelvin waves (black, curly lines) coming from the mixing with hydrophoton and phonon. The Kelvin waves ``live'' on the worldsheet, whereas phonons and hydrophotons can also propagate in the bulk.}}
\label{fig:kelvin}
\end{figure}

To see how that is possible, consider the relevant diagrams, which are depicted in Fig.~\ref{fig:kelvin}: a world-sheet $\pi$ field mixes with the bulk modes, and their propagation off the world-sheet corrects the propagator of $\pi$. The couplings responsible for the mixing are still of the form \eqref{JA JB}, but now with sources 
\begin{subequations}
\begin{align}
 J_A ^a & = \big(\bar n \lambda  \, \partial_z  \pi^a + 2T_{(10)} \, \epsilon^{ab} \partial_z  \dot \pi^b \big)\, \delta^2(\vec x_\perp) \; , \\
 J_A^z  & = \big(-  \bar n \lambda \, \pi^a + 2T_{(10)} \,  \epsilon^{ab} \dot \pi^b \big) \nabla_a \delta^2 (\vec x_\perp) \ , \\
J^a_B &=  2 T_{(01)} \, ( \pi^b  \nabla_b) \nabla_a\delta^2(\vec x_\perp) + \big(\bar n \lambda \, \epsilon^{ab} \dot \pi^b - 2T_{(10)}\, \ddot \pi^a\big)\,   \delta^2 (\vec x_\perp) \; ,  \\ 
J^z_B   &= 2 T_{(01)}\partial_z \pi^a \nabla_a \delta^2 (\vec x_\perp) \ ,
\end{align}
\end{subequations}
(the indices $a,b$ run over the directions transverse to the string, and we take the convention that $\epsilon^{12} = 1$), which we get by expanding eqs.~\eqref{KR expanded} and \eqref{NG' expanded} to first order in the $\vec \pi$ field.  Integrating out $A$ and $B$ yields the correction \eqref{Seff} to the effective action, now with these sources. It is straightforward to guess the result.

First of all, the bulk propagators are the same as we used for the straight-string energy, eq.~\eqref{propagators}. There, we formally evaluated them at zero momentum and zero frequency. However, we had to cut off the momentum integral for $\vec A$ in the IR at the inverse size of the container, thus effectively introducing an external momentum of order $1/L$. In this case, the IR cutoff will be set by the wavelength of the $\vec \pi$ perturbation.   And, as far as frequency goes, the $\vec A$ propagator does not depend on frequency at all, whereas the $\vec B$ propagator  goes schematically as $1/k^2 + \omega^2/k^4 + \dots$ at low frequencies. Working at first order in time derivatives, we can neglect the frequency-dependence for the $\vec B$ propagator as well. In the limit of vanishing external momentum and frequency, the transverse momentum integrals we should perform are thus of the form \eqref{dim-reg}, \eqref{dim-reg2}, and generalizations thereof with higher powers of $p_\perp$, coming from the higher derivatives of the $\delta^2(x_\perp)$ in the sources. In dim-reg, the only non-trivial one is $\eqref{dim-reg}$, with $1/L$ replaced by the external $\pi$ momentum $k$.

Since such an integral is associated with the non-derivative $\delta^2(x_\perp)$ terms in the sources, we reach the conclusion that---to the order we are working---the only effect of integrating out $\vec A$ and $\vec B$ in the presence of perturbations is to induce a logarithmic running of the coefficients of several local terms in the action: $(\partial_z  \pi)^2$,  $\epsilon \cdot \partial_z \pi  \partial_z  \dot \pi$, $(\partial_z  \dot \pi)^2$, $\dot \pi^2$, $\epsilon \cdot \dot \pi  \ddot \pi$, and $\ddot \pi^2$. To lowest order in time-derivatives, we can neglect all of them but the first one, which  is related by symmetry to other local terms in the action, all those that come from expanding $|\partial_\sigma \vec X|$ in \eqref{NG' expanded}. In particular, the logarithmic running of the gradient energy $(\partial_z \vec \pi)^2$ is nothing but the logarithmic running of the tension we discussed above.
Looking at the result of the energy per unit length of the straight string in the form \eqref{energy mu}, we can interpret the running tension $T(\mu)$ as the sum of all UV contributions, which include those associated with the microphysics of the string (our original $T$) as well as those coming from the exchange of bulk modes up to scales of order $1/\mu$. Then, for our new computation, if we choose the renormalization scale $\mu$ to be at the typical momentum scale of our process---the wavenumber $k$ of the perturbation---we can neglect the contribution of bulk modes of wavelengths larger than $1/\mu$, thus effectively concentrating all effects of bulk mode exchange in the running tension.

%Then the computation that we set out to perform involves exactly the same ingredients as the computation above, and integrating out the hydrophoton field gives the following correction to the effective action:
%\begin{equation}
%\begin{split}
%S \supset &\int dt \frac{d^{3}\vec{p}}{(2\pi)^3}\Big[\frac{1}{2}\bar{n}^2 \lambda^2 p_{z} \pi^i(-p_z) i G^{ij}_{A}(\vec p)p_z \pi^{j}(p_z) \Big]\\
%& = \frac{1}{2} \frac{\bar{n}^2\lambda^2}{\bar{w}}\int dt dz\, \Big[-(\partial_z \pi^i \partial_z \pi^j)\Big]\int \frac{d^2 \vec{p}_{\bot}}{(2\pi)^2}\frac{1}{\vec{p}^2_{\bot} + p^2_z}(\delta^{ij}-\hat{p}^i_{\bot}\hat{p}^{j}_{\bot})\\
%& = -\frac{\bar{n}^2 \lambda^2}{8 \pi \bar{w}} \int dt dz\, (\partial_z \vec{\pi})^2 \log(\Lambda/p_z)
%\end{split}
%\end{equation}
%This contribution is a gradient energy for the Kelvin waves, and the strength of the interaction will depend on scale and can be absorbed into a running coupling.  In fact, this is just the running tension found in the previous subsection.
%The diagram evaluates to
%\be
%k^2 (\bar{n}\lambda)^2 \int \frac{d^2 \vec{k}_{\bot}}{(2\pi)^2}\frac{i}{\bar{w}(\vec{k}^2_{\bot} + k^2)} = -i \frac{k^2 \bar{n}^2 \lambda^2}{2\pi \bar{w}}\log(k/\Lambda)
%\ee
%which comes from the following term in the effective action:
%\be
%-\frac{\bar{n}^2\lambda^2}{8 \pi \bar{w}}\int dt dz \, |\partial_z \vec{\pi}|^2 \log(k/\mu) 
%\ee
%{\bf BH: Don't we also need to consider the term $\epsilon B \dot{X} X'$?  This is of the same order, according to the power counting in section \ref{powercounting}.} 

In conclusion, we can consistently describe string excitations of typical momentum $k$  by the simple  world-sheet effective action
\begin{align} 
S_{\rm eff} & = \int d t d \sigma \big[ - \sfrac13   \bar n \lambda  \, \epsilon_{ijk} \, X^k \d_t X^i \d_\sigma X^j - T(k) \big| \d_\sigma \vec X \big| \, \big] \label{local Seff}\\
& \to \int d t d z \big[ - \sfrac12   \bar n \lambda \,  \epsilon_{ab} \, \pi^a \d_t \pi^b -  T(k) \sqrt{1+(\d_z \vec \pi)^2} \, \big] \; ,
\end{align}
where in the second line $\epsilon_{ab}$ is restricted to the $xy$ plane.  Note that although we have only checked this result explicitly to quadratic order in the perturbation $\vec \pi$, the symmetries of the action allows us to extend it to the higher order terms as well.  

Given that rotations about the unperturbed string are unbroken, our excitations will carry a conserved `quantum' number associated with that symmetry, which is nothing but the $z$ component of angular momentum. To make use of this, it is convenient to combine the two components of $\vec \pi$ into a complex world-sheet scalar describing circularly polarized waves,
\be
\phi \equiv \sfrac{1}{\sqrt 2}(\pi_x + i \pi_y) \; ,
\ee
in terms of which the effective action becomes
\be \label{Seff phi}
S_{\rm eff}  = \int d t d z \big[ \bar n \lambda \, \phi^* i \d_t \phi -   T(k) \sqrt{1+ 2 |\d_z  \phi|^2} \, \big] \; .
\ee
The action is now symmetric under a global $U(1)$ symmetry, and $\phi$ and $\phi^*$ carry opposite charge under it.

%%%%%%%%%%%%%%%%%%%%%%%%
%%%%%%%%%%%%%%%%%%%%%%%%

\subsection{Other running couplings}
\begin{figure}[b!]
\centering%
\includegraphics[scale = 0.2]{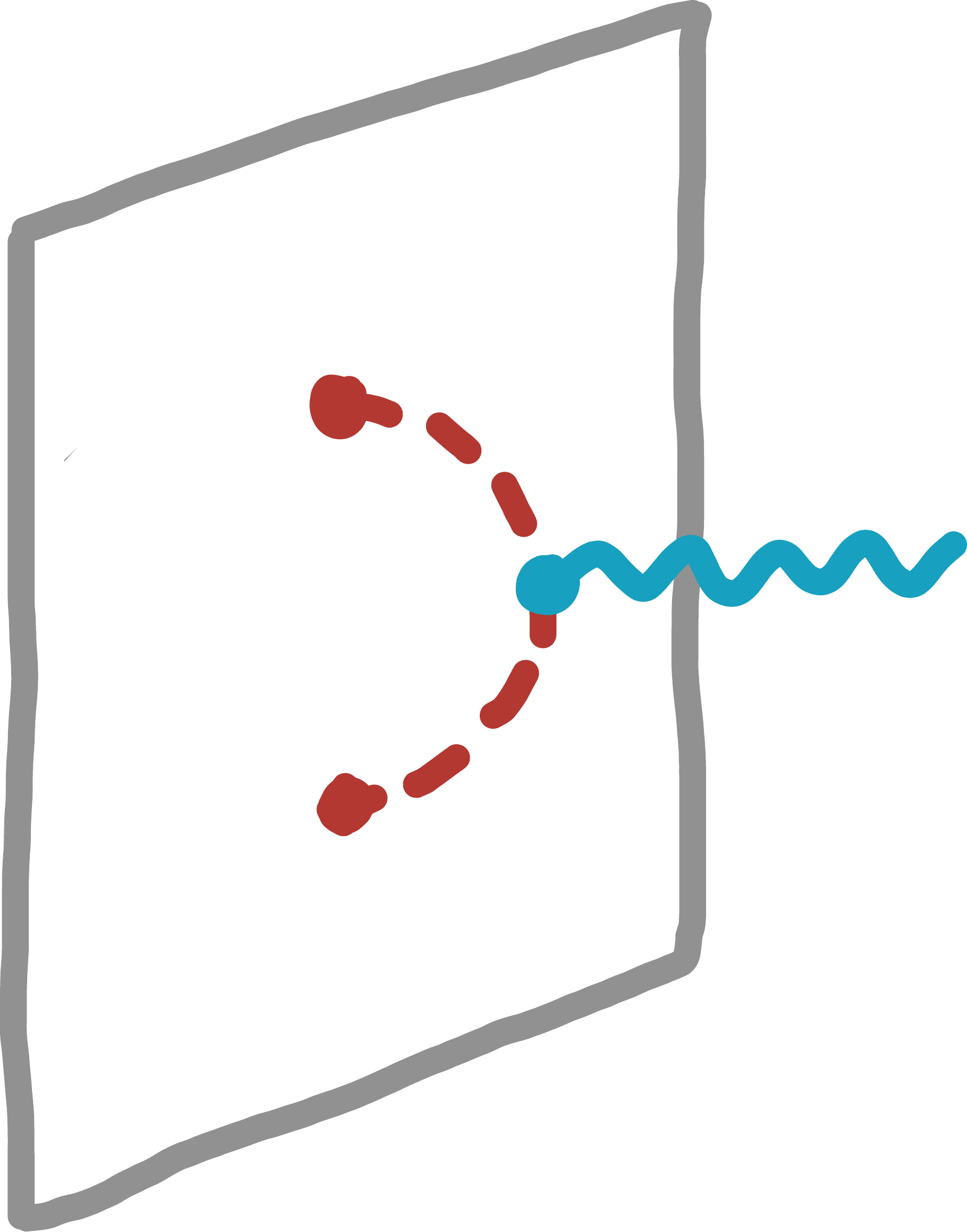}
\caption{\emph{Hydrophoton correction to the coupling of sound to the worldsheet.}}
\label{fig:BAA}
\end{figure}

We can apply the same techniques to processes that involve external $\vec A$ and $\vec B$ fields as well. Consider for instance the way an external sound mode $\vec B$ can couple to the string: there is a direct world-sheet interaction in \eqref{NR limit}, as well as a more non-local one mediated by $\vec A$, as depicted in Fig.~\ref{fig:BAA}. 
As before, it is useful to phrase the physical effect of the latter in terms of an effective action, now for $\vec X$ {\it and} $\vec B$,
\be
e^{i S_{\rm eff}[X, B]} = \int {\cal D}A  \, e^{iS[X, A, B]} \; .
\ee
Using standard perturbation theory for the path integral, we get a correction to the effective action which to leading order is given by
\begin{align}
S_{\rm eff} [\vec X, \vec B] \supset  \sfrac12 \bar w(1- c_s^2)\int & \frac{d^4 p}{(2\pi)^4}  \frac{d^4 q}{(2\pi)^4}  \, J_A^i(-p) J_A^j(-q) 
(\vec \nabla \cdot \vec B)(-p-q) \nonumber \\
& \times \big[-p\cdot q \, G_A^{ik}(p) G_A^{kj}(q)+ q^k p^l G_A^{ik}(p) G_A^{lj}(q) \big] \; ,
\end{align}
This is definitely a non-local correction to the action. However, we can get a sense of the kind of non-locality involved by working at very low momenta for the string excitations and for the external $\vec B$ field. To this end, let us take the simplified configuration in which the string is straight, so that the source $\vec J_A$ is simply that given in \eqref{JA JB straight}.
After changing the $q$ integration variable, $-(p+q) \to q$, we  get
\be
S_{\rm eff} \supset \frac12 \frac{\bar n^2 \lambda^2}{\bar w} (1 - c_s^2) \int \frac{d^2 p_\perp}{(2\pi)^2}  \frac{d^2 q_\perp}{(2\pi)^2}
(\vec \nabla \cdot \vec B)(\vec q_\perp) \frac{\vec p_\perp \cdot (\vec q_\perp +\vec p_\perp )}{p^2_\perp  ( \vec q_\perp +\vec p_\perp )^2} \; ,
\ee
where now $(\vec \nabla \cdot \vec B)(\vec q_\perp)$ is evaluated at zero frequency and zero $q_z$.

The integral in $\vec p_\perp$ diverges logarithmically in the UV. We can first isolate and compute the divergent piece, and then recover with logarithmic accuracy the finite piece  by dimensional analysis. To do so, recall that the external $\vec B$ field is concentrated at low momenta:\ expanding the integrand in powers of $\vec q_\perp$, to zeroth order we get
\begin{align}
S_{\rm eff} & \supset \frac12 \frac{\bar n^2 \lambda^2}{\bar w} (1 - c_s^2) \bigg[ \int \frac{d^2 q_\perp}{(2\pi)^2} (\vec \nabla \cdot \vec B)(\vec q_\perp) \bigg] \times  \bigg[ \int\frac{d^2 p_\perp}{(2\pi)^2} \frac{1}{p_\perp^2} \bigg]  \; .
\end{align}
The $p_\perp$ integral is the same as \eqref{dim-reg}, but now  by dimensional analysis the IR cutoff $1/L$  has to be taken to be of the order of the typical transverse momentum of the external $\vec B$ field, because that was the only momentum scale appearing in the original $(q_\perp, p_\perp)$ integral.

Rewriting the $q_\perp$ integral in real space and including now the local contribution from \eqref{NG' expanded} we finally get
\begin{align}
S_{\rm eff}  \supset \int dt dz \, \Big\{ & 2 T_{(01)}     - \frac1{8\pi} \frac{\bar n^2 \lambda^2}{\bar w} (1 - c_s^2) \Big[ \frac{2}{d-2}+ \gamma_E - \log 4 \pi \Big] \nonumber \\
&  + \frac1{4\pi} \frac{\bar n^2 \lambda^2}{\bar w} (1 - c_s^2)  \log(\mu/q_{\bot} )\Big\} \, \vec \nabla \cdot \vec B \ .
\end{align}
Following the same RG logic we adopted for the renormalization of the tension, we see that the leading non-local effect of the diagram in Fig.~\ref{fig:BAA} is to make the local coupling $T_{(01)}$ run with scale:
\be \label{running T01}
\frac{d}{d \log \mu} T_{(01)}(\mu) = -  \frac{\bar n^2 \lambda^2}{\bar w} \frac{1}{8 \pi} (1-c_s^2) \; , \qquad T_{(01)}(\mu) =  -  \frac{\bar n^2 \lambda^2}{\bar w} \frac{1}{8 \pi}(1- c_s^2)  \log (\mu/\mu'_0) \; ,
\ee
where the reference UV scale $\mu_0'$ is in general numerically different from that appearing in the running tension, although we expect both to be of order of $1/a$---the inverse string thickness.
 
We thus conclude that, even when we include perturbations of the string, the leading  effects of integrating out $\vec A$ are conveniently parameterized by a running $T_{(01)}$ vertex, 
\be \label{T01}
S_{\rm eff} \supset \int dt d \sigma   \big| \d_\sigma \vec X \big| \, 2 T_{(01)} (\mu) \, \vec \nabla \cdot \vec B(\vec X, t) \; ,
\ee
to be evaluated at $\mu$ of the order of the relevant momentum scale of the process under consideration, to minimize the contributions coming from $\vec A$ modes with momenta between the scale of the process and $\mu$. 

In the computation above, we set the string excitations to zero, and so the only external momentum that could play the role of an IR cutoff was the transverse momentum of the external $\vec B$ field.  But we could have done the opposite: we could have worked at zero momentum for $\vec B$, but in the presence of string excitations. In that case, we would have seen that the IR cutoff  would have been of order of the typical momentum of these excitations. For a more generic process, we have both an external momentum for $\vec B$, and momenta for the string excitations. In such a case, the leading order contribution to the momentum integrals comes from cutting them off in the IR at the {\it largest} of these momentum scales. Thus, for a generic process involving $\vec B$ and string excitations, the appropriate choice for $\mu$ to minimize the error is the largest of the momenta involved.

\begin{figure}[htb!]
\centering%
\includegraphics[scale = 0.2]{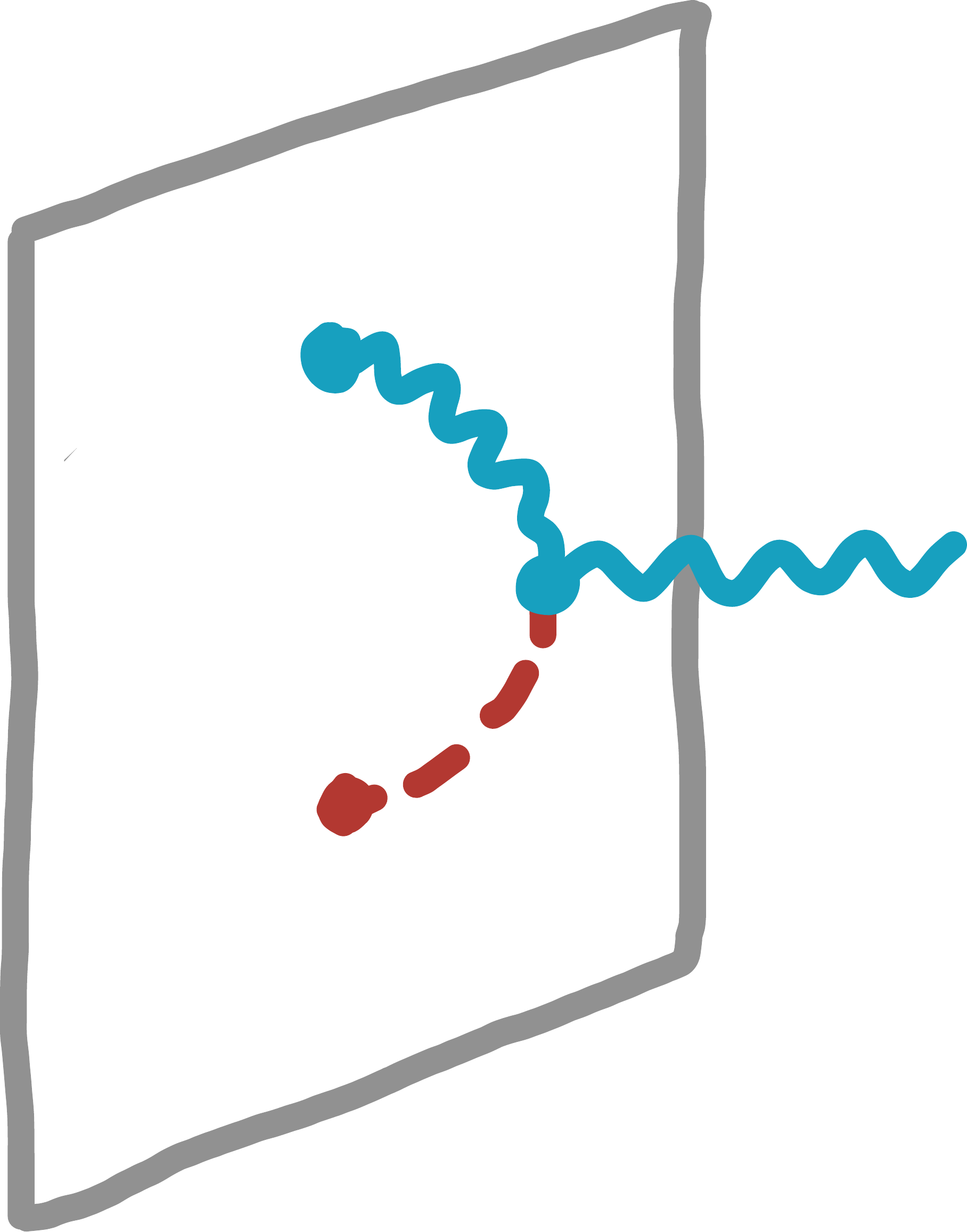}
\caption{\emph{Leading contribution to the running of $T_{(10)}$.}}
\label{fig:BBA}
\end{figure}

We can continue this process to higher orders in perturbation theory, and for terms of higher order in our general action.  Before closing this section, we consider the running of the coupling $T_{(10)}(\mu)$, which accompanies the term in the action \eqref{NG' expanded} of order $\dot{{B}} \cdot \dot{X}$,  %In addition to the action in eq.~\eqref{actionslowstring, KR expanded} we will need the following worldsheet couplings, which are of higher order in the slow-string limit: 
\begin{equation}
%\begin{split}
S_{\rm NG'} \supset \int dt d\sigma |\partial_{\sigma}\vec{X}|\Big[2T_{(10)}(\dot{{\vec B}}_{\bot} \cdot {\vec v}_{\bot})\Big] \ ,%\\
%S_{\rm KR} &\supset \int dt d\sigma \Big[\epsilon_{ijk} B^k \partial_t X^i \partial_{\sigma}X^j\Big]  
%\end{split}
\end{equation}
which will be renormalized by the diagram in Fig.~\ref{fig:BBA} via the $\dot{\vec{B}}\cdot(\vec \nabla \times \vec A)(\vec \nabla \cdot \vec B)$ bulk coupling in eq. \eqref{actionslowstring}.
The corresponding contribution to the effective action is given by
\begin{equation}
\begin{split}
S_{\rm eff}[\vec{X},\vec{B}] \supset &- \bar{w}(1-c^2_s)\int \frac{d^4 p}{(2\pi)^4}\frac{d^4 q}{(2\pi)^4} \, J^{i}_{B}(-p)J^{j}_{A}(-q)\dot{B}^{k}(-p-q)\\
& \qquad \qquad \times \epsilon^{kab}(-i q_a)G^{jb}_{A}(q)(-i p_l)G^{il}_{B}(p) \ .
\end{split}
\end{equation}
Here we are taking the configuration in which the string is straight but moving with constant transverse velocity ${\vec v}_{\bot}$, which is off-shell and does not satisfy the equations of motion, since an infinite straight string does not move unless we change the boundary conditions at infinity:
\be
\vec X = ({\vec v}_{\bot} t, \sigma)\,,
\ee
and then we have
\begin{subequations}
\begin{align}
\vec{J}_{A} & = \big(\bar{n}\lambda  + 2T_{(10)}v_a \epsilon^{ab}\nabla_b \big)\hat{z}\cdot\delta^2(\vec{x}_{\bot} - {\vec v}_{\bot}t) \ , \\
\vec{J}_{B} & = \big(\bar{n}\lambda (\vec{v}\times\hat{z}) - 2T_{(01)}\vec{\nabla} + 2T_{(10)}{\vec v}_{\bot} v^{j}_{\bot}\nabla_j \big)\cdot\delta^{2}(\vec{x}_{\bot} - {\vec v}_{\bot}t) \ .
\end{align}
\end{subequations}
As in the previous subsection, $a,b$ run over the transverse directions, and we ignore the terms with a derivative on the Dirac delta function in the low-momentum limit.  Furthermore, we keep only the leading terms in $\vec v = {\vec v}_{\bot}$ in the small-velocity limit.  In this case $\vec{J}_A$ is given by \eqref{JA JB straight}, and
\be
\vec{J}_{B}(\vec{x}) = \bar{n}\lambda (\vec{v}\times\hat{z})\delta^{2}(\vec{x}_{\bot})\,.
\ee
Changing the integration variable $-(p+q) \to q$, we have
\be
S_{\rm eff} \supset  \frac{\bar{n}^2 \lambda^2}{\bar{w} c^2_s}(1-c^2_s) \int \frac{d^2 p_{\bot}}{(2\pi)^2}\frac{d^2 q_{\bot}}{(2\pi)^2}\,\dot{B}^{k}(\vec{q}_{\bot})\epsilon^{imn}\epsilon^{kaj}\hat{z}^j \hat{z}^n v^m \frac{p^{i}_{\bot} (p_{\bot} + q_{\bot})^a}{p^2_{\bot}(p_{\bot} + q_{\bot})^2} \ .
\ee
Performing the UV part of the $\vec{p}_{\bot}$ integral using dimensional regularization and rewriting the $\vec{q}_{\bot}$ integral in real space, we have
\begin{equation}
\begin{split}
S_{\rm eff} \supset &\int dt dz\, \Big\{2T_{(10)} + \frac{\bar{n}^2\lambda^2}{8\pi c^2_s \bar{w}}(1-c^2_s)\Bigg[\frac{2}{d-2} + \gamma_{E} - \log 4\pi\Bigg]\\
& \qquad \qquad - \frac{\bar{n}^2 \lambda^2}{4\pi \bar{w}c^2_s}(1-c^2_s)\log(\mu/q_{\bot})\Big\}(\dot{\vec{B}}_{\bot} \cdot \vec{v}_{\bot}) \ ,
\end{split}
\end{equation}
and therefore the running of $T_{(10)}(\mu)$ is given by
\be\label{running T10}
\frac{d}{d \log \mu}T_{(10)}(\mu) = \frac{\bar{n}^2 \lambda^2}{8 \pi \bar{w}}\frac{(1-c^2_s)}{c^2_s}\, , \qquad T_{(10)}(\mu) = \frac{\bar{n}^2 \lambda^2}{8 \pi \bar{w}}\frac{(1-c^2_s)}{c^2_s} \log(\mu/\mu''_0)\,.
\ee
As before, the most convenient choice of $\mu$ is the largest of the external momenta involved in the process.

%%%%%%%%%%%%%%%%%%%%%%%%%%
%%%%%%%%%%%%%%%%%%%%%%%%%%

\subsection{Localized effective action for string and sound} \la{sec:localized effective action}

We can now put everything together and write an effective action for the string and its interactions with sound, with only world-sheet localized interactions. Expanding the worldsheet couplings out to leading order in powers of $\vec B$, this is
\begin{align}
S_{\rm eff}[\vec X, \vec B] & \simeq      \int d^ 4x \, \sfrac12 {\bar w} \big( \dot{\vec{B}} \, ^2 - c^{2}_{s} (\vec{\nabla} \cdot \vec{B})^2 \big)  \nonumber \\ 
&  +   \int d t d \sigma \Big[ - \bar n \lambda  \sfrac13  \epsilon_{ijk} \, X^k \d_t X^i \d_\sigma X^j -T(\mu)  \big| \d_\sigma \vec X \big| \la{Seff B phi}  \\
& \qquad   +  \epsilon_{ijk} B^k \partial_t X^i \partial_{\sigma}X^j + 2 T_{(01)}(\mu) \big| \d_\sigma \vec X \big| \,  \vec \nabla \cdot \vec  B (\vec X, t )  \nonumber \\
& \qquad + 2T_{(10)}(\mu)|\partial_{\sigma} \vec X|\Big(\dot{\vec B}\cdot \partial_t \vec X - \frac{(\dot{\vec B}\cdot \partial_{\sigma} \vec X)(\partial_t \vec X \cdot \partial_{\sigma} \vec X)}{(\partial_{\sigma}\vec X)^2}\Big)\Big] \; . \nonumber
\end{align}
%{\bf [RP: shouldn't we also add the non-derivative coupling to $B$? In fact, since we have integrated out off-shell phonons the $B$'s above should be radiation phonons, no? In which case the non-derivative coupling is even more important (according to our power counting) than the one we are writing down.]}
The first line describes the free propagation of sound in the bulk of the superfluid, the second describes the dynamics of the string, and the third and fourth describe its interactions with sound. 
The $T$-couplings run with momentum scale according to \eqref{running T}, \eqref{running T01} and \eqref{running T10}. Their running encodes the physical effects of  exchanging  bulk $\vec A$ modes, as detailed above. %On the other hand, to this order, the exchange of off-shell $\vec B$ modes only renormalizes the tension $T$ by an additive power-law divergent piece, with no long-distance consequences.

Being linear in $\vec B$, the string-sound interactions above are the leading ones contributing to sound emission or absorption by the string; on the other hand, to consider sound scattering off the string, one should extend our analysis beyond leading order and consistently compute the contributions to the effective interactions up to quadratic order in $\vec B$.  Note also that this is not the same expansion as was used in \eqref{actionslowstring}; we will discuss how to organize the terms in perturbation theory systematically in Section \ref{sec: power counting}.
 
For a string that is approximately straight, we can choose the gauge $\sigma = X^3 \equiv z$,  parametrize the perturbations as above,
\be
\vec X(t, \sigma) = \vec X_0(\sigma) + \vec \pi(t, \sigma) \; , \qquad \vec X_0(\sigma) \equiv (0,0,\sigma) \; ,  \qquad \vec \pi \equiv \sqrt 2 ({\rm Re} \, \phi, {\rm Im} \, \phi, 0) \; ,
\ee 
and, if needed, expand our effective action in powers of $\phi$,
\begin{align}
-\sfrac13  \epsilon_{ijk} \, X^k \d_t X^i \d_\sigma X^j  \quad \to \quad  & \phi^* i \d_t \phi \\
\big| \d_\sigma \vec X \big|  \quad \to \quad & \sqrt{1+ 2 |\d_z  \phi|^2} = 1 + |\d_z  \phi|^2 + \dots \\
B^i(\vec X)  \quad \to \quad & B^i(\vec X_0) + \sfrac1{\sqrt 2}\big[ \phi \cdot \big(\partial_x B^i(\vec X_0) - i \partial_y B^i(\vec X_0) \big) \nonumber \\
& +\phi^* \cdot \big(\partial_x B^i(\vec X_0) + i \partial_y B^i(\vec X_0) \big) \big] + \dots \; ,
\end{align}
and so on.

%\begin{align} \label{Seff B phi}
%S_{\rm eff}  & \simeq \int d^ 4x \, \sfrac12 {\bar w} \big( \dot{\vec{B}} \, ^2 - c^{2}_{s} (\vec{\nabla} \cdot \vec{B})^2 \big)   \\
%& + \int d t d z \Big[   \bar n \lambda \cdot \phi^* i \d_t \phi + \sqrt{1+ 2 |\d_z  \phi|^2} \Big( -T(\mu) + 2 T_{(01)} (\mu) \vec \nabla \cdot \vec B(\vec X_0 + \vec \pi, t) \Big)\, \Big] \; , \nonumber
%\end{align}
%to be expanded in powers of $\phi$.
%

%%%%%%%%%%%%%%%%%%%%%%%%%%
%%%%%%%%%%%%%%%%%%%%%%%%%%

\subsection{A non-renormalization theorem}\label{non-renormalization}

Before turning our attention to some concrete applications of our formalism, we would like to conclude this section by briefly discussing one more formal aspect of our effective theory: a non-renormalization theorem for the Kalb-Ramond coupling $\lambda$.
%
% As we already mentioned in sect. \ref{action for lines}, the coupling $\lambda$ in front of the Kalb-Ramond term is related to the vorticity $\Gamma$ by equation (\ref{lambda}), which we reproduce here for the reader's convenience:
%%
%\be
%\lambda = \frac{\bar w}{\bar n} \Gamma \; .
%\ee
%%
%In superfluids, the vorticity $\Gamma$ is known to be quantized in units of $2 \pi/m$  (with $\hbar =~1$)~\cite{landau9}. Therefore, since $\Gamma$ and in turn $\lambda$ are not allowed to change continuously in superfluids (as long as we keep the background quantities $\bar w$ and $\bar n$ fixed), they cannot receive quantum corrections because these would be continuous functions of the other coupling constants of the theory.

Let us consider the general relation \eqref{lambda} between $\lambda$ and the circulation, and let us apply it to the case of a non-relativistic superfluid such as  liquid helium. The enthalpy density $\bar w$ then simply reduces to the number density $\bar n$ times the mass $m$ of a single particle. Moreover, $\Gamma$  is quantized in units of $2 \pi/m$  (for $\hbar =1$)~\cite{landau9}. So, in this case, our coupling $\lambda$ is quantized in units of $2\pi$, with no dependence on any parameter of the theory. This means that, in this case, $\lambda$ cannot get renormalized continuously---a statement that we should be able to prove within our effective field theory, and that would then apply to more general cases as well.

Further evidence for the non-renormalization of $\lambda$ comes from the fact that the Kalb-Ramond term \eqref{KR}, being the integral of a two-form over a two-dimensional world-sheet, is invariant under generic spacetime diffeomorphisms.\footnote{More precisely, since the integral does not depend on the metric, this term is diffeomorphism invariant even without considering the transformation of the metric tensor.} This part of the action then has a highly enhanced symmetry compared to the rest; it is unlikely (though not impossible) that interactions that violate this symmetry yield contributions to the effective action that respect it.

To prove such a non-renormalization theorem within our effective theory, we first notice that  $\lambda$ cannot receive quantum corrections that depend on the other couplings. The reason is that the Kalb-Ramond term is the only term in the action that is invariant under gauge transformations $A_{\mu\nu} \to A_{\mu\nu} + \d_{[\mu} \xi _{\nu]}$ only \emph{up to a total derivative}.
Since all the other terms in the action are gauge invariant with no need to integrate by parts, they would remain so even if their coupling ``constants" were in fact arbitrary functions of the coordinates. On the other hand, the Kalb-Ramond term is gauge invariant if and only if $\lambda$ is a constant. This means that the renormalization of $\lambda$ cannot depend on the other couplings, because that would make $\lambda$ spacetime dependent whenever the other couplings are.

Although very powerful and probably familiar to many readers, this argument is not sufficient by itself to rule out the possibility that the Kalb-Ramond term renormalizes itself.
%Then, it remains to be proven that $\lambda$ cannot renormalize itself. 
%This is obvious when the vortex lines are treated as external sources, because then the Kalb-Ramond term is just a linear coupling between $A_{\mu\nu}$ and an external source, and  there is no vertex proportional to $\lambda$ with more than one external leg.  
%When instead the vortex lines are treated as dynamical objects, the Kalb-Ramond term yields some non-trivial interactions among phonons, hydrophotons and Kelvons. The easiest way to show that $\lambda$ cannot renormalize itself in this case either is through a concrete example. 
To prove that this is impossible, let us consider for simplicity perturbations around a perfectly straight vortex line oriented along the $\hat z$ axis, but formally keeping all terms in the expansion in perturbations. 
In the gauge $ \tau = X^0 \equiv t, \sigma =X^3 \equiv z$, we can write
\be
X^\mu = (t, \pi^1(t,z), \pi^2(t,z), z) \; , \qquad A_{ij} = -\sfrac13 \bar n \,  \epsilon_{ijk} x^k + \delta A_{ij} \; , \qquad A_{0i} = \delta A_{0i} \; ,
\ee
and thus the Kalb-Ramond term reduces to 
\begin{align} \la{KR spelled out}
S_{\rm KR} =  \int dt dz \, \big[ & \sfrac{1}{2} \bar{n}\lambda \,  \epsilon_{ab} \pi^b \d_t \pi^a + \lambda \, \delta A_{0z} \\
& + \lambda \big(\delta A_{0a} \partial_z \pi^a + \delta A_{az} \d_t \pi^a + \delta A_{ab} \d_t \pi^a \d_z \pi^b + \cdots  \big) \big] \; , \nonumber
\end{align}
where the  dots stand for all the terms that arise when we Taylor-expand $\delta A_{\mu\nu}$ in powers of $\vec \pi$ around $ X^\mu = (t, 0,0, z)$, and so now all the $\delta A_{\mu\nu}$'s are evaluated on the unperturbed string. 

Gauge invariance and the non-linearly realized spacetime symmetries demand that the terms in eq. (\ref{KR spelled out}) always appear in this specific combination. This means that, for our purposes, it is enough to show that {\em one} particular term cannot get renormalized. It is easy to realize that the second term in (\ref{KR spelled out})---the tadpole for $\delta A_{0z}$---cannot get renormalized: it is the only term in the whole Lagrangian with an undifferentiated $\delta A_{0z}$, and it is a linear term that cannot yield vertices with more than one external leg, so there are no interactions anywhere in the theory that can contribute to a loop with an undifferentiated $\delta A_{0z}$.
This implies that its coefficient---our $\lambda$---cannot receive loop corrections.

%\footnote{As a matter of fact, the second term in (\ref{KR spelled out}) is the only term \emph{in the whole Lagrangian} with an undifferentiated $A_z$. Therefore, our last argument would have been sufficient to prove that $\lambda$ does not get renormalized. However, we wanted to explicitly make the point that the usual argument based on turning the couplings into functions of the space-time coordinates, although very powerful and probably familiar to many readers, is not by itself sufficient to rule out the possibility that the Kalb-Ramond term renormalizes itself.} 

As mentioned previously, this non-renormalization theorem is a low-energy manifestation of the fact that our effective theory admits at least one UV completion where the coupling $\lambda$ turns out to be quantized.

%%%%%%%%%%%%%%%%%%%%%%%%%%
%%%%%%%%%%%%%%%%%%%%%%%%%%
\section{Applications}

We will now use  the effective action we have just derived for a number of sample computations. Some just reproduce classic results in a new language; others actually lead to new results.

%%%%%%%%%%%%%%%%%%%%%%%%%%
%%%%%%%%%%%%%%%%%%%%%%%%%%

\subsection{Kelvons, nonlinear Kelvin waves, and the self-pipe}\label{kelvons}

To begin with, consider the free propagation of small perturbations on a straight string---the famous Kelvin waves. The effective action \eqref{Seff phi} expanded to quadratic order reads
\be\label{Seff2 kelvon}
S_{\rm eff}  \simeq \int d t d z \big[ \bar n \lambda \, \phi^* i \d_t \phi -   T(k) |\d_z  \phi|^2 \, \big] \; .
\ee
Plugging a plane-wave ansatz $\phi \sim e^{-i (\omega t - k  z)}$ into the corresponding equations of motion, we immediately get 
the dispersion law
\be
\omega =  \frac{1}{\bar n \lambda} T(k) \, k ^2 \; .
\ee
Using our formula for the running tension, eq.~\eqref{running T}, we finally get
\be \label{omega kelvin}
\omega = \frac{1}{4\pi} \frac{\bar n \lambda}{\bar w} \,  \log(\mu_0/k) \, k ^2 \; ,
\ee
which, upon relating our coupling $\lambda$ to the circulation via eq.~\eqref{lambda}, matches the classic result~\cite{donnelly1991quantized}.
%Notice that this result is usually derived in the incompressible limit. Here we proved that sound waves cannot correct it:\ at this order in the expansion in small fields and velocities, their only effect is to renormalize the tension by an additive constant term, without affecting its running; that is, they only contribute to the value of $\mu_0$.

In the literature, our fixed scale $\mu_0$ is taken to be  the inverse string thickness $1/a$ times order-one numerical factors that depend on the model one adopts for the core of the string. Although by dimensional analysis we do expect $\mu_0$ to be of order of $1/a$, we will refrain from being too specific about their  relation: as we hope we have made clear, the truly robust prediction of our effective field theory is the dependence on the IR scale $k$, 
\be
\frac{d(\omega/k^2)}{d \log k} = - \frac{1}{4\pi} \frac{\bar n \lambda}{\bar w} \; .
\ee
If the actual value of $\mu_0$ is needed, it is better left as something to be fit for, for instance by measuring the energy it takes to set up a string in a container of given size $L$.

Upon quantization, the Kelvin waves describe gapless excitations traveling along the string, with energies quantized in units of \eqref{omega kelvin}. These excitations are called `kelvons'. Notice that, due to the one-derivative nature of our kinetic term, $\phi$ only has positive frequency modes and $\phi^*$ only negative frequency ones. This implies that, in the quantum theory, $\phi$ can only destroy quanta and $\phi^\dagger$ can only create them:
\be
\phi(x)  = \frac{1}{\sqrt{\bar n \lambda}} \int \frac{dk}{2\pi}  \, a_k \, e^{-i (\omega_k t - kz)} \; , \qquad\quad \phi^\dagger(x) = \frac{1}{\sqrt{\bar n \lambda}}  \int \frac{dk}{2\pi} \, a^\dagger_k \, e^{+i (\omega_k t - kz)}  \; ,
\ee
with canonical commutation relations $[a_k, a^\dagger_q] = (2\pi) \delta(k-q)$.
In particular, there are no antiparticles. Yet, the kelvons carry a conserved $U(1)$ quantum number, which is nothing but the $z$-component of the angular momentum. This means that kelvons cannot annihilate nor be created unless some angular momentum is exchanged with the bulk modes. So, for instance, consider the expansion of the term proportional to $T_{(01)} (\mu)$ in\eqref{Seff B phi}:\ a scalar coupling like $|\partial _ z \phi|^2 (\vec \nabla \cdot \vec B)_0$ cannot trigger kelvon creation or annihilation---indeed, it contains $a a^\dagger$ only---whereas a tensor coupling like $\pi^i \pi^j \, \big(\d_i \d_j \vec \nabla \cdot \vec B\big)_0$ can.

Going back to classical wave solutions, notice that there is no reason why we should stop the expansion of the effective action at quadratic order in $\phi$: the action \eqref{Seff phi} is valid at non-linear order as well, as long as the perturbations of the string are characterized by a single characteristic wavenumber $k$, and the velocities are small.
So, we can look for non-linear wave solutions as well.
Consider then a plane-wave ansatz of momentum $k$,
\be
\phi (t,z) = \sfrac{1}{\sqrt 2} R \, e^{-i (\omega t - k z)} \; ,
\ee
where we are parameterizing the amplitude directly in terms of the radius $R$ of the helix described by the perturbed string.
Plugging this into the non-linear equations of motion,
\be
i \d_t \phi + \frac{T(k)}{\bar n \lambda} \d_z \bigg( \frac{\d_z  \phi}{\sqrt{1+2|\d_z  \phi|^2} } \bigg) = 0 \; ,
\ee
we get
\be
\omega = \frac{T(k)}{\bar n \lambda} \frac{k^2}{\sqrt{1+ (R k)^2}} \; .
\ee
For amplitudes much smaller than the wavelength, we get back the correct dispersion law for linear Kelvin waves.  In the opposite limit, we get an approximately linear dispersion law:
\be \la{disp-rel self pipe}
\omega \simeq \frac{T(k)}{\bar n \lambda R} \cdot k\; , \qquad R \gg 1/k \; .
\ee 

It is interesting to think of this large-amplitude Kelvin wave as a tightly wound solenoid:\ recall that there is a formal analogy between vortex lines and magnetostatics, with the lines playing the roles of current-carrying wires, and the velocity field that of the magnetic field \cite{donnelly1991quantized, Endlich:2013dma}. Then, for a configuration like ours, we must have a negligible fluid-flow outside the helix, and an approximately uniform one inside it, given by
\be \la{v in}
\vec v \simeq \hat z \, \frac{\Gamma k}{2\pi} \; ,
\ee
where $\Gamma = \bar n \lambda/ \bar w$ is the line's circulation. That is, this is a solution of the hydrodynamical equations in which the fluid arranges itself into a flow-carrying {\it pipe}. We call this a ``self-pipe". Notice that the pipe itself moves, with a velocity given by the propagation velocity of our non-linear wave,
\be
v(k) \simeq \frac{\Gamma}{4\pi R} \log(\mu_0/k) \; .
\ee
Our non-linear solution is self-consistent, in the sense that one can show that---on this solution---all the terms  we have neglected to arrive at eq.~(\ref{Seff phi}) are smaller than  the ones we have kept. 
%In fact, it is easy to check that all the interactions coming from the generalized Nambu-Goto term contain at least one time-derivative $\d_t \phi$. Based on the dispersion relation (\ref{disp-rel self pipe}) and the fact that the velocity $v$ in (\ref{v in}) must be small, we have 
%%
%\be
%\d_t \phi \sim \omega R \sim \fr{\Gamma k^2R }{\sqrt{1 + (k R)^2}} \sim \fr{v}{\sqrt{1 + (k R)^2}} \times kR \ll kR \sim \d_z \phi.
%\ee
%%
%Thus, in our regime it is consistent to neglect the generalized Nambu-Goto term while keeping an infinite number of powers of $\d_z \phi$.

Notice also  that in the non-linear regime we cannot take linear combinations of our plane-waves to construct a localized wave packet. So, our non-linear solutions are completely delocalized monochromatic plane waves. Perhaps this makes them uninteresting from a physical standpoint. Still, it would be interesting to check their existence via numerical simulations: to the best of our knowledge, they have not appeared in the literature. In fact, in the standard vortex-filament model the local induction approximation would break down for $R \gtrsim 1/ k $---because the typical distance between the spires of the helix $1/k$ would become smaller than the local radius of curvature $R$---thus making it difficult to extend the analysis of Kelvin waves to non-linear amplitudes.\footnote{We thank Claudio Barenghi for pointing this out to us.} 
On the other hand, our RG considerations are still valid and show that the tension has to be evaluated at $k$ rather than $1/R$: the field theory does not know that the amplitude $R$ is physically a length scale; thanks to its own ignorance, like Sikorsky's famous bumblebee,  it can fly farther.

%%%%%%%%%%%%%%%%%%%%%%%%%%
%%%%%%%%%%%%%%%%%%%%%%%%%%
\subsection{Vortex rings and their interactions} \la{vortex interactions}

As another application of the same set of ideas, consider a perfectly circular vortex ring of radius $R$ in an otherwise unperturbed superfluid. As is well known~\cite{donnelly1991quantized}, such a  ring moves at a constant speed of order
\be
v \sim \frac{\Gamma}{R} \log R/a \; ,
\ee
where $\Gamma$ is the circulation, and $a$ is typically taken to be on the order of the inverse core size. We want to recover this result with our techniques.

Assuming the ring stays circular, we can parameterize its dynamics in terms of its radius $R(t)$, the position $\vec x_0(t)$ of its center of mass, and
its normal unit vector $\hat n(t)$. We thus have
\be
\vec X(t, \sigma) = \vec x_0 (t) + \Delta \vec X(t, \sigma) \; ,
\ee
where, as a function of $\sigma$  at fixed $t$, $\Delta \vec X$ spans a circle with orientation $\hat n(t)$ and radius $R(t)$. Choosing $\sigma$ to be the angle for such a circle and performing the $\sigma$ integral in our effective action \eqref{local Seff}, we get
\be \label{ring action}
S_{\rm eff} [\vec x_0, R, \hat n] \simeq \int dt \big[ \lambda \bar n \, \pi R^2 \hat n \cdot \dot{\vec x} _0 - 2\pi R \, T(1/R) \big] \; ,
\ee
where  we now chose $\mu = 1/R$ as renormalization scale. Recalling that $\hat n$ has to be varied while preserving its unit norm,
the equations of motion are 
\begin{align}
\delta \vec x_0  :& \qquad  \frac{d}{dt} (R^2 \hat n) = 0 \\
\delta \hat n :  & \qquad \big(  \dot{\vec x} _0 \big)_\perp = 0 \\
\delta R  : & \qquad \hat n \cdot \dot{\vec x} _0 = \frac{1}{\lambda \bar n}  \, \frac{1}{R} T(1/R) \; .
\end{align}

The first equation says that the ring must preserve its size and orientation; the second says that the ring can only move along $\hat n$; the third says that it does so at a constant speed
\be \label{ring speed}
v = \frac{1}{\lambda \bar n}  \, \frac{1}{R} T(1/R) =  \frac{\bar n \lambda}{\bar w} \, \frac{\log(R \mu_0)}{4\pi R}  \; .
\ee
This matches the standard result~\cite{donnelly1991quantized}, again after identifying the prefactor $\bar n \lambda/ \bar w$ with the circulation $\Gamma$ (see eq.~\eqref{lambda}).

From the effective action above, we can also derive the momentum and energy of the ring, either by Noether's theorem, or most simply by the canonical relations,
\begin{align}
\vec p & = \frac{\d {\cal L}_{\rm eff}}{\d \dot{\vec x}_0} = \lambda \bar n  \, 2\pi R^2 \, \hat n \\
E & = \vec p \cdot \dot{\vec x}_0 - {\cal L}_{\rm eff} = \frac{\bar n^2 \lambda^2}{\bar w} \frac{R}2  \log (R \mu_0) \; ,
\end{align}
in agreement with the classic results~\cite{donnelly1991quantized}.

%Again, in the literature these results are derived in the incompressible limit. Our techniques show that, to first order in time derivatives, the inclusion of sound cannot affect their $R$-dependence, but  only the precise value of the reference UV scale $\mu_0$. In particular, the formula for $\vec p$ is completely unaffected.

We can use our formalism to keep track of couplings to the bulk fields as well.  The leading couplings are the terms in $S_{\rm KR}$---expanding out $\vec X(t, \sigma) = \vec x_0 (t) + \Delta \vec X(t, \sigma)$ and integrating $\sigma$ over the circle as before, we find
\begin{equation} \label{AB ring}
\bar{n}\lambda \pi R^2 \int dt \, \l\{ -\partial_t{\vec B}  + \vec \nabla \times \vec A - \dot{\vec{x}} \, \big( \vec \nabla \cdot \vec B \big) \r\} \cdot \hat{n} \,.
\end{equation}
Note that this expression contains only gauge invariant combinations of $\vec{A}$ and $\vec B$, as it must. Moreover, even though $\partial_t{\vec B}  - \vec \nabla \times \vec A$ and $\vec \nabla \cdot \vec B$ are separately gauge invariant, the relative coefficient is not arbitrary but fixed by the non-linearly realized symmetries. By integrating out $\vec A$ and $\vec B$, we can compute  the long-distance interaction potential between two vortex rings. To leading order in the $v/c_s$ expansion, such a potential is dominated by the exchange of a single $\vec A$. The general formalism is still that of sect.~\ref{running tension}, but now with the $\vec J_A$ source of eq.~\eqref{JA JB} given  by
\be
\vec J_A(\vec x , t) =  -\sum_{I=1,2}  \vec \mu_I \times \vec \nabla \delta^3(\vec x -\vec x_I ) \; , \qquad\qquad  \vec \mu_I \equiv \bar n \lambda_I (\pi R_I^2)  \hat n_I \ ,
\ee
where $I$ labels the two vortex rings. Plugging in the $\vec A$ propagator and recalling that the potential appears in the action as $S_{\rm eff} \supset - \int dt \, V_{\rm eff}$, we find the effective potential \cite{Endlich:2013dma}
\be
V_{\rm eff} \simeq \frac{1}{4\pi \bar{w}}\cdot \frac{3(\vec \mu_1 \cdot \hat{r})(\vec \mu_2 \cdot \hat{r}) - \vec \mu_1 \cdot \vec \mu_2}{r^3} \; ,
\label{hydrophotonpotential}
\ee
where $\vec r$ is the vector connecting the two rings. Notice that this potential has precisely the same structure as a dipole-dipole interaction in magnetostatics. In fact, the interaction of a ring with $\vec A$ in \eqref{AB ring} is formally the same as that of a magnetic dipole with the vector potential, with dipole moment proportional to our $\vec \mu$.

It should be recalled that this effective potential is physically an energy, but it does not lead to a ``force" in any standard sense, because of the unconventional kinetic action for a free vortex ring \eqref{ring action}---in particular, because of its single time derivative \cite{Endlich:2013dma}. It is thus better to think of (minus) the effective potential as a term in the effective action for two rings, which corrects the equations of motion as implied by the variational principle.

The subleading contributions to the  two-ring effective action coming from sound exchange are also easy to compute. The coupling in \eqref{AB ring} corresponds to a source for~$\vec B$
\be
J_B^i = \sum_{I=1,2} \big( - \mu^i_I \,  v^j_I  + (\vec \mu_I \cdot \vec v_I) \delta^{ij}  \, \big) \nabla ^j\delta^{3}(\vec x - \vec x_I) \; . 
\ee
Plugging in the $\vec B$ propagator, we find the correction to the effective Lagrangian
\begin{equation}
\begin{split}
{\cal L}_{\rm eff} \supset &  \frac{1}{8 \pi \bar{w} } \, \frac{(\mu_1 v_1) (\mu_2 v_2)}{c_s^2 r^3} \big\{-1 + 2(\hat{n}_1 \cdot \hat{n}_2)^2 - 12(\hat{n}_1 \cdot \hat{n}_2)(\hat{n}_1\cdot \hat{r})(\hat{n}_2 \cdot \hat{r}) \\
& \qquad \qquad \qquad \qquad \qquad + 15(\hat{n}_1 \cdot \hat{r})^2(\hat{n}_2 \cdot \hat{r})^2 \big\}  \ . \label{Bdot Bdot}
\end{split}
\end{equation}
Recalling that, up to logs, $ v \sim \Gamma / R \sim (\bar n \lambda/ \bar w) R$, we see that this is suppressed by a factor of $(v/c_s)^2$ compared to the leading order result \eqref{hydrophotonpotential}. Then, at this order, there is another ring-sound coupling that we should consider,
\be \label{ring div B}
\int dt \, (4\pi R) \, T_{(01)}(1/R) \, \vec \nabla \cdot \vec B \; ,
\ee
which comes from taking the integral of \eqref{T01} around the ring. Using this vertex in conjunction with the previous one, we get a further correction to the two-ring effective Lagrangian,
\be
{\cal L}_{\rm eff} \supset \frac{(\mu_1 v_1) (4 \pi R_2 T_{(01)}(\mu)_2 )}{8 \pi c_s^2 r^3} \big(1 - 3(\hat{n}_1 \cdot \hat{r})^2 \big) + 
( 1 \leftrightarrow 2 ) \label{Bdot div B}
\ee
(the vertex \eqref{ring div B} used twice only yields a contact interaction, proportional to $\delta^3(\vec x_1 - \vec x_2)$, with no physical consequences at large distances.)

Upon using the free ring eom \eqref{ring speed}, the sum of \eqref{Bdot Bdot} and \eqref{Bdot div B} matches the sound-mediated long-distance interaction computed in \cite{Endlich:2013dma} via somewhat different methods. There, two variables ($\alpha$ and $\beta$) parametrized the core structure of the ring for the case of somewhat ``fat" rings, which is the realistic case for vortex rings in ordinary fluids like water. Here, the same role is played
by the two UV reference scales ($\mu_0$ and $\mu_0'$) appearing logarithmically in $v$ and $T_{01}$. In the limit of very thin rings, the logs are large and  the difference between these two scales can be neglected.

%%%%%%%%%%%%%%%%%%%%%%%%%%
%%%%%%%%%%%%%%%%%%%%%%%%%%

%%%%%%%%%%%%%%%%%%%%%%%%%%
%%%%%%%%%%%%%%%%%%%%%%%%%%
\subsection{Phonon absorption} 
\begin{figure}[htb!]
\centering%
\includegraphics[scale = 0.2]{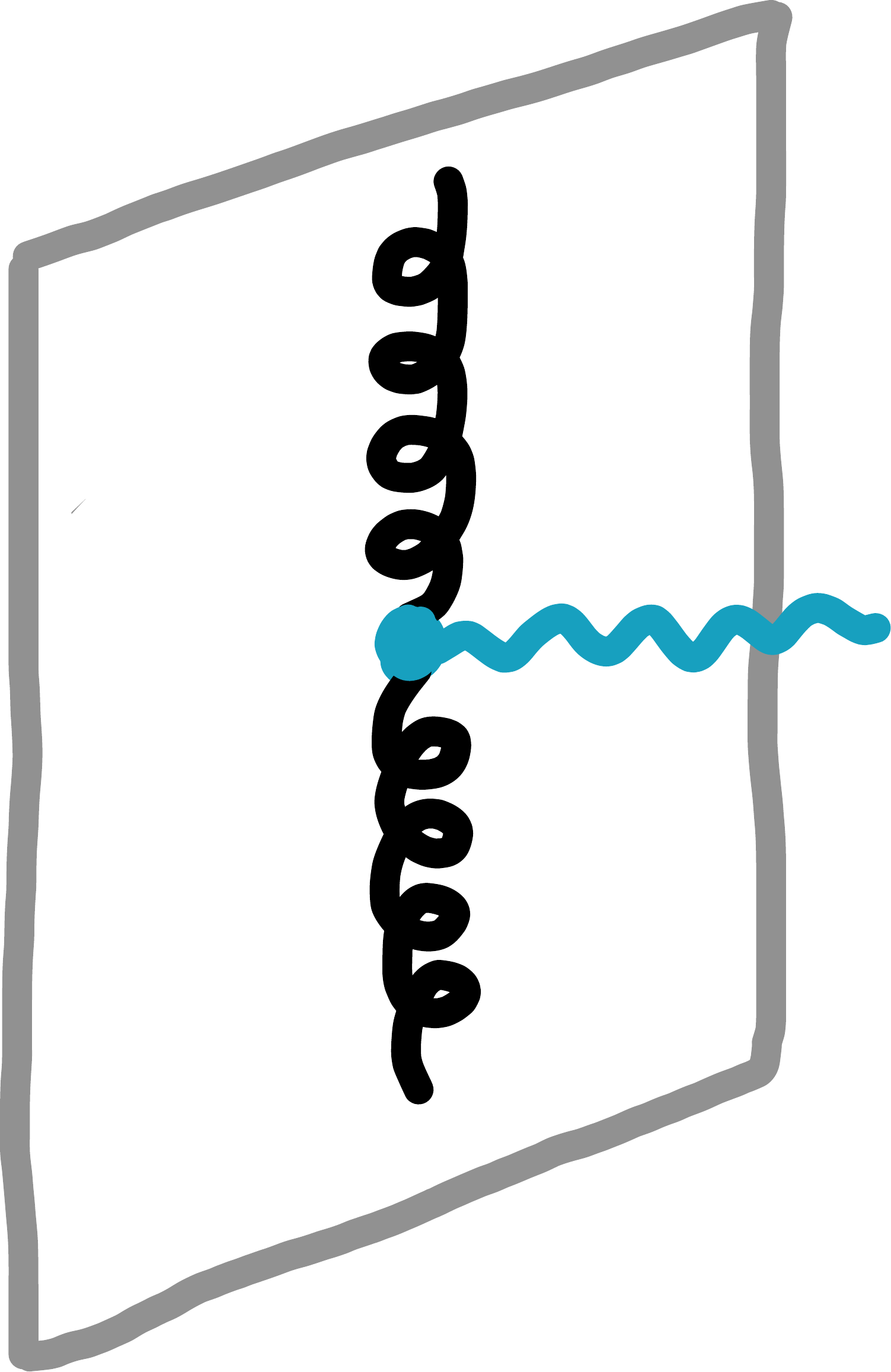}
\caption{\emph{Absorption of a phonon into a pair of kelvons.}}
\label{fig:phonon to kelvons}
\end{figure}

\la{sec:phonon absorption}
As a final application of our methods, we now study a process in which a phonon gets absorbed by a straight string. 
For simplicity, let us consider first the case in which the initial phonon propagates perpendicularly to the string, with momentum $k$ and energy $\omega=c_s k$. The string breaks translations perpendicularly to itself, which means that the momentum of the phonon in the transverse direction is not conserved. However, momentum along the string {\it is} conserved, and so is energy.
The leading contribution to absorption thus comes from converting the energy of the phonon into two quanta of Kelvin waves (kelvons) with equal energies $E = \omega/2$ and opposite momenta $\pm q$ along the string, as in the diagram of Fig.~\ref{fig:phonon to kelvons}. 

Notice that the final kelvon momenta are much larger than the initial phonon momentum. To see this, observe that, for given initial energy $\omega = c_s k$, the momentum $q$ of the two final kelvons is given implicitly by their dispersion law:
\be \label{E q}
E = {\omega}/{2}=\frac{1}{\bar n \lambda} \, T(q) \, q^2 \; .
\ee
Up to logarithmic corrections, $q$  scales as $\sqrt{\omega}$, while the phonon momentum  $k$ scales as $\omega$, which makes $q$  much bigger than $k$ in the low-frequency limit.
This implies that, if we now consider the more generic case in which the initial phonon propagates at an angle $\theta$ relative to the string, the kinematics of the process are essentially unaltered. The reason is that even though now there is a mismatch in the final kelvon momenta, it is only $\Delta q = k \cdot \cos \theta \ll q$ and  can thus be neglected in first approximation.\footnote{Another consequence of the large $q/k$ hierarchy---which  turns out not to be relevant for the computation at hand---is that one should choose $\mu \sim q$ as renormalization scale for all string-sound running couplings.} For a generic initial angle $\theta$ we thus have
\be
q_1 \simeq -q_2 \equiv q \; , \qquad E_1 \simeq E_2 \equiv E = \omega/2 \; .
\ee

Keeping in mind the remarks of section~\ref{kelvons}, the leading contribution to kelvon pair production by sound comes from expanding  the $\epsilon \cdot B \,  \d X \d X$ coupling in \eqref{Seff B phi}:
\be
\bar n \lambda  \cdot \epsilon_{ijk} B^k(\vec X, t) \, \d_t X^i \d_z X^j \supset \sfrac14 {\bar n \lambda} \cdot  \phi^* {}^2 \big[ \d_x \dot B_y + \d_y \dot B_x + i \big( \d_y \dot B_y - \d_x \dot B_x \big) \big] \ .
\ee
Without loss of generality, we can take the direction of propagation for the phonon to lie in the $yz$-plane. Since $\vec B$ is a longitudinal field, this kills all the terms apart from
\be
\sfrac14 {\bar n \lambda} \cdot \phi^* {}^2 \, i \d_y \dot B_y  \; .
\ee

We provide a quick derivation of  generic amplitude and cross-section formulae for mixed bulk/world-sheet processes like this in Appendix \ref{FeynmanKelvin}. After one takes into account that the single-time derivative nature of the $\phi$ kinetic term yields an extra factor of $\sqrt{2E}$ for each external $\phi$ line, the final formulae are what one expects from analogy with more standard relativistic cases: the scattering amplitude is
\be \label{iM}
i {\cal M} = \frac{2E}{\bar n \lambda \sqrt{\bar w}} \times \frac{\bar n \lambda}{2}\,  i k \omega \sin^2 \theta \; , 
\ee 
where the first factor collects all the normalization factors associated with the external lines, including the non-canonical normalizations of our fields, while the second  comes from the interaction Lagrangian.
The differential cross section is related to the amplitude by
\be
d \sigma = \frac{1}{c_s} \frac{1}{2 \omega} |{\cal M}|^2 d \Pi_2   \; .
\ee
The two-particle final phase-space integrates to
%\be
%d \Pi_2 = \frac{dq}{(2\pi) 2E} \frac{dq'}{(2\pi) 2 E'} \, (2 \pi)^2 \delta(\omega - (E+E')) \delta(q + q') \; ,
%\ee
%which integrates to 
\be
\Pi_2 \simeq \frac{1}{8 E^2 v(q)} \; ,
\ee
where $v(q)$ is the group velocity of the Kelvin waves,
\be
v(q) \equiv {dE}/{dq} \simeq 2 E/q \; .
\ee
We assumed that $\log \mu_0/q$ is very large, so that we can neglect terms that are not log enhanced.
Putting  everything together, we get a total cross-section 
\be \label{cross section}
{\sigma} \simeq \frac{1}{16\, \bar w c_s^3} \, \omega^2 q \sin^4 \theta\; ,
\ee
where $\omega$ and $q$ are related by eq.~\eqref{E q}. Notice that $\sigma$ has units of length, as befits the cross-sectional {\it width} of a string. It is the probability rate of absorption per unit string length and unit phonon incoming flux.

To the best of our knowledge, this result is new. We can express it in terms of the original phonon's frequency $\omega = 2 E$ only, by approximately inverting \eqref{E q} at low energies,
\be \label{q E}
q^2 \simeq - \frac{8 \pi \bar w}{\bar n \lambda} \cdot  \frac{E}{\log  \frac{8 \pi \bar w}{\bar n \lambda } \frac{E}{\mu_0^2} } \; ,
\ee
where we assumed once again that the running log is large. In such a case, one needs not be too precise about the value of $\mu_0$, and one can probably approximate it by $\mu_0 \sim 1/a$. However, the actual value of $\mu_0$ is completely well defined and physical, and can be fit for by measuring the string tension \eqref{running T} at any chosen scale $\mu$. Plugging \eqref{q E} into \eqref{cross section}, we see that the cross section scales as $\omega^{5/2}/ \sqrt{\log \omega}$ at low frequencies. Such a peculiar scaling is a robust prediction of the effective field theory.

%%%%%%%%%%%%%%%%%%%%%%%%%%
%%%%%%%%%%%%%%%%%%%%%%%%%%

\section{Power Counting} \la{sec: power counting}

The low-energy effective action (\ref{effS}) contains an infinite number of terms.  From an EFT viewpoint, however, only a finite number of these terms will contribute to observable quantities \emph{at any given level of precision}.  The precision is controlled by typical scales in the system: in the simplest relativistic theories these are usually ratios of energies and masses to some UV scales, whereas in our theory the expansion parameters will be ratios of lengths and velocities.  In this section  we develop a systematic power counting scheme, along the lines of \cite{Goldberger:2004jt,Goldberger:2007hy}, to estimate the sizes of various terms in the effective action. Since our action (\ref{effS}) should describe superfluids as well as fluids in which the vorticity is concentrated along lines, in what follows we will err on the side of generality and develop a power counting scheme that applies to both systems.

Our EFT approach treats the vortex lines as purely (1+1)-dimensional objects. Clearly, this is a good approximation only if the core radius $r_c$ is much smaller than the typical length scale $\ell$ over which the shape of the vortex line changes appreciably. In other words, one expansion parameter of our theory will certainly be the ratio $r_c / \ell$, since when this becomes of order one our effective theory stops being accurate. At the same time, in ordinary fluids the radius of a vortex core can be much larger than the inter-particle separation $a$, which is the scale suppressing interactions of the bulk hydrodynamical modes.\footnote{To be precise, for a weakly coupled  gas one should replace the inter-particle separation $a$ with the mean free path, which can be much bigger than $a$. For simplicity, in the following we will ignore this difference, and thus strictly speaking our analysis applies to liquids.}  Therefore the ratio $a/\ell$ is in principle a second parameter independent from $r_c / \ell$, in general smaller than it, and potentially much smaller. Finally, we have pointed out in section \ref{expansion} that the  velocity $v$ of vortex lines has to be much smaller than the speed of sound $c_s$, which in turn cannot be greater than the speed of light $c$.  This suggests the existence of two more expansion parameters given by the ratios $v / c_s$ and $c_s/c$. 

In the simplest case of non-relativistic superfluids like liquid helium, the parameter $c_s/c$ is negligibly small, while $r_c / \ell$, $a/\ell$, and $v/c_s$ are all of the same order, and thus the theory has really just one small parameter. This can be easily explained using dimensional analysis, keeping separate units for space and time (but still setting $\hbar =1$). Since the only microscopic scales that characterize a non-relativistic superfluid are the atomic mass $m$ and the inter-particle separation $a$, the vortex core size must be given by $r_c \sim a$, the speed of sound  by $c_s \sim (m a)^{-1}$ and the quantized circulation by $\Gamma \sim c_s a \sim 1/m$. Therefore, we must have
\be \la{superfluidparameters}
v \sim \fr{\Gamma}{\ell} \sim  \fr{c_s a}{\ell} \qquad  \Longrightarrow \qquad  \fr{v}{c_s}\sim \fr{a}{\ell} \sim \fr{r_c}{\ell}, \qquad\qquad \text{(NR superfluid}).
\ee
In more general cases, however, these three parameters are in principle all independent of each other as well as of $c_s/c$. For this reason, in what follows we will keep track of powers of $r_c / \ell, a/\ell, v/c_s$ and $c_s/c$ separately. Where appropriate, it is straightforward to express everything in terms of a single expansion parameter using equation (\ref{superfluidparameters}).

\subsection{Bulk fields}

In order to determine which terms in the action should be kept and which ones should be neglected to calculate a given observable up to a certain precision, we need to be able to estimate how each term scales with our expansion parameters. In order to do this consistently, we will follow~\cite{Goldberger:2007hy,Goldberger:2004jt} and use the \emph{method of regions}~\cite{Beneke:1997zp,Griesshammer:1997wz}. The main idea is that to come up with well defined power counting rules one must first identify the kinematical regions that are relevant for the problem at hand, and then decompose all fields into a sum of contributions that ``live'' in the different regions.  Let us see how this works in practice in the case of the bulk fields $\vec A$ and $\vec B$.

As we already emphasized in section ~\ref{two form}, with our gauge choice the field $\vec A$ is purely non-dynamical (see the $\vec A$ propagator in eq.~(\ref{propagators})). This means that its typical frequency and wave-number are completely determined by the sources, namely the vortex lines, and thus we have
\be
\vec{A}: \qquad \omega \sim \fr{v}{\ell}, \qquad k \sim \fr{1}{\ell}.
\ee
In keeping with the standard nomenclature~\cite{Goldberger:2007hy}, we will refer to this kinematical region as the \emph{potential region}. This terminology emphasizes that the fields are never  on-shell in this region. To capture instead effects such as the emission of on-shell phonons from vortex lines~\cite{Endlich:2013dma}, we need to introduce another kinematical region, which is traditionally called the \emph{radiation region}. Since the frequency of the emitted radiation is determined by the typical frequency of the sources, this will again be of order $v / \ell$. However, the relevant wave-number is now fixed by the phonon dispersion relation, and therefore for \emph{radiation phonons} we have
\be \la{radiation region}
\vec{B}_\text{rad}: \qquad \omega \sim \fr{v}{\ell}, \qquad k \sim \fr{v}{c_s \ell}.
\ee
Finally, consistency requires also the introduction of \emph{potential phonons}, whose frequency and wave-number scale as follows: 
\be \la{potential region}
\vec{B}_\text{pot}: \qquad \omega \sim \fr{v}{\ell}, \qquad k \sim \fr{1}{\ell}.
\ee
This is because potential phonons can be produced for instance when two $\vec{A}$'s interact in the bulk via the cubic vertex shown in eq.~(\ref{NR limit}). Potential phonons are not just necessary for consistency, however, but are in fact responsible for very physical effects such as the phonon-mediated interaction between vortex lines discussed in section \ref{vortex interactions} and in ref.~\cite{Endlich:2013dma}. To keep track of both potential and radiation phonons, we will simply perform the substitution 
\be
\vec B = \vec{B}_\text{rad} + \vec{B}_\text{pot}
\ee
in the effective action (\ref{effS}). As we will see in what follows, terms in the action that depend on $\vec{B}_\text{rad}$ will in general scale differently from terms that depend on $\vec{B}_\text{pot}$, and therefore we need to separate these two contributions explicitly.

Now that we have identified the relevant kinematical regions, we know how derivatives acting on $\vec A, \vec{B}_\text{pot}$ and $\vec{B}_\text{rad}$ will scale. Next, we need to determine how the fields themselves scale with $\ell, r_c, a, v$ and $c_s$ (for now we are setting $c=1$ for simplicity). This can be done by estimating the size of the propagators in momentum space. For instance, the Fourier transform of the field $\vec A$ scales schematically as follows:
\be \la{Ascaling1}
\langle \tilde A \tilde A \rangle \sim \frac{1}{\bar w}\frac{\delta(\omega)\delta^3(k)}{k^2} \sim c_s a^4 \times \fr{\ell}{v} \times \ell^3 \times \ell^2 \qquad \Rightarrow \qquad  \tilde A \sim \frac{c_s^{1/2}a^2 \ell^3}{ v^{1/2}}
\ee
To estimate the size of $\bar w$ we used the fact that it is a mass density, and thus it must scale like $\bar w \sim m / a^3 \sim (c_s a^4)^{-1}$. We can now use equation (\ref{Ascaling1}) and go back to position space to obtain the scaling 
\be \la{Ascaling2}
\vec A \sim \int d\omega d^3 k \, \tilde A \sim \fr{v}{\ell} \times \fr{1}{\ell^3} \times  \frac{c_s^{1/2}a^2 \ell^3}{ v^{1/2}} \sim  (v c_s)^{1/2}  \fr{a^2}{\ell}.
\ee
This scaling result should not be confused with the amplitude of a field $\vec A$ sourced by a typical vortex configuration. The latter can be estimated using the scaling (\ref{Ascaling2}) together with simple linear response theory; up to order one factors:
\be
\langle \vec A \rangle \simeq \int dt d^3 x \, \langle \vec A \vec A \rangle J \quad \text{with} \quad  J  = \bar{w}\Gamma \int  d\sigma \, \delta^{3}(x-X(\sigma, t)) \d_\sigma X \qquad
\Rightarrow \qquad \langle \vec A \rangle \simeq v \, \ell \; .   \qquad 
\ee
Instead, the scaling rule \eqref{Ascaling2} and the ones that follow should be interpreted as formal building blocks that we can combine to estimate the {\em relative} size of all the  terms in the action that can contribute to a given process, without any direct implication for the absolute size of those terms for that same process.

In a similar way, we can also find the scaling of potential and radiation phonon fields:
\begin{align}
\langle \tilde B_\text{pot} \tilde B_\text{pot} \rangle \sim \frac{1}{\bar w}\,  \frac{\delta(\omega)\delta^3(k)}{c_s^2 k^2} \sim c_s a^4 \times \fr{\ell}{v} \times \ell^3 \times \fr{\ell^2}{c_s^2} \quad & \Rightarrow \quad \vec B_\text{pot} \sim \bigg(\fr{v}{c_s}\bigg)^{1/2}  \fr{a^2}{\ell} \la{Bpscaling} \\
\langle \tilde B_\text{rad} \tilde B_\text{rad} \rangle \sim \frac{1}{\bar w} \,  \frac{\delta(\omega)\delta^3(k)}{\omega_k^2 - c_s^2 k^2} \sim c_s a^4  \times \fr{\ell}{v} \times \fr{c_s^3\ell^3}{v^3} \times \fr{\ell^2}{v^2} \quad & \Rightarrow \quad \vec B_\text{rad} \sim \fr{v}{c_s}  \fr{a^2}{\ell}   \; . \la{Brscaling}
\end{align}
Notice that in the Fourier transform of $\tilde B_\text{rad}$, the integration measure $d^3k$ scales like $v^3 /(c_s^3 \ell^3)$. The scalings (\ref{Ascaling2}), (\ref{Bpscaling}) and (\ref{Brscaling}) do not depend on the radius of the core: this makes sense, because our estimates are based on the bulk propagators, and the bulk part of the action does not know anything about the vortex lines.

\subsection{Vortex lines as external sources}

Let us now consider the vortex lines. If we treat them as external sources rather than dynamical objects, their scaling is entirely determined by simple geometric considerations to be:
\be
\vec X \sim \ell, \qquad\quad  \d_t \vec{X} \sim v, \qquad\quad  \d_\sigma \vec X \sim 1.
\ee
By combining these scalings with the ones for the hydrophoton and phonon fields, we can easily estimate the size of all the terms in the action. To this end, we will assume that the various coefficients in the Lagrangian take on a natural value which can be inferred from the logarithmic divergences studied in sect.~\ref{runningcouplings}.  What follows is a pedagogical derivation of the size of the most relevant terms in the action. We will repeatedly use the fact that $\lambda  =\Gamma \bar w /\bar n$.
\begin{itemize}
\item First, let us estimate the size of the kinetic term for the vortex lines. Schematically, we have
\be \la{kinsize}
\bar n \lambda \int dt d \sigma X \dot X X' \sim v\ell \times \fr{1}{c_s a^4} \times \fr{\ell^2}{v} \times \ell \times v \times 1 \sim \frac{v}{c_s}\fr{\ell^4}{a^4} \sim L,
\ee
where $L \sim \bar{ w} \ell^3 \times  v \times \ell$ is the typical angular momentum of the whole configuration, which has the right units to be compared to the Planck constant $\hbar$.  
We will always be interested in the regime $L/\hbar \gg 1$, which makes the dynamics of our strings essentially classical.
It is easy to check that the tension term is also of the same order: 
\be \la{tensize}
\Gamma^2 \bar{w} \int dt d\sigma |X'| \sim (v\ell)^2 \times \frac{1}{c_s a^4} \times \fr{\ell^2}{v} \times 1 \sim L.
\ee
\item Let us now estimate the size of the couplings to the hydrophoton.  The most relevant of such couplings is the non-derivative one, which scales as
\be \la{AX'}
\bar n \lambda \int d t d \sigma A X' \sim  (v \ell) \times \fr{1}{c_s a^4} \times \fr{\ell^2}{v} \times (v c_s)^{1/2}\frac{a^2}{\ell} \times 1 \sim  \sqrt{L}.
\ee
\item The leading non-derivative coupling with the potential phonon is suppressed compared to this by an extra power of $v/c_s$,
\be \la{BpX'Xdot}
\bar n \lambda  \int d t d \sigma \, B_{\rm pot} \dot X X' \sim v \ell \times \fr{1}{c_s a^4} \times \fr{\ell^2}{v} \times \left(\fr{v}{c_s}\right)^{1/2} \frac{a^2}{\ell} \times v \times 1 \sim \frac{v}{c_s} \, \sqrt{L},
\ee
and are of the same order as the leading derivative couplings:
\ba
\frac{\bar n^2 \lambda^2}{\bar w} \int d t d \sigma \, \nabla B_{\rm pot} X' \sim \fr{(v\ell)^2}{c_s a^4} \times \fr{\ell^2}{v} \times \fr{1}{\ell} \times \left(\fr{v}{c_s}\right)^{1/2} \frac{a^2}{\ell} \times 1 \sim \frac{v}{c_s} \, \sqrt{L}, \\
\frac{\bar n^2 \lambda^2}{\bar w c_s^2} \int d t d \sigma  \, \dot B_{\rm pot} \dot X  \sim \fr{(v\ell)^2}{c^3_s a^4} \times \fr{\ell^2}{v} \times \fr{v}{\ell} \times \left(\fr{v}{c_s}\right)^{1/2} \frac{a^2}{\ell} \times v \sim \frac{v}{c_s} \, \sqrt{L}.
\ea
\item Let us finally consider the couplings to the radiation phonon field. These are subdominant compared to the couplings with both hydrophoton and potential phonons. In particular, for the non-derivative coupling we have
\be\la{BrX'Xdot}
\bar n \lambda \int d t d \sigma  \, B_{\rm rad} \dot X X' \sim \fr{v \ell}{c_s a^4} \times \fr{\ell^2}{v} \times \left(\fr{v}{c_s}\right) \frac{a^2}{\ell} \times v \times 1 \sim \left(\frac{v}{c_s}\right)^{3/2} \, \sqrt{L},
\ee
while derivative couplings are even more suppressed:
\ba
\frac{\bar n^2 \lambda^2}{\bar w} \int d t d \sigma \, \nabla B_{\rm rad} X' \sim \fr{(v\ell)^2}{c_s a^4} \times \fr{\ell^2}{v} \times \fr{v}{c_s \ell} \left(\frac{v}{c_s}\right)\frac{a^2}{\ell} \times 1 \sim  \left(\frac{v}{c_s}\right)^{5/2} \, \sqrt{L} \ ,\\
\frac{\bar n^2 \lambda^2}{\bar w c_s^2} \int d t d \sigma  \dot B_{\rm rad} \dot X  \sim \fr{(v\ell)^2}{c^3_s a^4} \times \fr{\ell^2}{v} \times \fr{v}{\ell} \times \left(\frac{v}{c_s}\right)\frac{a^2}{\ell} \times v \sim  \left(\frac{v}{c_s}\right)^{7/2} \, \sqrt{L} \ .
\ea
\item
Finally, we should stress that at higher orders in the $v/c_s$ expansion, it becomes important to take into account  that the radiation field is to be evaluated on the perturbed string; since its spatial derivatives scale like $\d_i \sim (v/c_s) \, 1/\ell$, the expansion in powers of $v/c_s$ remains consistent only if we Taylor expand the radiation field around some geometric center of the vortex configuration, 
\begin{align}
\vec B_{\rm rad} (t, \vec X(t,\sigma))  & = \sum_{n=0}^{\infty} \frac{1}{n!}X^{k_1} \cdots X^{k_n}\partial_{k_1} \cdots \partial_{k_n}\vec B_{\rm rad}(t,0) \\
& \sim \sum_{n=0}^{\infty} \frac{1}{n!} \l(\fr{v}{c_s}\r)^n \vec B_{\rm rad}(t,0) \; ,
\end{align}
and keep only the terms up to the desired order in $v /c_s$. This expansion is equivalent to performing a multipole expansion~\cite{Ross:2012fc}, and is necessary only for radiation phonons. The reason is that the derivative of the potential phonons and of the hydrophotons are not suppressed by powers of $(v/c_s)$.
\end{itemize}

Notice that even though for a generic fluid the small parameters $({r_c}/{\ell})$ and $({a}/{r_c})$ are independent, for the leading terms considered above they always appear in combination and the parameter $r_c$ does not appear explicitly. This in turn allowed us to write our estimates only in terms of $v/c_s$ and the angular momentum $L$. Were we to keep terms with more than one derivative  per field, however, this would no longer be the case.  Consider for instance the term
\be
\frac{\bar n^2 \lambda^2}{\bar w}\int dt dz\, |X'|R \, r_c^2 \sim L \left(\frac{r_c}{\ell}\right)^2 \ ,
\ee
where $R$ is the curvature  constructed from the worldsheet metric. This kind of terms are known as \emph{finite-size terms}~\cite{Goldberger:2007hy,Goldberger:2004jt}, in that they account for the fact that the vortex core has actually a finite thickness.  From an EFT viewpoint, they are generated when the dynamics at scales of the order of core radius is integrated out, and that is why they are suppressed by powers of $r_c$.

Finally, notice also that all the terms we have considered above are independent of the ratio $c_s/c$. This is not the case for all the terms in our effective action, and in fact whenever $c_s /c \ll1$ one can safely neglect all the terms that scale with positive powers of this ratio and obtain the non-relativistic version of our effective action. We will explain how to take the non-relativistic limit in detail in the following section, but first we need to include power counting rules for terms involving dynamical perturbations of the vortex line.

\subsection{Vortex lines as dynamical fields}

Depending on the problem under consideration, it may be necessary to consider perturbations of the vortex geometry around some background configuration, and to assign power counting rules separately to the background and the perturbations (kelvons).  As an illustration, we will perturb around a long straight line and consider the interaction of kelvons with the hydrophoton and phonon fields. In this simple case, there is only one relevant region for the kelvons, namely the one in which they are on-shell. As we have seen in section \ref{kelvons}, the dispersion relation for the Kelvin waves is (up to a logarithm) $\omega \sim \Gamma k^2$. Since a local perturbation of the line with wave number $k \sim 1 /\ell$ will result in a local motion of the line with velocity $v \sim \Gamma /\ell$, we conclude that the region where the kelvons are on-shell is
\be
\vec \pi: \qquad \omega \sim \frac{v}{\ell} \, , \qquad k \sim \frac{1}{\ell}.
\ee
Notice that in this region the frequency and momentum scale like in the phonon potential region (\ref{potential region}), but in fact this region is more akin to the phonon radiation region (\ref{radiation region}) because it is where the field $\pi$ is on-shell.

As before, we can assign a definite scaling to the kelvon field $\pi$ but looking at its kinetic term. Expanding $\vec{X}(t, z) = (\vec{\pi}(t,z), z)$, the leading order kinetic term for $\vec \pi$ is (schematically)
\be
\bar{w}\Gamma \int dt dz \,(  \dot{\pi} \pi - \Gamma \, \pi' \pi').
\ee
This means that its propagator in Fourier space will scale like
\be
\langle \tilde{\pi}\tilde{\pi}\rangle \sim \frac{1}{\bar{w}\Gamma}\frac{\delta(\omega)\delta(k)}{\omega - \Gamma k^2} \sim \frac{c_s a^4}{v \ell} \times \frac{\ell^2}{v} \times \frac{\ell^2}{v \ell} \; ,
\ee
and, going back to real space, we can obtain the scaling of the field $\pi$:
\be
\pi \sim \int d\omega_k dk \, \tilde{\pi} \sim \frac{v}{\ell} \times \frac{1}{\ell} \times \frac{c^{1/2}_s}{v^{3/2}} a^2 \ell \sim \left(\frac{v}{c_s}\right)^{-1/2}\frac{a^2}{\ell}\,.
\ee
We can now use this scaling, together with the scalings for the bulk fields, to compare the relative sizes of various vertices involving kelvons.\footnote{The power counting rules developed in this subsection are valid in the regime where $k \pi \ll 1$, and as such they do not apply for instance to the self-pipe solution discussed in section \ref{kelvons}.}
\begin{itemize}
\item First, the non-derivative interaction term between a kelvon and a hydrophoton:
\be
\bar n \lambda \int dt d \sigma \vec{A} \cdot \vec{\pi} \,'  \sim \fr{v\ell}{c_s a^4} \times \fr{\ell^2}{v} \times (v c_s)^{1/2}\fr{a^2}{\ell}  \times \fr{1}{\ell} \times \left(\frac{c_s}{v}\right)^{1/2} \frac{a^2}{\ell} \sim 1.
\ee
Thus, this interaction term is of the same order as the kinetic term of the kelvons. This was to be expected, since it is the mixing between the hydrophoton and the kelvons at the level of the propagator that gives the kelvons their gradient energy.
\item The non-derivative interaction between kelvons and potential phonons scales like
\be
\bar n \lambda \int dt d \sigma \, \epsilon_{ab} B^{a}_{pot} \dot \pi^b \sim \fr{v\ell}{c_s a^4} \times \fr{\ell^2}{v} \times \left(\frac{v}{c_s}\right)^{1/2}\fr{a^2}{\ell}  \times \fr{v}{\ell} \times \left(\frac{v}{c_s}\right)^{1/2}\frac{a^2}{\ell} \sim \fr{v}{c_s} \; .
\ee
It is therefore safe to ignore the kelvon-phonon mixing in the propagator (to leading order in $v/c_s$).  

\item The analogous coupling to a radiation phonon is further suppressed by a factor of $(v /c_s )^{1/2}$. 

\item For the reasons discussed in sect.~\ref{kelvons},  the leading interaction that allows a radiation phonon to be absorbed by the string and converted into a pair of kelvons is  the one we used in sect.~\ref{sec:phonon absorption}, which scales like:
\ba
\bar n \lambda \int dt dz \, \epsilon_{ab} \, \partial_c B^{a}_{\rm rad} \, \pi^c \dot{\pi}^b \sim \left(\frac{v}{c_s}\right)^2 \frac{a^2}{\ell^2}.
\ea
\end{itemize}
%

%%%%%%%%%%%%%%%%%%%%%%%%%%
%%%%%%%%%%%%%%%%%%%%%%%%%%

\section{Small velocity approximations}\la{sec:small}

The effective action (\ref{effS}) is appealing to the eye of an high energy theorist, but it is probably overkill for describing the outcome of most experiments that can be carried out in the lab. In fact, ordinary media in the lab are highly non-relativistic, in the sense that their sound speed is much smaller than the speed of light.
In this limit, an infinite subset of the terms that appear in the action (\ref{effS}) becomes negligible (based on the power counting scheme developed in the previous section) because they are suppressed by powers of $c_s /c$, and it would be practical to dispose of  them from the very beginning. In section \ref{sec:non-relativistic limit} we do this following a somewhat bottom-up approach, by constructing a non-relativistic action that is manifestly Galilei-invariant. The same result can also be derived more systematically by taking the formal limit $c \to \infty$ of our action \eqref{effS}. We find this alternative derivation interesting not only because it provides an independent check of our non-relativistic result, but also because it highlights many of the subtleties involved in taking the non-relativistic limit. The analysis is quite technical though, and therefore has been relegated to an appendix (App. \ref{app: NR limit}).

Since we have argued that the typical velocity $v$ of a vortex line is also much smaller than $c_s$ within the regime of validity of the effective theory, it is often a useful approximation to also let $c_s \to \infty$. In this limit the phonons decouple from the vortex lines, and  almost all the terms in the effective action drop out, except for a finite number of them. This is the incompressible regime, which we will address in section~\ref{sec:Incompressible limit}.

\subsection{Non-relativistic limit} \la{sec:non-relativistic limit}

We want to find a consistent truncation of the action
\be \la{S_repeat}
S  = \int d^4 x \, G(Y)+  \int d \tau d \sigma \l\{  \lambda \, A_{\mu\nu} \, \d_\tau X^\mu  \d_\sigma X^\nu -   \sqrt{- \det g} \, \mathcal{T}\big( g^{\alpha\beta} h_{\alpha\beta}, Y \big) \r\}
\ee
that describes the interaction of sound and vortex lines in fluids with sound speed much smaller than the speed of light. In particular, the truncation we are after should be invariant under Galilei boosts,
\be \label{galilei}
\vec x \to \vec x = \vec x + \vec v_0 \, t \; , \qquad t \to t \; .
\ee
Our fields $A_{\mu\nu}$ and $X^\mu$ have standard transformation properties under Lorentz boosts:
\begin{align}
X^\mu(\tau, \sigma) & \quad \to \quad \Lambda^\mu {}_\nu X^\nu(\tau, \sigma) \\
A_{\mu\nu} (x) & \quad \to \quad \Lambda_\mu {} ^\rho \Lambda_\nu {}^\sigma A_{\rho\sigma} (\Lambda^{-1} \cdot x) \ . 
\end{align}
Taking the background value of $A_{\mu\nu}$ into account, in the Galilei limit these reduce to\footnote{In order to take the non-relativistic limit unambiguously, it is convenient to first gauge-fix $\tau$ reparametrizations on the world sheet, so that the notion of ``small'' vortex-line velocity becomes particularly transparent. We will therefore set $\tau = X^0 \equiv t$, as we already did in previous sections.}
\begin{align}
\vec X & \quad \to \quad \vec X + \vec v_0 t \label{galilei X}\\
\vec A & \quad \to \quad \vec A + \vec v_0  \times \big( -\sfrac13 \vec x + \vec B \big) \\
\vec B & \quad \to \quad \vec B + \sfrac13 \vec v_0 t \; , \label{galilei B}
\end{align}
with the understanding that the arguments of $\vec A$ and $\vec B$ transform as the inverse of \eqref{galilei}, so that the time-derivatives acquire an extra piece:
\be
\d_t \to \d_t - (\vec v_0 \cdot \vec \nabla) \label{galilei dt}\; .
\ee

We  find it convenient to discuss the bulk, Kalb-Ramond, and generalized Nambu-Goto terms in the action separately.
Let us start with the bulk part. At lowest order in the derivative expansion, the non-relativistic action for a fluid takes the well-known form
\be \la{S^nr_bulk}
S^{\rm nr}_{\rm bulk} = \int d^4 x \l\{ \sfrac12 {m n_{\rm nr}  \vec{u}^{\,2}} - U(n_{\rm nr}) \r\},
\ee
where $m$ is the mass of the elementary constituents of the fluid, $n_{\rm nr}$ is their  non-relativistic number density (see below), $\vec u$ is the local velocity of the fluid, and $U$ is a generic function, related to the equation of state, which can be interpreted as the internal energy density of the fluid.

From our discussion in section \ref{two form}, we know how to express these hydrodynamic variables in terms of our fields $\vec A$ and $\vec B$. The velocity $\vec{u}$ is given in equation (\ref{fluid3velocity}), which we reproduce here for convenience:
\be \la{u}
\vec{u} = \frac{\dot{\vec B}-\vec \nabla \times \vec A}{1 - \vec \nabla\cdot \vec B} \; .
\ee
The number density $n$ follows instead from eq. (\ref{mun}) and the expression for $Y$ in terms of $\vec A$ and $\vec B$ given above eq. (\ref{S2}). Notice however that, based on simple dimensional analysis, the term $\dot B$ in the expression for $Y$ must be suppressed by a factor of $c$ compared to $\vec \nabla \cdot \vec B$ and thus becomes negligible in the non-relativistic limit. At the same time, gauge invariance requires that the quantity $\vec \nabla \times \vec A$ also becomes negligible, because only the combination $\dot{\vec B}-\vec \nabla \times \vec A$ would be gauge invariant (see discussion at the end of sec. \ref{two form}). We therefore conclude that in the non-relativistic limit the number density reduces to:
\be \la{n_nr}
n \simeq \bar n \, (1 - \vec \nabla \cdot \vec B) \equiv n_{\rm nr} \; .
\ee
By plugging the expressions (\ref{u}) and (\ref{n_nr}) into equation (\ref{S^nr_bulk}) we get the non-relativistic bulk action for $\vec A$ and $\vec B$.

We can check explicitly for Galilean invariance. Given the transformation laws (\ref{galilei X}) -- (\ref{galilei dt}), we have
\begin{align}
\vec \nabla \cdot \vec B & \quad \to \quad \vec \nabla \cdot \vec B \\
\big(\dot{\vec B}-\vec \nabla \times \vec A \, \big) & \quad \to \quad  \big(\dot{\vec B}-\vec \nabla \times \vec A \, \big)+ \vec v_0 \, (1 - \vec \nabla \cdot \vec B \, ) \; .
\end{align}
We thus see that $n_{\rm nr}$ is invariant, $\vec u$ transforms as expected, $\vec u \to \vec u + \vec v_0$, and the action \eqref{S^nr_bulk} is thus invariant up to a total derivative.

Let us now turn our attention to the worldsheet part of the action. In $\tau = X^0 \equiv t$ gauge, the first term---the Kalb-Ramond one---reads
\be \la{KR-nr}
S^{\rm nr}_{\rm KR} = \bar n \lambda  \int dt d \sigma \l\{- \sfrac{1}{3} \vec X \cdot \d_t \vec{X} \times \d _\sigma\vec X +  \vec A \cdot \d_\sigma \vec X + \vec B  \cdot \d_t \vec X \times \d_\sigma \vec X \r\}.
\ee
Because $\vec \nabla \times \vec A$ and $\dot{\vec B}$ are of the same order in the expansion in $c_s /c$ due to gauge invariance, it means that schematically we have $A \sim  B \dot X$. Therefore, the $A$ and $B$ terms appearing in (\ref{KR-nr}) are of the same order in $c_s /c$, and both belong in the non-relativistic action\footnote{Notice however that these two terms are \emph{not} of the same order in $v/c_s$, as we have shown for instance in equations (\ref{AX'}), (\ref{BpX'Xdot}) and (\ref{BrX'Xdot}) using our power counting rules.}.

It is immediate to check that \eqref{KR-nr} is invariant under the Galilean transformations (\ref{galilei X})--(\ref{galilei B}).
However, since \eqref{KR-nr} is an exact rewriting of the original relativistic expression \eqref{KR}, it is still invariant under {\it Lorentz} transformations as well. In fact, being the integral of a two-form over a two-dimensional world-sheet, it is invariant under generic spacetime diffeomorphisms, which admit Lorentz and Galilei as finite-dimensional subgroups, even without considering the transformation of the metric.\footnote{We thank Rachel Rosen for this remark.} As we saw in sect.~\ref{non-renormalization}, this is related to the non-renormalization property of $\lambda$: from either viewpoint---relativistic or non-relativistic---this part of the action has an enhanced symmetry compared to the rest.

The Kalb-Ramond term did not simplify in the non-relativistic limit, but fortunately, the generalized Nambu-Goto term will.
To begin with, notice that (\ref{KR-nr}) only depends on the components of the velocity $\d_t \vec X$ perpendicular to $\d_\sigma \vec X$. This is no accident: velocities along the vortex line are not physical, because the action is invariant under (time-dependent) reparametrizations of $\sigma$. For the very same reason, the generalized Nambu-Goto term will also depend only on the perpendicular components of the velocity. Interestingly, this prevents us from writing down for a vortex line the standard non-relativistic kinetic term $(\d_t\vec{X})^2$. One could be tempted to consider instead the term $(\d_t\vec{X}_\perp)^2$, but it is easy to check that this would not be invariant under Galilean boosts $\vec X \to \vec X + \vec v_0 t$,  even up to a total derivative\footnote{Recall that usual non-relativistic kinetic terms are invariant under boosts only up to a total derivative.}, because
\be
\d_t X^i_\perp \to \d_t X^i_\perp + \bigg[ \delta^i_j - \fr{ \d_\sigma X^i \d_\sigma X_j}{|\d_\sigma \vec{X}|^2} \bigg] v_0^j.
\ee

There is however one Galilean invariant that we can build using $\dot{\vec{X}}_\perp$ provided we also use the perpendicular velocity of the fluid $\vec u_\perp$, and that is the difference $\dot{\vec{X}}_\perp - \vec u_\perp$. Since this difference is an exact Galilean invariant (as opposed to up to a total derivative), the action will in general be an arbitrary function of its square. In fact, in appendix \ref{app: NR limit} we show explicitly that the combination $g^{\alpha\beta} h_{\alpha\beta}$ that appears in (\ref{S_repeat}) reduces precisely  to $\big(\dot{\vec{X}}_\perp - \vec u_\perp \big)^2$ in the non-relativistic limit. Since we have already argued that in this limit $Y \to n_{\rm nr}$, we conclude that the non-relativistic version of the generalized Nambu-Goto term must be
\be
S^{\rm nr}_{\rm NG'} = - \int dt d \sigma |\d_\sigma \vec X| \, \mathcal{T} \Big( \big(\dot{\vec{X}}_\perp - \vec u_\perp \big)^2, n_{\rm nr} \Big),
\ee
where the (Galilean invariant) overall factor of $|\d_\sigma \vec X|$---which is in fact the non-relativistic limit of $\sqrt{-g} \, $---is there to maintain invariance under reparametrizations of $\sigma$. 

In conclusion, the full non-relativistic action for vortex lines coupled to sound is given by:
\be \la{S_nr}
S_{\rm nr} = S^{\rm nr}_{\rm bulk} + S^{\rm nr}_{\rm KR} + S^{\rm nr}_{\rm NG'}.
\ee
For a top-down derivation of this result, we refer again the reader to appendix \ref{app: NR limit}.

\subsection{Incompressible limit} \la{sec:Incompressible limit}

Let us now turn our attention to the incompressible limit. In this limit, the vortex lines are moving so slowly compared to the sound speed that for all practical purposes there is no sound emission associated with their motion. This limit corresponds to an even more dramatic truncation of the effective action (\ref{S_repeat}), in which only the \emph{leading} terms in the expansion of $v /c_s$ are kept. 
%Since the sound speed is always smaller than the speed of light, there is no loss of generality in starting directly from the non-relativistic action (\ref{S_nr}) {\bf (AN: ?)}. \
For simplicity, we will start directly from the non-relativistic action (\ref{S_nr}), but our final result applies unaltered to the incompressible limit of relativistic fluids as well.

Based on the power counting rules developed in section \ref{sec: power counting}, we see from eq. (\ref{n_nr}) that to leading order in $v/ c_s$ the number density remains constant, i.e. $n_{\rm nr} = \bar n$. This should not come as a surprise: it is the very reason why this small velocity limit is known as ``incompressible''. Thus, the leading terms in the bulk action are simply
\be \la{Sbulkinc}
S_{\rm bulk}^{\rm inc} = \sfrac12 {\bar w} \int d^4 x \, \l\{ (\vec \nabla \times \vec A )^2 - c^2_s (\vec \nabla \cdot \vec B)^2 \r\},
\ee
where we used the fact that in the non-relativistic limit $\bar w \approx m \bar n$, and the power counting rules of section \ref{sec: power counting} imply that the $\dot{\vec{B}}$ terms are subleading.

Moreover, from eqs. (\ref{kinsize}), (\ref{tensize}) and (\ref{AX'}) it follows that the leading worldsheet terms are 
\be
S_{\rm worldsheet}^{\rm inc} =  \int dt d\sigma \l\{- \sfrac13 {\lambda \bar n} \,  \vec X \cdot \d_t \vec{X} \times \d _\sigma\vec X   - T |X'| + \lambda \bar n  \, \vec A \cdot \d_\sigma \vec X  \r\} \; .
\ee
Notice that we have to include the $A \cdot \d X$ hydrophoton interaction, even though according to our estimates it scales as $\sqrt L$, a factor of $1/\sqrt L $ down with respect to  $X$'s kinetic terms. 
The reason is that such an interaction can be used several times in a diagram, to yield contributions to any process that can scale as higher powers of $L$.
In particular, the hydrophoton can be integrated out exactly (since it appears quadratically in the action) to yield a $1/r$ potential energy between vortex line elements~\cite{Endlich:2013dma} that is of the same size as the kinetic terms: 
\begin{align} 
S_{\rm bulk}^{\rm inc} + S_{\rm worldsheet}^{\rm inc}  \quad \to \quad S_\text{inc} = \; &  \bar w \int d t d \sigma \bigg\{  -\sfrac13{\Gamma} \vec X \cdot \d_t \vec{X} \times \d _\sigma\vec X - ({T}/{\bar w}) | \vec X '| \nonumber \\
& - \frac{\Gamma^2}{8 \pi} \int d\sigma' \, \fr{\d_\sigma \vec X \cdot \d_{\sigma'} \vec X'}{|\vec X -\vec X'|}  \, \bigg\} \; ,
\la{Sincompressible}
\end{align}
where $\vec X'$ is shorthand for $\vec X(\sigma', t)$, and we have used eq. (\ref{lambda}) to trade the coupling $\lambda$ for the circulation $\Gamma$ and match the standard notation in the literature. Notice that $\vec B$ is completely decoupled from the vortex lines, and therefore we have omitted the bulk term $(\nabla \cdot B)^2$ appearing in (\ref{Sbulkinc}).

Using our power counting rules it is easy to see that all three terms in (\ref{Sincompressible}) are of the same order. We have already shown in  (\ref{kinsize}) and (\ref{tensize}) that the first two terms are of order of the total angular momentum $L$, and now we also have
\be
\Gamma^2 \bar{w} \int dt d\sigma d{\sigma}' \frac{\d X \cdot \d' {X}'}{|X-X'|} \sim (v \ell)^2 \times \fr{1}{c_s a^4} \times \frac{\ell^3}{v} \times \frac{1}{\ell} \sim \left(\frac{v}{c_s}\right)\frac{\ell^4}{a^4} \sim L.
\ee

Eq. (\ref{Sincompressible}) is the effective action in the strictly incompressible limit.  Next-to-leading order (in $v/c_s$) corrections to this action   were studied systematically in~\cite{Endlich:2013dma}.  Notice that, in the incompressible limit, the tension term is the only counterterm needed to renormalize the UV-divergence in the $\sigma'$ integral of the $1/r$ potential.

%%%%%%%%%%%%%%%%%%%%%%%%%%
%%%%%%%%%%%%%%%%%%%%%%%%%%

\section{Outlook}

In this paper we have constructed an effective theory for vortex lines in superfluids, valid at distances much larger than the core size of the vortex.  For classical phenomena, it would be interesting to investigate potential experimental signatures of our results for vortex lines and vortex rings in ordinary fluids like water, such as, for instance, their interactions with sound modes.

Perhaps more interesting would be to use our formalism to investigate quantum phenomena for vortex lines in superfluid helium-4. 
We are particularly interested in understanding {\em rotons}---elementary gapped excitations that correspond to a finite-momentum minimum in the energy-momentum dispersion relation, and that are believed to be some microscopic cousins of vortex rings \cite{feynman1955application, donnelly1991quantized}. Their size is not much bigger than atomic scale, and they are thus outside the regime of validity of our effective theory. However, as a preliminary step, we can use our formalism to study quantum effects like virtual phonon exchange for smaller and smaller vortex rings---but still of size big enough that we can trust our computations---and see whether the resulting quantum-corrected energy-momentum dispersion relation is consistent with having a roton minimum for smaller sizes. Or, more responsibly, we should generalize what we have done here and develop the general effective theory for a roton-like point particle coupled to the superfluid bulk modes, and study quantum effects there.

%While our effective theory is not directly applicable to atomic-scale vortex configurations, which would be the regime directly relevant for studying microscopic excitations such as the roton, this picture already allows a rich structure of vortex behavior, such as running tension and retrograde motion of vortex rings, as investigated in \cite{Gubser:2014yma}.  For larger vortex rings, the couplings to sound written here may be directly accessible to experiment.  Although we have worked at the classical level, it would be interesting to use the action here to construct a first quantized treatment of Kelvin waves and their interaction with phonons.
%

It would also be interesting to analyze the Kelvin wave spectrum at high energies, and whether such study can shed light on the relation between the superfluid transition and the Hagedorn transition in free bosonic string theory. This would probably lead us outside the regime of validity of our effective theory, but the integrability of the system in the local induction approximation (which was noticed in \cite{Hasimoto}) may be of help here.

\section*{Acknowledgements}
We would like to thank Claudio Barenghi, Solomon Endlich, Sebastian Garcia-Saenz, Ermis Mitsou, and Rachel Rosen for illuminating discussions. We are particularly grateful to Steve Gubser, Revant Nayyar, and Sarthak Parikh for sharing insights on related work in progress, as well as for collaboration on related issues.  This work was supported in part by the United States Department of Energy under contracts DE-FG02-11ER41743 and DE-FG02-92-ER40699.
\appendix

\section{Sound and hydrophoton propagator} \la{appa}

We want to derive the propagators for the $\vec A$ and $\vec B$ fields in the presence of a generic gauge fixing term of the form (\ref{gf}). Our starting point will be the quadratic action (\ref{S2}), which we reproduce here for convenience:
\begin{align} 
S_{(2)} = \bar w \int \! d^4 {x}  \, & \Big\{ \sfrac12  (\vec{\nabla} \times \vec{A})^2 
+  \sfrac12  \big[ \dot{\vec{B}}^2 - c^{2}_{s} (\vec{\nabla} \cdot \vec{B})^2 \big] \\
& -  \dot{\vec{B}}\cdot(\vec{\nabla} \times \vec{A}) - \sfrac{1}{2\xi} (\vec{\nabla} \times \vec{B})^2 
+ \sfrac{1}{2\xi} (\vec{\nabla} \cdot \vec{A})^2 \Big\} \; .
\nonumber
\end{align}
This action can also be written more succinctly by switching to Fourier space and introducing the doublet of vector fields $\Phi^i = (A^i, B^i)$, in which case we find
\be
S_{(2)} =  \frac{1}{2}  \int \! \fr{d^4 k}{(2 \pi)^4} \, \Phi^i (-k) \cdot M_{ij} (k) \cdot \Phi^j (k)  \ ,
\ee
with
\begin{equation}
M_{ij} (k)  = \bar w \begin{pmatrix} k^2 \, \delta_{ij} +\big(\sfrac1\xi-1 \big)  k_i  k_j  & \omega k^l \, \epsilon_{ilj} \\ -\omega k^l \,\epsilon_{ilj} & \omega^2 \, \delta_{ij} - k^2 \, \sfrac1\xi \delta_{ij}+ \big(c^2_s - \sfrac1\xi \big) k_i  k_j  \end{pmatrix} .
\end{equation}
The matrix can be inverted using the ansatz
\be
(M^{-1})^{ij} (k) \equiv \fr{1}{\bar w} \begin{pmatrix} c_1\delta^{ij} + c_2 \hat k^i \hat  k^j & c_3 \hat k_l \epsilon^{ilj} \\ -c_3 \hat k_l \epsilon^{ilj} & c_4 \delta^{ij}+ c_5 \hat k^i \hat k^j \end{pmatrix}
\ee
where the $c$'s are homogeneous functions of $\omega$ and $k$ of degree $-2$.
We get

\begin{subequations}
\ba
c_1 &=& \frac{1}{k^2} - \xi \, \frac{ \omega^2}{k^4}   \\
c_2 &=& (\xi-1)\frac{1}{k^2}  + \xi \,\frac{\omega^2}{k^4} \\
c_3 &=& \xi \, \frac{\omega}{k^3} \\
c_4 &=& -\xi \, \frac{1}{k^2} \\
c_5 &=& \frac{1}{\omega^2 - c^2_s k^2} + \xi \,\frac{1}{k^2}\,.
\ea
\end{subequations}
The matrix of propagators is then equal to $i (M^{-1})^{ij} (k)$. In the particular case $\xi =0$, we see that this matrix becomes block diagonal and we recover the propagators in equation~(\ref{propagators}).

\section{Generalized Nambu-Goto term from the coset}

In this appendix we provide an alternative derivation of our effective action for vortex lines based on the coset construction~\cite{Callan:1969sn} for spontaneously broken space-time symmetries~\cite{Volkov:1973vd,ogievetsky:1974ab}. This technique has been recently applied to a variety of systems (see e.g.~\cite{Nicolis:2013lma}), and we refer the reader to~\cite{Delacretaz:2014oxa} for a nimble review of this formalism. Our goal here is to use this technique to confirm that the generalized Nambu-Goto action (\ref{NG'}) is indeed the most general worldsheet action one can write with one derivative acting on each field. Along the way, our analysis will also provide a nice illustration of how coset calculations can be simplified by temporarily introducing a fictitious hierarchy between symmetry breaking scales.

For simplicity, we will base our discussion on the symmetry breaking pattern that arises in the scalar field language (where time translations are broken, but spatial translations are not). Since the generalized Nambu-Goto term (\ref{NG'}) depends on the bulk fields only through the 4-velocity $u^\mu$, the scalar field is as good as the 2-form when it comes to this part of the worldsheet action.
A more responsible approach would be perhaps to take as starting point the symmetry breaking pattern of the 2-form language (where spatial translations are broken, time translations are not), because this would also allow us to recover  the non-derivative Kalb-Ramond coupling between vortex line and sound. However, since the 2-form theory is gauge invariant, this would require dealing with an infinite number of non-linearly realized symmetries. Such a construction is feasible~\cite{Goon:2014ika} but more involved than the one we will present here.

Let us consider a superfluid with a perfectly straight vortex line embedded in it, and let us work in the limit of infinite volume and infinite vortex length. Such a configuration spontaneously breaks several symmetries. 
Some of these symmetries would be broken even in the absence of the vortex line, namely boosts, because the superfluid as a whole admits a preferred reference frame in which it is at rest, and also particle number $Q$ and time translations, which are  broken down to the diagonal linear combination~ (this is in fact the defining property of a superfluid~\cite{Son:2002zn,Nicolis:2013lma}). 
Some other symmetries are instead broken only because of the vortex line, and these are the translations in the directions perpendicular to the vortex and rotations around these same directions. The only symmetries that are left unbroken are translations along the vortex line, rotations around it and, as already mentioned, a linear combination of particle number and time translations.

To simplify our calculations, we will exploit the fact that in principle there could be a hierarchy between the interatomic length scale at which boosts and particle number are broken, and the vortex core size, which sets the scale where the additional translations and rotations get broken.  It just so happens that in superfluids these two length scales are comparable, but in general the former can be much smaller than the latter, as is the case for vortex lines in ordinary fluids. For our purposes, it will actually be convenient to work in the limit where the core size is much larger than the interparticle separation. In a first approximation, this amounts to consider a system where boosts and particle number are \emph{explicitly} broken, but there is still an effective notion of unbroken time translations. This follows from a universal property of systems featuring spontaneous symmetry breaking: when the symmetry breaking energy scale is raised, the associated Goldstone bosons become more and more weakly coupled, among themselves as well as to other sectors; in the limit in which the Goldstones become invisible, the corresponding symmetry can be thought of as being explicitly broken.

The symmetry breaking pattern we are interested in thus is
\ba \la{pattern2}
\mbox{unbroken} =  \l\{
\begin{array}{l}
P_0 \\
P_3  \\
J_3 \\
\end{array}
\r.
\qquad \qquad 
\mbox{broken} =  \l\{
\begin{array}{l}
P_1, P_2 \equiv P_n \\
J_1, J_2 \equiv J_n \; .
\end{array}
\r.
\ea
Only at the very end  will  we restore Lorentz and reparameterization invariance, by reintroducing the 4-velocity of the medium $u^\mu$  as a spurion field. 
We will base our construction on the coset parameterization
\be \la{Omega}
\Omega = e^{i (t P_t + z P_z)} e^{i \pi^n P_n} e^{i \xi^n J_n},
\ee
where $(t,z)$ are the coordinates on the worldsheet and $\pi^n, \xi^n$ are Goldstone fields. Starting from (\ref{Omega}), we can calculate the Maurer-Cartan form\footnote{Our convention for the indices is such that $\alpha,\beta,\gamma ... = 0,3$ whereas $m, n, p, ... = 1,2$.} 
\be
\Omega^{-1} \d_\alpha \Omega \equiv i e_\alpha{}^\beta (P_\beta + \nabla_\beta \pi^n P_n + \nabla_\beta \xi^n J_n + A_\beta J_z).
\ee
The explicit form of the coefficients $e_\alpha{}^\beta, \nabla_\beta \pi^n,\nabla_\beta \xi^n $ and $ A_\beta$ can be easily calculated using the algebra of translations and rotations. For our purposes, the most important quantities are going to be $e_\alpha{}^\beta$ and $\nabla_\beta \pi^n$. The former plays essentially the role of a vielbein, in that $dt dz \det e$ is an invariant integration measure on the worldsheet, whereas the latter are the covariant derivatives of the Goldstones $\pi^n$. Because $[ P_3, J_n ] \sim P_n$, we can impose the inverse Higgs constraints~\cite{Ivanov:1975zq} $\nabla_3 \pi^n \equiv 0$ and solve them to express the Goldstones $\xi^n$ in terms of the $\pi^n$'s. The fully nonlinear result reads:
\be \la{ihsol2}
\xi_n = \epsilon_{nm} \d_3 \pi^m \l( \fr{\arctan \sqrt{\d_3 \pi_p \d_3 \pi^p}}{\sqrt{\d_3 \pi_q \d_3 \pi^q}} \r) \ ,
\ee
where $\epsilon_{12} = - \epsilon_{21} = 1$ and $\epsilon_{11} = \epsilon_{22} = 0$.

By using this result, we can express $\nabla_0 \pi^n, \nabla_\beta \xi^n$ and $e_\alpha{}^\beta$ solely in terms of the Goldstones $\pi^n$. In particular, since $\nabla_\beta \xi^n \approx \d_\beta \d_z \pi^n$, these quantities are of higher order in the derivative expansion and thus negligible at low energies. Following the usual coset mantra~\cite{Weinberg:1996kr}, at lowest order in derivatives the most general invariant Lagrangian can be written by taking $J_3$-invariant contractions of $\nabla_0 \pi^n$. The corresponding action is
\be \la{S3} 
S =  \int d t d z \det e \, \mathcal{L} (\nabla_0 \pi_n \nabla_0 \pi^n ),
\ee
with
\begin{subequations}
\ba
\det e &=& \sqrt{1+ \d_3 \pi_n \d_3 \pi^n} \\
\nabla_0 \pi_n \nabla_0 \pi^n &=& \fr{\d_0 \pi_n \d_0 \pi^n + (\epsilon^{nm} \d_0 \pi_n \d_3 \pi_m)^2 }{1 + \d_3 \pi_p \d_3 \pi^p }.
\ea
\end{subequations}

In order to connect our result (\ref{S3}) with the generalized Nambu-Goto term (\ref{NG'}), it is sufficient to notice that the quantities above can be rewritten in terms of the  4-velocity at rest $u_\mu = \delta_\mu^0$ and the gauge-fixed embedding $X^\mu = (t, \pi^n, z)$ as follows:
\begin{subequations}
\ba
\det e &=& \sqrt{ \epsilon^{\alpha\beta} \d_\beta X^\lambda  \d_\delta X_\lambda \epsilon^{\gamma\delta} \d_\alpha X^\mu  \d_\gamma X^\nu u_\mu u_\nu} =\sqrt{- g^{\alpha\gamma} h_{\alpha\gamma}  \det g} \\
\nabla_0 \pi_n \nabla_0 \pi^n &=& 1 + \fr{\det \d_\alpha X^\mu \d_\beta X_\mu}{ \epsilon^{\alpha\beta} \d_\beta X^\lambda  \d_\delta X_\lambda \epsilon^{\gamma\delta} \d_\alpha X^\mu  \d_\gamma X^\nu u_\mu u_\nu} = 1 - \fr{1}{g^{\alpha\gamma} h_{\alpha\gamma}} \ .
\ea
\end{subequations}
Thus, we conclude that the action (\ref{S3}) can be rewritten in a manifestly covariant and reparameterization-invariant form as
\be
S = \int d\tau d\sigma \sqrt{- \det g} \, F (g^{\alpha\gamma} h_{\alpha\gamma}) \ .
\ee
In fact, this is the \emph{only} way to rewrite this equation (\ref{S3}) in a way that restores both Lorentz and reparameterization invariance at the same time. Up to an additional dependence on $Y$, which we missed because we worked in the limit of where the $U(1)$ baryon number is explicitly broken, this result agrees with equation (\ref{NG'}).

\section{Feynman rules for the kelvon field}\label{FeynmanKelvin}

In this section we derive the Feynman rules for the canonically normalized nonrelativistic scalar action 
\be
\int d^d {x} \, \Big[\phi^{*}_c \, i \partial_t \phi_c  - \frac{|\vec \nabla \phi_c|^2}{2m} + \cdots \Big]
\ee
which includes our quadratic kelvon action eq.~\eqref{Seff2 kelvon} as a special case with $d = 2$, $\sqrt{\bar{n}\lambda} \, \phi \to \phi_c$, and $m = \frac{\bar{n}\lambda}{2T(k)}$. It is worth emphasizing that, despite the formal similarity between the kelvon case and the non-relativistic massive particle one, for the former the parameter $m$ cannot be interpreted as a mass.  We use the so-called relativistic convention for the normalization of the one-particle states,
\be
\langle \vec{p} \, |\vec{q} \, \rangle = (2E_p)(2\pi)^{(d-1)}\delta^{(d-1)}(\vec{p}-\vec{q})\, , \qquad 1 = \int \frac{d^{d-1}{ p}}{(2\pi)^{d-1}}\, |{\vec p} \, \rangle \frac{1}{2 E_{p}}\langle{\vec p} \, |
\ee 
which, when combined with the mode expansion
\be
\phi_c(x) = \int \frac{d^{d-1} k}{(2\pi)^{d-1}}a_{\vec k} \, e^{-i(\omega_k t - \vec{k}\cdot \vec{x})}\, \qquad \phi_c^{*}(x) = \int \frac{d^{d-1} k}{(2\pi)^{d-1}}a^{\dagger}_{\vec k} \, e^{+i(\omega_k t - \vec{k}\cdot \vec{x})} \; ,
\ee
gives
\be
\langle 0 | \phi_c(\vec{x} )|\vec{p} \, \rangle = \sqrt{2E_p} \, e^{-i(E_p t -{\vec p}\cdot {\vec x})} 
\ee
and so every external scalar line has a factor of $\sqrt{2E_p}$.  Note that this is the (non-relativistic) energy associated with the Lagrangian above, $E_p = p^2/2m.$

The infinitesimal cross section for absorption of a single phonon into $n$ kelvons is 
\be
d\sigma = \frac{1}{2 E_p \, c_s}|{\cal M}(p \to n)|^2 d\Pi_{n}
\ee
where ${\cal M}$ is given by
\be\label{defM}
\langle {\vec p}_1, \cdots {\vec p}_n |(S-1)| \vec{p}\rangle = (2\pi)^{d}\delta^{d}(\vec p - \vec p_1 - \cdots \vec p_n) \times i {\cal M}
\ee
and the $n$-body phase space is
\be
d \Pi_{n} = (2\pi)^{d}\delta^{d}(\vec p - \vec p_1 - \cdots \vec p_n)\left(\prod_{j=1}^n \int\frac{d^{d-1} p_j}{(2 \pi)^{d-1}}\frac{1}{2 E_j}\right)
\ee
% Note that this energy is still the one given by the non-relativistic expression.  
For a $D$-dimensional bulk field interacting with a $d$-dimensional kelvon field, the mass dimension of the matrix element and the $n$-body phase space are given by
\be
\Big[{\cal M}\Big] = d + \frac{n}{2}- n\left(\frac{d-1}{2}\right) - \frac{(D - 2)}{2}\,, \qquad \Big[\Pi_{n}\Big] = n(d-2)
\ee
where we have made use of \eqref{defM}, keeping in mind the factors of $\sqrt{2E}$ for each external kelvon leg. 
%when going from the amplitude to the scattering amplitude, as per the LSZ formula.  
The final dimension of the cross section is therefore
\be
\Big[\sigma\Big] = -1 + n(d-1) + 2\left(d + \frac{n}{2}- n\left(\frac{d-1}{2}\right) - \frac{(D - 2)}{2}\right) - d = d - D + 1\,.
\ee
For $d=2$, $D = 4$, this has the dimensions of length, as expected.

\section{Non-relativistic limit as the formal $c \to \infty$ limit}  \la{app: NR limit}

In this appendix, we rederive the non-relativistic effective action discussed in section \ref{sec:non-relativistic limit} by taking the formal limit $ c \to \infty$ of the relativistic action (\ref{effS}). In order to take this limit correctly, one must keep in mind that the arbitrary functions appearing in the effective action (\ref{effS}) have a non-trivial dependence on the speed of light. In order to illustrate this point, let us revert for a moment to the scalar field description of a superfluid and consider a very specific equation of state, namely $p = (c_s^2 / c^2) \rho$ with $c_s^2 =$ constant. Such a superfluid is described by the effective action (\ref{F(X)}) with
\be \la{LX}
P(X) = \bar p \, (\sqrt{X} / \bar \mu )^{1+ c^2/ c_s^2},  \qquad\qquad \sqrt{X} = \sqrt{ \bar \mu^2 + 2 \bar \mu \dot \pi + \dot \pi^2 - c^2 (\nabla \pi)^2},
\ee
where $\bar p$ and $\bar \mu$ are once again the pressure and chemical potential at equilibrium. Since we are eventually interested in taking the $c \to \infty$ limit, we have explicitly reintroduced all factors of $c$. As a consistency check, one can plug this particular form of $P(X)$ into equations (\ref{hydroX}) and see that indeed $d p / d \rho = c_s^2 / c^2$.  The factor of $c^2$ is there because $\rho$ is an energy density, and not a mass density. Then, from equation (\ref{LX}), one immediately sees that the background values of the derivatives of $P(X)$ with respect to $\sqrt{X}$ scale for $c \gg c_s$ like
\be \la{scalingX}
\fr{d^n P}{d \sqrt{X}^{\,n}} \propto \fr{\bar p}{\bar \mu^n} \l( \fr{c^2}{c_s^2} \r)^n, \qquad \qquad (c \gg c_s).
\ee
Therefore, the quadratic Lagrangian for the Goldstone $\pi$ in the $c \gg c_s$ limit reduces to 
\be \la{L2X}
\mathcal{L}_2 = \fr{d P}{d \sqrt{X}} \l[- \fr{c^2}{2 \bar \mu} (\nabla \pi)^2 \r] + \fr{1}{2} \fr{d^2 P}{d \sqrt{X}^{\,2}}  \dot \pi^2 \quad \stackrel{c \gg c_s}{\longrightarrow} \quad \fr{\bar p}{2\bar \mu^2 }  \l[ - \fr{c^4}{c_s^2}  (\nabla \pi)^2 + \fr{c^4}{c_s^4}  \dot \pi^2\r].
\ee

As usual, the coefficient in front of the $\dot \pi^2$ term sets the magnitude of the field fluctuations with a given wavelength. Since we want the size of these fluctuations to remain finite while we take $c \to \infty$, we introduce the canonically normalized field
\be
\pi_c \equiv \fr{\bar p^{1/2} }{\bar \mu} \, \fr{c^2}{c_s^2} \, \pi
\ee
and take the $c \to \infty$ limit while keeping $\pi_c$ constant. If we now express the effective Lagrangian in terms of $\pi_c$, expand it in series around $\sqrt{X} =\bar \mu$ and keep in mind that the derivatives of $P(X)$ scale with $c/c_s$ as shown in equation (\ref{scalingX}), we~find 
\be
\mathcal{L} = \sum_{n=0}^\infty \fr{1}{n!} \fr{d^n P}{d \sqrt{X}^{\,n}} (\sqrt{X} - \bar \mu)^n = \sum_{n=0}^\infty \fr{ \bar p^{1 -n/2}}{n!} \l[\dot \pi_c   - c_s^2\fr{(\nabla \pi_c)^2}{2 \bar p^{1/2}} + \mathcal{O} (1/c^2) \r]^n.
\ee
Thus, we see that in the non-relativistic limit $c \to \infty$ the low-energy effective Lagrangian depends only on the Galilean-invariant combination $\dot \pi_c   - c_s^2 (\nabla \pi_c)^2 /(2 \bar p^{1/2})$. 

The key input that allowed us to derive this result was the scaling behavior (\ref{scalingX}), which we obtained starting from the particular $P(X)$ shown in (\ref{LX}).  However, our result---namely the fact that the nonrelativistic action must be a function of the combination $\dot \pi   - (\nabla \pi)^2$ (schematically)---is completely general and it applies to nonrelativistic superfluids with an arbitrary equation of state~\cite{Greiter:1989qb}. This is because the scaling (\ref{scalingX}) itself can be derived from very general considerations. In fact, from the quadratic Lagrangian (\ref{L2X}) we see that the speed of sound for a generic equation of state is given by $ c_s^2 = c^2 P' / (\bar \mu P'')$ evaluated on the background $\sqrt{X} = \bar \mu$, with each prime denoting a derivative with respect to $\sqrt{X}$.\footnote{Notice that in sect.~\ref{two form} we used primes to indicate derivatives with respect to $X$ or $Y$, whereas in this appendix  they will denote derivatives with respect to the square roots $\sqrt{X}$ or $\sqrt{Y}$. This change of notation will help to streamline the discussion in this section, hopefully without generating confusion.} Now, we would like the sound speed $c_s$ to be much smaller than $c$ over a large range of values of chemical potential, and not just for one particular value $\bar \mu$. This can be arranged provided $P' /(\bar \mu P'')$ is not only small, but also varies sufficiently slowly with $\bar \mu$, which means
\be
\bar \mu \fr{d}{d \bar \mu} \l( \fr{P'}{\bar \mu P''} \r) \approx 0  \quad \Longrightarrow \quad \fr{P'''}{P''} \propto \fr{P''}{P'}  \propto \fr{1}{\bar \mu} \, \fr{c^2}{c_s^2}.
\ee
Similarly, it is easy to show that $ \bar \mu P''''/P''' \sim c^2 / c_s^2$, and so on. The scaling relation (\ref{scalingX})---and therefore the non-relativistic limit---is completely determined within the effective theory once we require that the speed of sound be smaller than the speed of light for generic boundary conditions. Let us now see how all this works in the 2-form language.

We can repeat a similar analysis and determine how the derivatives of $G(Y)$  depend on the ratio $c/c_s$. We will proceed in three steps. First, we can use the relation between $\rho$ and $p$ given in (\ref{hydroY}) (remember that there the prime stood for $d / dY$) together with the fact that when $c_s^2$ is small and approximately constant we have $\bar \rho \approx \bar p \, c^2 / c_s^2 \gg \bar p$ to obtain
\be \la{Yscaling1}
\fr{d G}{d \bar n} = - \fr{\bar \rho + \bar p}{\bar n} \approx - \fr{\bar p}{\bar n} \, \fr{c^2}{c_s^2}. 
\ee
Second, we can combine this result with the definition of the sound speed, $c_s^2 = c^2 \bar n \, G'' / G'$, to determine the size of the second derivative of $G$:
\be
\fr{d^2 G}{d \bar n^2} = \fr{1}{\bar n}  \fr{d G}{d \bar n}  \fr{c_s^2}{c^2}  \approx - \fr{\bar p}{\bar n^2}.
\ee
Finally, we  impose that $c_s^2$ be small (compared to $c^2$) over a large range of values of $\bar n$. This means that $\bar n \, G'' / G'$ must have only a mild dependence on the $\bar n$, or equivalently that 
\be \la{Yscaling2}
\bar n \fr{d}{d \bar n} \l( \fr{\bar n G''}{G'} \r) \approx 0  \quad \Longrightarrow \quad  \fr{d^3 G}{d \bar n^3} \propto \fr{1}{\bar n} \fr{d^2 G}{d \bar n^2} \propto \fr{\bar p}{\bar n^3},
\ee
and similarly $ G'''' \propto G'''  / \bar n$ and so on. As we can see, the derivatives of $G(Y)$ do not scale like the derivatives of $P(X)$:  $G'$ behaves like $P'$ in that they both grow like $c^2 /c_s^2$, but all other derivatives of $G$ are of the same order (in units of $\bar n$) and remain constant when $c \to \infty$. This should be contrasted with the scaling for the derivative of $P$ in equation (\ref{scalingX}).

We are now in a position to determine the canonical normalization of our fields $\vec A$ and $\vec B$. To this end, let us expand the action up to quadratic order to get
\be
\mathcal{L}_2 = - \fr{G' \bar n}{2}  (\vec \nabla \times \vec A - \tfrac{1}{c}\dot{\vec{B}})^2 + \fr{G'' \bar n^2}{2}  (\vec \nabla \cdot \vec B)^2 \propto \bar p\l[ \fr{c^2}{c_s^2}  (\vec \nabla \times \vec A - \tfrac{1}{c}\dot{\vec{B}})^2 -  (\vec \nabla \cdot \vec B)^2\r] . 
\ee
This shows that the canonically normalized fields are 
\be \la{canAB}
\vec A_c \equiv  \bar p^{1/2} \fr{c}{c_s} \vec A  , \qquad \quad \vec B_c \equiv \bar p^{1/2} \fr{\vec B}{c_s}.
\ee
These are the fields that we need to keep fixed while taking the limit $c \to \infty$. In fact, by doing so $\sqrt{Y}$ admits a well defined nonrelativistic limit, as we can see if we rewrite in terms of the canonically normalized fields:
\be
\sqrt{Y} = \bar n\l(1 - \fr{c_s \vec  \nabla \cdot \vec B_c }{ \bar p^{1/2}} \r)\l\{ 1- \fr{c_s^2}{c^2} \fr{ (\vec \nabla \times \vec A_c - \dot{\vec{B}}_c)^2}{2 \bar p (1 - c_s \vec  \nabla \cdot \vec B_c/ \bar p^{1/2})^2} + \mathcal{O}(1/c^4) \r\}
\ee
Keeping in mind the equations (\ref{Yscaling1})--(\ref{Yscaling2}), we find that in the $c \to \infty$ limit the bulk action (\ref{Sbulk}) reduces to
\be \la{NRL}
S_\text{bulk} \to \int d^4 x\l[ \fr{ (\vec \nabla \times \vec A_c - \dot{\vec{B}}_c)^2}{2 (1 - c_s \vec  \nabla \cdot \vec B_c/ \bar p^{1/2})} - \bar p \, V \l(  c_s \vec  \nabla \cdot \vec B_c/ \bar p^{1/2}\r) \r],
\ee
where we have introduced a new dimensionless function $V$ whose derivatives are all of order 1. In particular, with our parameterization we have $V'=0$ and $V''=1/2$. 

Notice that all interactions are suppressed by the same quantity, namely $ \bar p^{1/2} / c_s$. Since in the nonrelativistic limit the energy density is dominated by the rest mass, we have $\bar p = \bar \rho c_s^2 /c^2 \to m \bar n c_s^2$, which means that $ \bar p^{1/2} / c_s$ is just the square root of the mass density $\bar \rho_m \equiv m \bar n$. After all, given that sound waves are perturbations of the medium that change the local mass density, it should not come as a surprise that their self-interactions are suppressed by the  mass density at rest: local compressions are small (large) when they lead to a small (large)  \emph{relative} change in density. This argument suggests that the ratio $c_s \vec  \nabla \cdot \vec B_c/ \bar p^{1/2}$ must be related to the relative density fluctuation $\delta \rho_m / \bar \rho_m$. To find the precise relation between these two quantities, we can express the 4-velocity of the superfluid given in (\ref{UY}) in terms of the canonically normalized fields and then take the $c \to \infty$ to find that the 3-velocity~is 
\be \la{3v}
\vec u \equiv \fr{(\dot{\vec{B}}_c - \vec \nabla \times \vec A_c)}{\sqrt{\bar \rho_m}\, (1 -  \vec \nabla \cdot \vec B_c/\sqrt{\bar \rho_m}\, )}.
\ee
Using this result, we find that the Lagrangian (\ref{NRL}) takes the usual form for a nonrelativistic fluid, i.e.
\be
\mathcal{L} = \fr{\rho_m \vec{u}^2}{2} - \bar \rho_m c_s^2  V(\rho_m / \bar \rho_m),
\ee
provided we identify the mass density with
\be
\rho_m \equiv \bar \rho_m (1 - \vec \nabla \cdot \vec B_c/ \sqrt{\bar \rho_m}) \equiv \bar \rho_m + \delta \rho_m.
\ee
We have therefore recovered the non-relativistic bulk action in eq. (\ref{S^nr_bulk}).

Now that we have figured out that the correct way to implement the nonrelativistic limit is to send $c \to \infty$ while holding $\vec A_c$ and $\vec B_c$ fixed, it is easy to repeat the same procedure for the worldsheet part of the action. Let us start with the Kalb-Ramond type interaction. Keeping in mind that the coupling $\lambda$ can be expressed in terms of the circulation $\Gamma$ as shown in equation  (\ref{lambda}), our starting point becomes
\be
S_{KR} = \frac{(\bar \rho + \bar p) \Gamma}{\bar n c^2} \int d\tau d \sigma A_{\mu\nu} \d_\tau X^\mu \d_\sigma X^\nu
\ee
where we have $X^\mu = (c \, t, \vec X)$, and we have added the overall factor of $1/c^2$ based on dimensional analysis. Let us now rewrite this equation in terms of the canonically normalized fields $\vec A_c$ and $\vec B_c$ defined in (\ref{canAB}).  If we now fix the time reparameterization by imposing that $\tau \equiv t$ and use once again the fact that $\bar \rho \approx\bar \rho_m c^2$ in the nonrelativistic limit, we get
\be\la{NRKR}
S_{KR} \to \bar \rho_m \Gamma \int d t d \sigma \l[ - \fr{1}{3} \vec X \cdot \d_t \vec{X} \times \d _\sigma\vec X + \fr{1}{\sqrt{\bar \rho_m}} \l( \vec A_c \cdot \d_\sigma \vec X + \vec B_c \cdot \d_t \vec{X} \times \d _\sigma\vec X \r) \r].
\ee

\begin{figure}[htb!]
\centering%
\includegraphics[scale = 0.2]{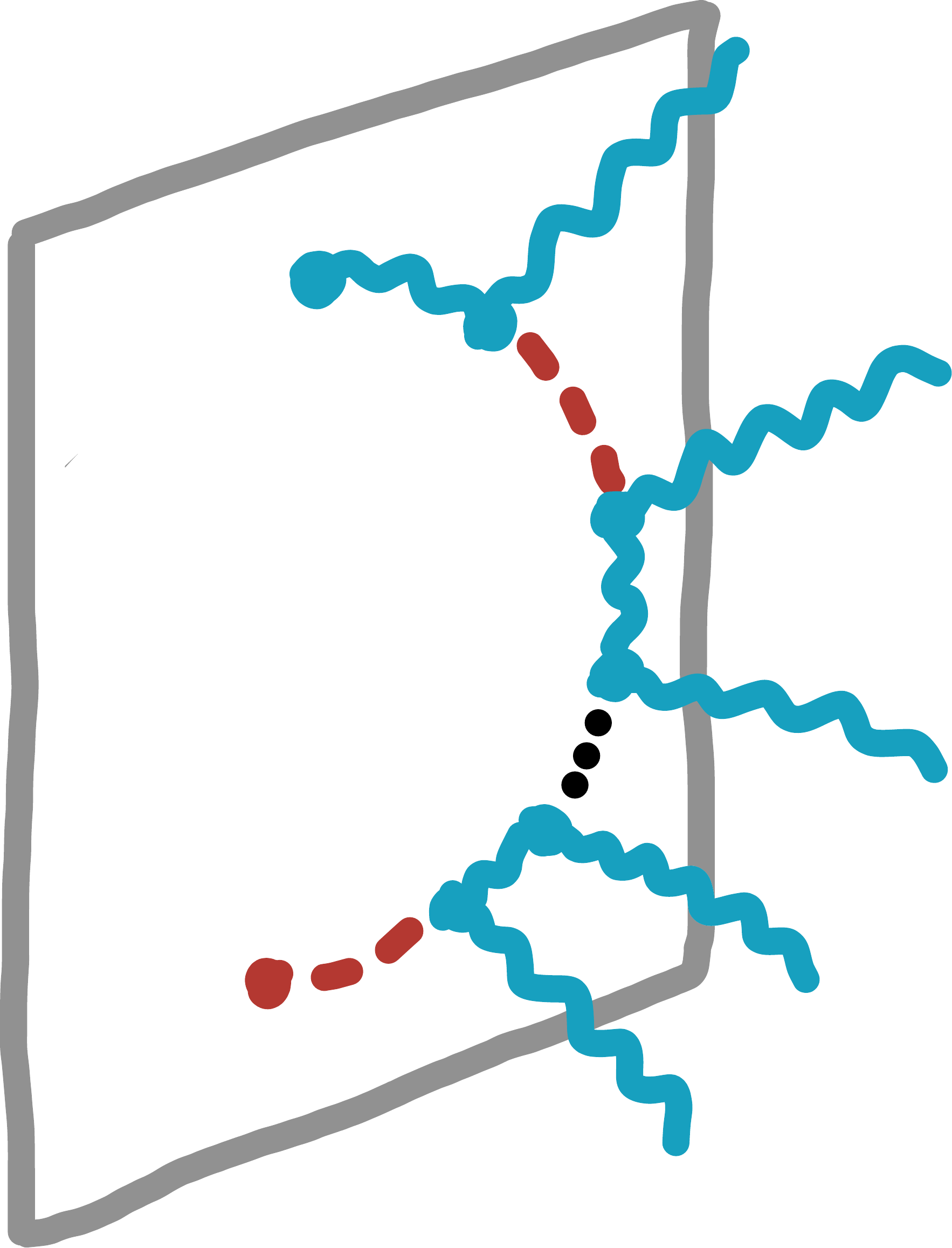}
\caption{\emph{Diagrams contributing to $T_{(n0)}$.}}
\label{fig:Tn0figure}
\end{figure}

As we can see, any dependence on the speed of light drops out and therefore the nonrelativistic limit does not bring about any simplification. Fortunately, this is not true for the rest of the worldline action, which we will see simplifies considerably when we let $c \to \infty$. We start from the generalized Nambu-Goto term
\be \la{NG'repeat}
S_{\rm NG'}  = - \int d \tau d \sigma \,  \sqrt{- \det g} \, \mathcal{T} \big( g^{\alpha\beta} h_{\alpha\beta}, Y \big) \; .
\ee
As we have seen in section \ref{runningcouplings}, this term is not only compatible with all the symmetries, but it is in fact necessary to absorb the logarithmic divergences that arise when vortex lines interact with the bulk fields. In turn, calculating the coefficients in front of these logarithmic divergences gave us a sense of what is the natural size of the function $\mathcal{T}$ and its derivatives. More specifically, we have seen that the order of magnitude of the background value of $\mathcal{T}$---which, up to a factor of $c$, is just the tension $T$---and its derivatives with respect to $Y/ \bar n$---which we denoted with $T_{(0n)}$---is
\be\la{T scaling}
T \sim T_{(0n)} \sim \fr{\bar w \Gamma^2}{c^3}.
\ee
We calculated the case of $T_{(01)}$ in detail, and the general result follows by considering the diagram where we generalize to have $n$ insertions of $(\vec \nabla \cdot \vec B)$ couple to the hydrophoton loop.
Once again, we have appropriately restored the powers of $c$ based on dimensional analysis. The derivatives of $\mathcal{T}$ with respect to $g^{\alpha\beta} h_{\alpha\beta}$ have instead a nontrivial dependence on the ratio $c_s /c$ and scale like 
\be \la{Tn0 scaling}
T_{(n0)} \sim \fr{\bar w \Gamma^2}{c^3} \l( \fr{c^2}{c_s^2}\r)^n.
\ee
Strictly speaking, in section \ref{runningcouplings} we only considered a diagram with the emission of a single phonon, and thus we only determined the magnitude of $T_{(10)}$. However, it is easy to convince oneself that a diagram with emission of $2n + 1$ phonons such as in Fig.~\ref{fig:Tn0figure}
\noindent is logarithmic divergent and must scale parametrically like in equation (\ref{Tn0 scaling}).  The running vertex gives a contribution of the form
\be
\bar{w}\Gamma^2 \int dt dz \frac{(\dot{\vec{B}}\cdot \vec{v})}{c^2_s}\left(\frac{\dot{\vec{B}}^2}{c^2_s}\right)^n
\ee
to the Lagrangian, and while the argument $\dot{\vec{B}}^2$ appears in both $g^{\alpha \beta}h_{\alpha \beta}$ and $Y$, the extra factor of $1/c^2_s$ together with the result in eq.~\eqref{T scaling} indicates that the derivative must be taken with respect to the first argument.

The scaling relations (\ref{T scaling}) and (\ref{Tn0 scaling}) suggest that we extract an overall factor of $ \bar w \Gamma^2 /c^3$ from the function $\mathcal{T}$. Then, we can expand the quantities appearing in (\ref{NG'repeat}) for large values of $c$ as
\begin{subequations}
\ba
\sqrt{- \det g} &=& c | \vec X '| + \mathcal{O}(1/c) \\
 g^{\alpha\beta} h_{\alpha\beta} &=& - 1 - \fr{(\dot{\vec X}- \vec u)^2}{c^2} + \fr{[(\dot{\vec X} - \vec u )\cdot \vec X']^2}{c^2 |\vec X' |^2}   + \mathcal{O}(1/c^4)\\
 Y / \bar n^2 &=& (1 - \vec  \nabla \cdot \vec B_c/\sqrt{\bar \rho_m})^2 + \mathcal{O}(1/c^2),
\ea 
\end{subequations}
where $\vec u$ is the 3-velocity of the superfluid defined in (\ref{3v}), and then combine these results with $\bar w \approx \bar \rho_m c^2$ to find that when $c \to \infty$ the generalized Nambu-Goto term reduces to
\be \la{NRNG'}
S_{NG'} \to  \bar \rho_m \Gamma^2 \int d t d \sigma | \vec X '| \, f \l(  \fr{(\dot{\vec X}_\perp- \vec u_\perp)^2}{c_s^2} , \fr{\vec  \nabla \cdot \vec B_c}{\sqrt{\bar \rho_m}} \r),
\ee
where we have introduced a dimensionless function $f$ whose magnitude and derivatives are all of order $1$. We have also simplified the notation by denoting with $\dot{\vec X}_\perp$ and $\vec u_\perp$ the components of $\dot{\vec X}$ and $\vec u$ perpendicular to the vortex line.

%\bibliographystyle{ieeetr}
%\bibliography{biblio}

\end{fmffile}

\end{document}